\documentclass[12pt]{article}

\usepackage{graphicx}
\usepackage{epstopdf}
\usepackage{ccaption}
\usepackage[usenames]{color}
\usepackage[centertags]{amsmath}
\usepackage{amssymb}
\usepackage[colorlinks=true,
      linkcolor=black,
      citecolor=blue,
      urlcolor=blue,
      filecolor=blue,
      pdfstartview=FitV,
      pdftitle={},
        pdfauthor={},
        pdfsubject={black rings},
        pdfkeywords={black rings, extremal},
        pdfpagemode=None,
        bookmarksopen=true
      ]{hyperref}

\makeatletter
\@addtoreset{equation}{section}

\makeatletter
\renewcommand\section{\@startsection {section}{1}{\z@}%
                                   {-3.5ex \@plus -1ex \@minus -.2ex}
                                   {2.3ex \@plus.2ex}%
                                   {\normalfont\large\bfseries}}
\renewcommand\subsection{\@startsection{subsection}{2}{\z@}%
                                     {-3.25ex\@plus -1ex \@minus -.2ex}%
                                     {1.5ex \@plus .2ex}%
                                     {\normalfont\bfseries}}

  \captionnamefont{\bfseries}
  \captiontitlefont{\small\sffamily}
  \captiondelim{: }
  \hangcaption

\def\baselinestretch{1.2}
\def\sec#1{\S \;\ref{#1}}
\def\fig#1{Fig.\,\ref{#1}}
\def\req#1{(\ref{#1})}
\def\App#1{Appendix \ref{#1}}

\def\eg{{\it e.g.}}

\def\cf{{\it cf.}}
\def\ie{{\it i.e.}}

\def\viz{{\it viz.}}

\def\etc{{\it etc.}}

\def\thus{\Longrightarrow}

\def\CA{{\cal A}}

\def\CE{{\cal E}}

\def\CH{{\cal H}}

\def\CN{{\cal N}}
\def\CO{{\cal O}}

\def\be{\begin{equation}}
\def\ee{\end{equation}}
\def\bea{\begin{eqnarray}}
\def\eea{\end{eqnarray}}

\newcommand{\const}{\,{\rm const}\,}

\def\Sp{{\bf S}}
\def\R{{\bf R}}
\def\ads#1{AdS$_{#1}$}

\def\AH{\CA_{\textrm H}}
\def\aH{a_{\textrm H}}

\title{\bf{Extremal vacuum black holes in \\ higher dimensions}}

\author{Pau Figueras$^{a}$\footnote{pau.figueras@durham.ac.uk}, \
 Hari K. Kunduri$^b$\footnote{h.k.kunduri@damtp.cam.ac.uk }, \ James Lucietti$^a$\footnote{james.lucietti@durham.ac.uk }, \ Mukund Rangamani$^a$\footnote{mukund.rangamani@durham.ac.uk} \\ \\
\small \sl $^a$Centre for Particle Theory, Department of Mathematical Sciences, \\ \small \sl University of Durham, South Road, Durham, DH1 3LE, UK\\ \\
\small \sl $^b$DAMTP, University of Cambridge \\
\small \sl Centre for Mathematical Sciences, Wilbeforce Road, Cambridge, CB3 0WA, UK  }

\date{}

\begin{document}

\setlength{\baselineskip}{16pt}
\begin{titlepage}
\maketitle
\begin{picture}(0,0)(0,0)
\put(350, 320){DCPT-08/17}
\put(350, 305){DAMTP-2008-22}
\end{picture}
\vspace{-36pt}

\begin{abstract}
We consider extremal black hole solutions to the vacuum Einstein
equations in dimensions greater than five. We prove that the
near-horizon geometry of any such black hole must possess an $SO(2,1)$
symmetry in a special case where one has an enhanced rotational
symmetry group. We construct examples of vacuum near-horizon
geometries using the extremal Myers-Perry black holes and boosted
Myers-Perry strings. The latter lead to near-horizon geometries of black ring topology, which in odd spacetime dimensions have the correct number rotational symmetries to describe an asymptotically flat black object. We argue
that a subset of these correspond to the near-horizon limit of
asymptotically flat extremal black rings. Using this identification we provide a
conjecture for the exact ``phase diagram'' of extremal vacuum black rings with a connected horizon in odd spacetime dimensions greater than five.

\end{abstract} \thispagestyle{empty} \setcounter{page}{0}
\end{titlepage}

\renewcommand{\baselinestretch}{1.2}  
\tableofcontents
\section{Introduction}
 \label{intro}

Exact black hole solutions to Einstein's equations are notoriously
difficult to find. In recent years remarkable  progress has been
made in five dimensions, starting with the important work of
\cite{Emparan:2001wk} on Weyl solutions.  This classification
inspired the discovery of the black ring \cite{Emparan:2001wn}, an
exact solution describing a black hole with horizon topology
$\Sp^1\times \Sp^2$ (\cf\ \cite{Emparan:2006mm} for a review).
Subsequently, for vacuum gravity, the existence of more systematic
techniques has led to the construction of a number of new solutions.
In particular, the Belinsky-Zhakarov method
\cite{Belinsky:1971nt,Belinsky:1979mh}, which employs ideas from
integrable systems such as inverse-scattering, has been exploited to
find many novel black hole configurations. These methods have been
used to construct a generalization of the black ring that carries
two independent angular momenta \cite{Mishima:2005id,Figueras:2005zp,Tomizawa:2005wv,Pomeransky:2006bd}, the `black Saturn'
\cite{Elvang:2007rd} (a multi-black hole solution consisting of a
rotating black ring with a spinning black hole at its centre) and
other solutions with multiple horizons such as the di-ring
\cite{Iguchi:2007is,Evslin:2007fv}, the bi-ring \cite{Izumi:2007qx,
Elvang:2007hs}, \etc. The recent reviews \cite{Emparan:2008eg,
Obers:2008pj} summarize our current understanding of higher
dimensional black holes.\footnote{See also \cite{Harmark:2004rm} for generalizations  of the Weyl technique to non-static solutions.}

Unfortunately such techniques do not extend beyond five spacetime
dimensions, if one is interested in stationary asymptotically flat spacetimes.
This is due to the fact that these techniques rely on the
integrability of Einstein's equations when restricted to $D$-dimensional
stationary spacetimes with a spatial $U(1)^{D-3}$ symmetry group.
For $D=4,5$ this coincides with the dimension of the maximal abelian
rotational group (which must be a subgroup of $SO(D-1)$). However, for $D>5$ this is too much (abelian)
symmetry to describe a localized object. As a result, very little is
known about higher dimensional asymptotically flat black hole
solutions. In particular, the only exact black hole solutions known
in the vacuum are the Myers-Perry (MP) black holes \cite{Myers:1986un} which have a
spherical horizon topology. Therefore black hole non-uniqueness
still remains to be proved for $D>5$; however, see \cite{Emparan:2007wm} for compelling arguments for non-uniqueness. Moreover, one expects other possible horizon topologies to exist; in particular, higher dimensional versions of black rings, \ie, black holes with horizon topology $\Sp^1 \times \Sp^{D-3}$. In fact, in the analysis of balanced black rings of \cite{Hovdebo:2006jy,Elvang:2006dd} it is argued that such solutions exist. Further, the recent
perturbative analysis of \cite{Emparan:2007wm} constructs approximate
solutions and infers properties of the `phase diagram' to argue for
non-uniqueness in dimensions $D \ge 6$ with one non-zero angular
momentum.

It is of course well known that a useful tool to aid construction of
exact solutions to Einstein's equations is supersymmetry. Indeed,
the exploration of supersymmetric black holes/branes has a rich
history and in recent years the systematic classification programme
has led to the discovery of many interesting new solutions
\cite{Gauntlett:2002nw,Reall:2002bh,Gauntlett:2003fk,Gutowski:2004ez,Gutowski:2004yv,Gutowski:2004bj,Gaiotto:2005gf,Elvang:2004rt,Gaiotto:2005xt,Bena:2004de,Gauntlett:2004wh,
Elvang:2004ds,Gauntlett:2004qy,Elvang:2005sa,Bena:2005ni,Kunduri:2006ek,Kunduri:2006uh,Kunduri:2007qy}. A supersymmetric black hole is necessarily extremal. It has
emerged that in fact it is this latter property which is responsible
for the black hole attractor mechanism (\cf, \cite{Sen:2007qy} for a
review of recent developments in this area). This is essentially due
to the fact that the near-horizon limit of an extremal black hole in
$D=4,5$ always has an $SO(2,1)$ symmetry group
\cite{Bardeen:1999px,Kunduri:2007vf}.\footnote{Intuitively, extremal
black holes have a long throat-like region, with the horizon located
at an infinite proper distance from the any point outside the black
hole. This suffices to erase the memory of the asymptotic boundary
conditions as one approaches the near-horizon. Hence one expects the
attractor mechanism to also hold in $D\ge6$.} While vacuum black
holes can never be supersymmetric, they can be extremal (\eg,
extremal Kerr in four dimensions). Therefore, a natural simplifying
assumption to study higher dimensional vacuum black holes is to
focus on extremal solutions. This will be the strategy employed in
this paper.

One
might be concerned that the assumption of extremality is too strong to reveal
any interesting aspects of black hole solutions in higher
dimensions. In five dimensions, where we have exact solutions, this
is not the case -- one still has both topologically spherical black holes and
black rings. In particular, the  known extremal vacuum black holes
consist of the extremal MP solution (which must have two non-zero
angular momenta) as well an extremal black ring (which also must have
two non-zero angular momenta). Curiously, black hole uniqueness is
not actually violated for these extremal solutions, in contrast to
the non-extremal case \cite{Elvang:2007hs}.  In any case the restriction to extremal black holes is not overly constraining. Another motivation for studying
extremal, non-supersymmetric black holes, comes from the fact that
recently there has been progress in providing exact microscopic
counting of their entropy \cite{Emparan:2004wy,Emparan:2006it,Emparan:2007en,Horowitz:2007xq,Reall:2007jv,Emparan:2008qn}.\footnote{The underlying
reason responsible for this seems to be the attractor mechanism,
which has been argued to guarantee that the entropy of the black
hole should not vary as we move in the moduli space (for instance by tuning the string coupling) \cite{Dabholkar:2006tb} (see also \cite{Astefanesei:2006sy}). Indeed, the assumption of $SO(2,1)$ symmetry in the near-horizon limit (subsequently proved in \cite{Kunduri:2007vf}) has been used to establish an attractor mechanism for rotating extremal black holes in $D=4,5$ \cite{Astefanesei:2006dd} (see also \cite{Astefanesei:2007bf}).}

In five dimensions, it turns out that the near-horizon geometry of
the vacuum extremal black ring is isometric to a special case of the
near-horizon geometry of the boosted Kerr-string
\cite{Kunduri:2007vf} (the boost is such that the string is
tensionless). It should be emphasized that this is a statement
regarding  {\it exact solutions} to the vacuum Einstein equations.
The usual relation between boosted black strings and black rings
occurs only in the limit of large thin rings. This is due to the
obvious qualitative expectation that a large thin ring locally
``looks like'' a black string. However, the statement about
near-horizon geometries of the extremal solutions is actually valid
at the full non-linear level. It is this intriguing connection which
we will exploit to make progress in higher dimensions.

One of our results
will be to give the explicit near-horizon geometries of the general extremal MP black holes in $D>5$ and the boosted MP black string in $D=2n+3$. These solutions possess the typical
$SO(2,1)$ symmetry which occurs for $D=4,5$ black holes \cite{Bardeen:1999px,Kunduri:2007vf}. Spatial
sections of the horizon of the MP string near-horizon geometry are of $\Sp^1 \times \Sp^{2n}$ topology, thus providing a candidate black ring near-horizon geometry
in higher dimensions. Indeed, we argue that the near-horizon geometry of the
MP black string is isometric to that of an asymptotically flat extremal black ring in $D=2n+3$
spacetime dimensions, provided one takes a particular value for the boost
parameter. This value must be such that the MP string is tensionless. Our construction only works in odd dimensions, since a $d=2n+2$ dimensional MP solution times a line has the correct number of abelian symmetries to describe an asymptotically flat  black object with a compactly generated horizon in $D=d+1$ dimensions (this is not the case for an odd dimensional MP black holes).

To support this proposal we will also prove a technical result
regarding the symmetries of near-horizon geometries of $D>5$
dimensional black holes. We expect that the $SO(2,1)$ symmetry which
occurs for $D=4,5$ black holes generalizes to higher dimensional
black holes. However, this general statement seems difficult to
prove. As a step towards this we establish that, in the special case
where the rotational symmetry group is enhanced to $U(1) \times
SU(n)$ (in $2n+2$ dimensions) or $U(1)^2 \times SU(n)$ (in $2n+3$
dimensions), one must have an $SO(2,1)$ symmetry. The proof directly
generalizes the one used in $D=4,5$ \cite{Kunduri:2007vf}, and uses dynamical information (\ie, certain components of the Ricci tensor must vanish). The
result is valid for vacuum solutions (even with a cosmological
constant) and we expect it can be generalized to include matter as
was done in $D=4,5$. Such an enhancement of rotational symmetry
typically occurs when one is dealing with black holes with equal
angular momenta in even dimensions, or all but one angular momenta equal in odd
dimensions.

The identification of the near-horizon of the black ring in odd $D >
5$ will be incomplete unless we can use it to provide a detailed
analysis of the physical parameters of the solution. The general
picture of the attractor mechanism would suggest that we should be
able to recover all the conserved charges from the near-horizon
geometry \cite{Suryanarayana:2007rk,Hanaki:2007mb}.\footnote{In
these analyses, the authors considered supersymmetric black holes,
in which case one also has the BPS relation fixing the mass. For
black holes with spherical topology, this will suffice to compute
the charges. Our interest, however, lies in generic extremal
situations where the near-horizon geometry is the only available
data.} As we discuss this expectation is not true in general; one
needs more information than the near-horizon geometry to identify
the conserved quantities of a black hole solution. We will clarify
exactly what data one should expect to be able to compute from a
near-horizon geometry. For example, obtaining the mass requires
knowledge of the asymptotic stationary Killing field, which is not
contained in the near-horizon geometry. Likewise knowledge of the
asymptotic stationary frame is required to identify the angular
velocities of the black hole/ring.

While this presents an obstacle to identifying the physical
parameters of the putative extremal black rings in $D>5$, we argue
that it is nevertheless possible to deduce the necessary quantities
by employing extra information at hand. For instance, to determine
the mass we use knowledge of the mass of the corresponding MP
string. We are able to recover certain components of the conserved
angular momenta from the near-horizon geometry, \ie, the components
orthogonal (at asymptotic infinity) to the $\Sp^1$ of the ring.
However, the angular momentum along the $\Sp^1$ is not as
straightforward. There is an important difference between a black
ring and a black string -- the former carries angular momentum,
while the latter carries linear momentum. Heuristically, one must
``fold" the latter (in the thin ring limit) to convert it to a black
ring. Here we encounter a crucial issue; the direction along the
string is not necessarily the $\Sp^1$ direction of the ring. To
complete the specification of the physical parameters, we need to
infer some aspects regarding the asymptotic geometry, in particular,
the plane in which the ring rotates. Using the fact that thin black
rings look like black strings, we argue that in $D>5$ certain
regularity conditions force the direction along the string to be the
same as the direction along the ring. This allows us to complete the 
determination of all the physical parameters of the extremal black
rings. Based on this identification we present results for the phase
diagram of these solutions in various dimensions.

The organization of this paper is as follows: we begin in
\sec{nhgeoms} with a general analysis of near-horizon geometries for
extremal solutions. Apart from proving (under certain restrictions) that the near-horizon geometry has an enhanced symmetry, we present examples of vacuum near-horizon geometries in $D>5$ and we also discuss in detail the
parameters we can extract from the knowledge of the near-horizon geometry alone. In \sec{extrmpbh} we discuss properties of extremal MP black holes in various dimensions focusing on the ``moduli space'' of extremal solutions. We extend this analysis to extremal MP black strings in \sec{extrmpbstr}. In \sec{fivedextr} we discuss the known extremal doubly spinning black ring. In \sec{hdrings} we argue that we have the exact near-horizon geometries of the yet to be found asymptotically flat extremal black rings in $D=2n+3$ and analyze their physical properties. Finally, we conclude in \sec{discuss} with a discussion. In \App{Adetnh} we
collect some technical details that are useful in extracting the
near-horizon geometries of known extremal solutions.

\section{Near-horizon geometries}
\label{nhgeoms}

We begin in \sec{gnull} by introducing a useful coordinate chart valid in the neighbourhood of the event horizon of a stationary black hole solution: the {\it Gaussian null coordinates} \cite{Isenberg:1983cc,Reall:2002bh}.\footnote{These are the analog of Gaussian normal coordinates used in the neighbourhood of a spacelike hypersurface.} We will then prove
in \sec{adssym} a general result regarding the symmetries of
near-horizon geometries in higher than five dimensions. Following this technical result, we will  present the near-horizon metrics for MP black holes and
black-strings in \sec{mpbhnh} and \sec{mpbsnh} as examples of
near-horizon geometries with spherical and ring-like topology.
Finally, in \sec{nhphys} we discuss what physical quantities can be
expected to be extracted from the knowledge of a near-horizon
geometry alone. A short word on notation: We will use $D$ to denote
the total spacetime dimension for black rings and black strings and
$d$ to denote the dimension of the black holes used to construct the strings, so that $D
= d+ 1$. For convenience, we
will also use a variable $n$, related to the spacetime dimension by
$D =2 n +3$.

\subsection{Gaussian null coordinates and the near-horizon limit}
\label{gnull}

In the neighbourhood of the event horizon of a stationary black hole
one can introduce coordinates $x^{\mu}=(v,r,x^a)$ such that the
metric takes the form \cite{Reall:2002bh}:
\be ds^2= r\, f(r,x) \,
dv^2 +2\, dv\, dr+ 2\,r \,h_a(r,x)\,dv\,dx^a +\gamma_{ab}(r,x)\, dx^a
dx^b. \label{bhnhc} \ee
The horizon is located at $r=0$, and
$\frac{\partial}{\partial v}$ is a Killing field normal to the
horizon. The exterior of the black hole is the region $r>0$. Spatial sections of the horizon consist of a
compact manifold $\CH$, with coordinates $x^a$, equipped with a
Riemannian metric $\gamma_{ab}(0,x)$. The surface gravity of the
horizon is given by
\begin{equation}
\kappa= -\frac{1}{2}\, f\big|_{r=0} \; .\label{surfgrav}
\end{equation}

By definition, an extremal black hole is one with vanishing surface
gravity. Therefore, assuming analyticity, it follows that
$f(r,x)=r\, F(r,x)$; this implies $g_{vv}=\CO\left(r^2\right)$. From
this it follows that the horizon is an infinite proper distance from any point outside the black hole. This allows us to zoom into the vicinity of
the horizon in a particular way and define a consistent near-horizon
limit; the limiting geometry explicitly solves Einstein's equations \cite{Reall:2002bh}.

For an extremal black hole one may define the near-horizon limit\footnote{Note this is analogous to the concept of a double scaling limit.} by: $r\to \epsilon \, r$, $v \to v/\epsilon$ where we take $\epsilon \to 0$. The resulting geometry (in the rescaled coordinates) is:
\be
ds^2= r^2\, F(x)\, dv^2 +2\, dv\,dr + 2\, r\, h_a(x) \, dv\,dx^a +\gamma_{ab}(x)\, dx^a \,dx^b
\label{nhgeom}
\ee
where $F(x)=F(0,x)$ \etc, and will be referred to as the
near-horizon geometry of the black hole. It is guaranteed to solve
the same equations of motion as the full black hole solution (since
it is just a well-defined limit of the original metric). We should
emphasize that the near-horizon limit only exists because
$g_{vv}=\CO\left(r^2\right)$ (\ie, from the extremality condition).

Similar scalings do not exist for the non-extremal solutions,
although we note that if $\kappa \neq 0$ then to leading order
``near" the horizon (\ie, for small $r$) one has
\be ds^2 \sim
-2\, \kappa \, r \, dv^2 +2\,dv\, dr +2\,r\,h_a(x) \,dv \,dx^a +\gamma_{ab}(x)\,
dx^a \,dx^b .
\label{nonextnh}
\ee
The first two terms are Rindler space written in
$(v,r)$ coordinates, which is familiar from studies of black hole thermodynamics. However, observe that the metric \req{nonextnh} is not
a solution of the field equations as it is not obtained via a limit
of a sequence of metrics, but rather just truncating the Taylor
expansion in $r$ of the metric coefficients.

\subsection{Symmetry enhancement in the near-horizon limit:  $SO(2,1)$}
\label{adssym}

The near-horizon geometry \req{nhgeom} has an enhanced symmetry, relative to the black hole, generated by $v\to \lambda\, v$ and $r \to r/ \lambda$, which when combined with translations in $v$ forms a 2d non-abelian group $\mathcal{G}_2$.\footnote{The generators of $\mathcal{G}_2$ are of course $v\left(\frac{\partial}{\partial v}\right) -r\left(\frac{\partial}{\partial r}\right)$ and $\left(\frac{\partial}{\partial v}\right) $, respectively. } In $D=4,5$ if one assumes the presence of $U(1)^{D-3}$ rotational symmetries one can prove that the Einstein's  equations (in a fairly general class of theories) imply that the symmetry is further enhanced to $SO(2,1)\times U(1)^{D-3}$ \cite{Kunduri:2007vf}. The existence of the $U(1)^{D-3}$ symmetry allows us to introduce coordinates on the horizon $x^a=(\rho,x^i)$, where $ i=1, \cdots , D-3$, such that the $\frac{\partial}{\partial x^i}$ are Killing fields. The near-horizon geometry can then be written in a particularly simple form:
\begin{equation}
 ds^2 = \Gamma(\rho)\, \left(A_{0}\, r^2\, dv^2 + 2\, dvdr\right)+ d\rho^2 + \gamma_{ij}(\rho)\,\left(dx^{i} + k^i \,r \, dv\right)\, \left(dx^{j} + k^j \,r \, dv \right)
\label{nhsym}
\end{equation}
 where $\Gamma(\rho)$ is a positive function of the coordinate $\rho$ introduced on the horizon, while $A_{0}$ and $k^i$ are constants. Note that the generators of the rotational symmetries will in general be linear combinations of the $\frac{\partial}{\partial x^i}$ and thus these need not have closed orbits.

An open question is whether the same occurs for $D>5$ when one
assumes\footnote{We use $[x]$ to denote the integer part of
$x$.} $[\frac{D-1}{2}]$ rotational abelian symmetries;\footnote{In fact there is
the possibility of having black holes with less symmetry than this
since the rigidity theorem \cite{Hollands:2006rj} only guarantees the existence on one rotational isometry for generic stationary solutions.} this is the maximal abelian subgroup of $SO(D-1)$. We will not be able to address this general question here. However, we are able to prove the existence of the $SO(2,1)$
symmetry in a special case where the rotational symmetry group
enhances to $U(1) \times SU(n)$ in $2n+2$ dimensions and $U(1)^2
\times SU(n)$ in $2n+3$ dimensions. Such a symmetry typically\footnote{Note that there
is no general proof of this statement, it is merely based on the
observation it occurs for all known solutions.}  occurs
when all the angular momenta are set equal in $2n+2$ dimensions, or
all but one are equal in $2n+3$ dimensions. The rest of this
section will be dedicated to the proof of this assertion. Readers interested in the physical aspects of the extremal solutions can jump directly to the examples discussed in \sec{nhexamples}.

For the sake of generality, consider horizons $\CH$ equipped with a metric $\gamma_{ab}$ with an isometry group $U(1)^m \times G$ whose orbits are generically cohomogeneity-1 $T^m$ (torus) fibrations over a homogeneous space $M=G/H$. We will also assume that the rest of the near-horizon data $F,h_a$ is also invariant under this symmetry. Then the full near-horizon geometry has an isometry group $\mathcal{G}_2\times U(1)^m \times G$ with $D-1$ dimensional orbits, and hence is also cohomogeneity-1. We
restrict our attention to homogeneous spaces $M$ that admit a unique
(up to homothety) $G$-invariant metric,\footnote{This is true for
$CP^{N}=SU(N+1)/SU(N)$ and $\Sp^{N}=SO(N+1)/SO(N)$ for example.} which we denote
by $\bar{g}$. It follows that $M$ cannot admit any $G$-invariant one
forms, for if such a one-form $\alpha$ existed, then $\bar{g} +
\lambda \,\alpha^2$ would yield a one-parameter ($\lambda$) family
of $G$-invariant metrics contradicting our original assumption.
However, $G$-invariant closed two-forms $J$ may exist. Suppose $K$
such two forms $J^{I}$ (where $I \in\{0,1, \cdots , K\}$) exist and
denote their respective potentials by $A^{I}$, so $J^{I}=dA^{I}$.
Note that the $A^{I}$ are invariant under $G$ up to a gauge
transformation. Now, one can introduce coordinates on the horizon
$x^a=(\rho,x^i,y^p)$, where $x^{i}$ are adapted to $U(1)^m$
Killing fields and $y^p$ are coordinates on $M$. Then,  the most
general $U(1)^{m}\times G$ invariant metric on the horizon can be
written as:
\begin{equation}
\gamma_{ab}\, dx^a\, dx^b = d\rho^2 + \gamma_{ij}(\rho)\,\sigma^{i}\,\sigma^{j} + R(\rho)\,\bar{g}_{pq}(y)\,dy^{p}\,dy^{q}
\end{equation}
where $\sigma^{i}=dx^{i}+A^{i}(y)$ and $A^i$ are some linear
combination of the $A^{I}$ \ie, $A^i = \sum_I \, \alpha^i_I \, A^I$. Note the action of $G$ on the $A^{i}$ shifts them by a total derivative which can  be compensated by a shift
in the $x^{i}$  to ensure $G$-invariance.
Observe that the $\sigma^{i}$ are the only $G$-invariant one forms
on $\CH$. Hence the most general $U(1)^{m}\times G$ invariant one-form on the horizon can be written as
\begin{equation}
h = -\frac{\Gamma'}{\Gamma}\,d\rho +\Gamma^{-1}\,  k_{i}(\rho)\, \sigma^{i}
\end{equation}
 where $\Gamma = \Gamma(\rho) >0$.

Performing the coordinate transformation $r \to \Gamma(\rho)\, r$ in the near-horizon metric \req{nhgeom} allows us to write it as
\bea
ds^2 &=& A(\rho) \,r^2 \,dv^2 +2\,\Gamma(\rho) \,dv\,dr + d\rho^2+ R(\rho) \, \bar{g}_{pq}(y) \,dy^{p} dy^{q}\nonumber \\
& &  +\; \gamma_{ij}(\rho) \, \left(\sigma^i+ k^i(\rho)\, r\,dv\right) \, \left(\sigma^j+k^j(\rho)\,r\,dv\right)
\label{simpGmet}
\eea
where $A=\Gamma^2 \,F- k^i\,k_i$ and $k^i = \gamma^{ij} \, k_j$. We are now ready to
present our result:

\paragraph{Theorem:} Consider a cohomogeneity-1 near-horizon geometry with an isometry group  $\mathcal{G}_2
\times U(1)^m \times G$, whose orbits on the horizon are generically cohomogeneity-1 $T^{m}$ fibrations over a homogeneous
space $M=G/H$. Suppose $M$ admits a unique (up to homothety)
$G$-invariant metric.  Introduce coordinates $(v,r,\rho,x^i,y^p)$ as described
above in \req{simpGmet}. If $R_{\rho i}=0$ and $R_{\rho v}=0$, then $k^i$ are
constants and $A=A_0 \,\Gamma$ for some constant $A_0$. If $A_0 \neq
0$ the near-horizon geometry then possesses $O(2,1) \times U(1)^m
\times G$ symmetry, where the $O(2,1)$ has 3-dimensional orbits if
$k^i\neq 0$ and 2-dimensional orbits otherwise. If $A_0=0$ then
$O(2,1)$ is replaced by the 2d Poincare group.

\paragraph{Proof:} Direct calculation gives
\be
R_{\rho i}= \frac{1}{2\,\Gamma} \, \gamma_{ij} \, (k^j)', \qquad R_{\rho v}= \frac{r}{\Gamma}\, \left( A'-\frac{\Gamma' \,A}{\Gamma}+ k_i \,(k^i)' \right)
\ee
and thus the vanishing of these components implies $k^i$ are constants and $A=A_0 \, \Gamma$ for some constant $A_0$. The near-horizon metric \req{simpGmet} then simplifies to
\bea
\label{symNH}
ds^2 &=& \Gamma(\rho)\, \left[A_0 \, r^2 \, dv^2 +2\, dv\, dr\right] + d\rho^2+ R(\rho) \, \bar{g}_{pq}(y) \, dy^{p} \, dy^{q} \nonumber \\
& &  \; + \gamma_{ij}(\rho) \; \left(\sigma^i+ k^i\, r\, dv \right) \, \left(\sigma^j+ k^j\, r\, dv\right)
\eea
The 2d maximally symmetric space in the square brackets
is: (i) AdS$_2$ when $A_0<0$ or (ii) dS$_2$ when $A_0>0$, both of which have $O(2,1)$ isometry groups, or
(iii) $R^{1,1}$ if $A_0=0$ which has 2d
Poincare symmetry.
The volume form of this 2d space is $dv \wedge
dr$ and thus under all its isometries  $r\,dv \to \pm( r\,dv +
d\phi)$ where $\phi(r,v)$ is some function (the choice of sign
depends on whether the isometry is orientation preserving or not).
Thus if under the isometries of the 2d space we also shift $x^i \to
\pm (x^i-k^i\,\phi)$, then the full near-horizon metric is invariant
under $O(2,1)$ if $A_0 \neq 0$ or the 2d Poincare group if $A_0=0$.
Further, if $k^i \neq 0$ the isometry of the 2d space has 3d orbits due
to the shifts in $x^i$.
\paragraph{Remarks:}
\begin{itemize}
\item Clearly the conditions of the above theorem are fulfilled for vacuum solutions with or without a cosmological constant. Further it is easy to show it is also valid when one couples scalar fields with a potential. Presumably it can also be established in the presence of Maxwell fields but we shall not pursue this here.

\item We can prove that $A_0 \leq 0$ (which rules out the dS$_2$ case) and argue that only the AdS$_2$ ($A_0<0$) case is relevant to black holes as follows~\cite{Kunduri:2007vf}. The $vr$ component of the Ricci tensor of the near-horizon metric (\ref{symNH}) is $R_{vr}=A_0+\frac{k^ik_i}{2\Gamma}-\frac{1}{2}\hat{\nabla}^2\Gamma$ where $\hat{\nabla}$ is the metric connection on $\mathcal{H}$. Integrate this over $\CH$ to conclude that, for a vacuum solution, $A_0 \leq 0$ with equality iff $k^i \equiv 0$ (in which case  $\Gamma$ is harmonic on a compact space $\mathcal{H}$ and thus a constant). Therefore the case $A_0=0$ is a direct product of $\R^{1,1}$ and a Ricci flat metric on $\CH$ which is a static near-horizon geometry. In fact, the general analysis of static vacuum near-horizon geometries
in~\cite{Chrusciel:2005pa} shows that they are always (\ie\ without the assumption of rotational symmetry) a direct product of $\R^{1,1}$ and a Ricci flat
metric on $\CH$. In $D=4,5$ it follows $\CH$ is a torus with a flat metric which one discounts in view of the horizon topology theorems. In $D>5$ a new feature arises as one can have non-trivial $D-2$ dimensional Ricci flat compact manifolds. Generically Ricci flat metrics on compact spaces do not have continuous isometries, except for flat torii factors. However, such a situation is incompatible with our cohomogeneity-1 assumption, as the only way to have the correct number of abelian rotational symmetries (\ie, $[(D-1)/2]$) would be to have higher cohomogeneity metrics.\footnote{If one relaxes the cohomogeneity-1 assumption these static cases can be made compatible with the correct number of rotational symmetries (for high enough $D$). However, none of these symmetries would have fixed points. While we are not aware of a theorem stating such horizons are incompatible with asymptotic flatness, we expect such horizons not to occur for asymptotically flat black holes.}. Note that with a negative cosmological constant one must have $A_0<0$.

\item The case relevant to vacuum black holes in $2n+2$ and $2n+3$ dimensions is when $m=1$ and $m=2$ respectively (corresponding to $\Sp^1$ or $T^2$ fibrations) and $M=CP^{n-1}$ so $G=SU(n)$. From cohomology we know there is a unique $G$-invariant closed 2-form, the K\"ahler form $J$, and hence $J^{i}=C^{i}\,J$ for some constants $C^{i}$. All known black hole solutions with these symmetries have equal angular momenta in even dimensions ($m=1$) or all but one equal in odd dimensions
($m=2$).

\item The case $m=1,2$ with $M=\Sp^{2n-2}$ is relevant to black holes spinning in a single plane in $2n+2$ and $2n+3$ dimensions respectively. While in the vacuum there are no known extremal solutions of this type, in the presence of charge one can have near-horizon geometries of this type. It is easy to see from cohomology that there are no closed two-forms when $n>2$; in this case $J^I=0$.

\item The staticity conditions of the near-horizon metric (\ref{symNH}) imply either: (i) $k^i=0$, or (ii) $A_0\, \Gamma=-k^i\, k_i$,  and $k_i=\Gamma \,\bar{k}_i$ with $\bar{k}_i$ constant and $J^i\bar{k}_i=0$. Case (i) leads to near-horizon geometries which are warped products of AdS$_2$ with $\CH$. Case (ii) is more interesting as one can show that in this case the near-horizon geometry is a warped product of AdS$_3$ with a compact space. This will be of relevance if one is considering non-vacuum black ring near-horizon geometries since these can be static~\cite{Reall:2002bh,Elvang:2004rt}.
\end{itemize}

\subsection{Examples of near-horizon metrics}
\label{nhexamples}

 In this section we will write down examples of vacuum near-horizon geometries and separate them according to horizon topology. We will in fact present some solutions with ring-like topology.

\subsubsection{Spherical topology horizons}
\label{mpbhnh}

The only known vacuum solutions in this class are the MP black holes. 
Thus, consider the extremal MP solution in $2\,n+1$ and $2\, n+2$ dimensions. \footnote{In the initial part of the section alone will we discuss black holes in $2n+1$ odd dimensions, for notational convenience. Later on we will exclusively focus on $D = 2n+3$ dimensions.} These solutions are stationary and have $n$ rotational isometries, so the isometry group is $\R_t \times U(1)^n$. The solution, in both cases, is specified by $n$ independent rotation parameters $a_i$, with $i \in \{1, \cdots , n\}$ and a mass parameter $\mu$. For convenience, we write down the explicit metrics in \App{Adetnh} in Boyer-Lindquist type coordinates $(t,r,\mu_i,\phi_i)$, where the $\mu_i$ are the direction cosines and the $\phi_i$'s are angles in each two plane with Cartesian coordinates $\{x^i,y^i\}$. In even spacetime dimensions one has an extra direction cosine $\alpha$ since there are an odd number of spatial dimensions. The direction cosines $\mu_i$ and $\alpha$ (which take values in the range $0 \leq \mu_i \leq 1$ with $-1\leq \alpha \leq 1$) satisfy
\begin{equation}
\sum_{i=1}^{n} \mu_i^2=1  \ ,\qquad \sum_{i=1}^{n}\mu_i^2+\alpha^2=1\ .
\label{dircosconst}
\end{equation}
in odd and even dimensions respectively.
The location of the horizon is determined by the largest positive number $r_+$ such that\footnote{The mass parameter $\mu$, which we will later have occasion to refer to as $\mu_D$ and $\mu_d$ should not be confused with the direction cosines.}\begin{equation}
\Pi(r_+)-\mu \, r_+^2=0 \qquad  \Pi(r_+)-\mu\,  r_+=0 ,
\label{mphor}
\end{equation}
in odd and even dimensions respectively, where
\begin{equation}
\Pi(r) =\prod_{i=1}^{n}\, (r^2+a_i^2)\,. \qquad
\label{Pidef}
\end{equation}
The extremal limit of these black holes is given by
\begin{equation}
\Pi'(r_+)=2\, \mu \, r_+, \qquad  \Pi'(r_+)=\mu
\label{extcond}
\end{equation}
in odd and even dimensions respectively. The conditions \req{mphor} and \req{extcond} can simultaneously hold only when the black hole is spinning in all the two planes available \ie, we need $a_i \neq 0 \; \forall \; i = 1, \cdots , n$.
Without loss of generality we will henceforth assume $a_i>0 \; \forall \; i = 1, \cdots , n$ and use \req{extcond} to eliminate $\mu$, the mass parameter.

The near-horizon geometry is obtained by taking the near-horizon
limit as described earlier such that $r \to r_+$ (\ie, the horizon
in these coordinates is at $r=r_+$). The procedure requires one to first introduce coordinates valid on the horizon and is greatly
facilitated if one works with co-rotating coordinates as
discussed in \cite{Kunduri:2007vf}. We present some of the
details relevant to these higher dimensional examples in \App{Adetnh}. The upshot of these calculations is that
the near-horizon geometry of the $2n+1$ and $2n+2$ dimensional
extremal MP black holes can both be written as:
\begin{eqnarray}
ds^2 &=& F_+ \, \left( -\frac{\Pi''(r_+)}{2 \, \Pi(r_+)} \, r^2\,dv^2 + 2\,dvdr \right) + \gamma_{\mu_i \mu_j} \, d\mu^i d\mu^j \nonumber \\
& & + \, \gamma_{ij} \, \left( d\phi^i+\frac{2\, r_+\, a_i}{(r_+^2+a_i^2)^2}\, r\, dv \right)\, \left(d\phi^j+\frac{2\, r_+\,  a_j}{(r_+^2+a_j^2)^2}  \, r\, dv \right) ,
\label{mpbhmet}
\end{eqnarray}
where
\begin{equation}
F_+= 1- \sum_{i=1}^n\frac{a_i^2\, \mu_i^2}{r_+^2+a_i^2},
\label{fpiplus}
\end{equation}
\begin{equation}
\gamma_{ij}= (r^2_++a_i^2) \, \mu_i^2 \, \delta_{ij}+ \frac{1}{F_+} \, a_i\, \mu_i^2\, a_j\, \mu_j^2 \ .
\label{gammaij}
\end{equation}
and
\begin{equation}
\gamma_{\mu_i\mu_j}\, d\mu^i\,d\mu^j=  \sum_{i=1}^n \; (r_+^2+a_i^2)\, d\mu_i^2, \qquad \gamma_{\mu_i\mu_j}\, d\mu^i\, d\mu^j=r_+^2\, d\alpha^2 +  \sum_{i=1}^n \;(r_+^2+a_i^2)\, d\mu_i^2
\end{equation}
in odd and even dimensions respectively.

The coordinates on the horizon are $(\mu_i, \phi_i)$, where $\phi_i$
are $2\pi$-periodic and the direction cosines are subject to the
appropriate constraint (\ref{dircosconst}). Note that the
coordinates $r,\phi_i$, here are not the same as those in the
original MP solution (see \App{Adetnh}).  The horizon in the above
coordinates is at $r=0$ and spatial sections are of topology
$\Sp^{2n-1}$ in $2n+1$ dimensions and $\Sp^{2n}$ in $2n+2$
dimensions. The metric on the horizon has $U(1)^n$ symmetry in both
cases and describes a cohomogeneity-$(n-1)$ metric on $\Sp^{2n-1}$
and a cohomogeneity-$n$ metric on $\Sp^{2n}$, respectively. The full
near-horizon geometry has $SO(2,1) \times U(1)^n$ symmetry and is
also cohomogeneity-$(n-1)$ in $2n+1$ dimensions and
cohomogeneity-$n$ in $2n+2$ dimensions. Thus, for generic values of
the parameters $a_i$, the metric \req{mpbhmet} provides explicit
examples of near-horizon symmetry enhancement, which are outside the
assumptions of our theorem proved in \sec{adssym}. These geometries
generalize the discussion of the extremal Kerr black hole
near-horizon geometry and the five dimensional extremal MP black
hole near-horizon geometry discussed in \cite{Bardeen:1999px,
Kunduri:2007vf}.

While the near-horizon geometries given above are complicated for generic values of the angular momenta parameters $a_{i}$, there are special cases in which the solutions can simplify drastically. These cases arise when some subset of the $a_{i}$ are taken to be equal. As will be explained in subsequent sections, for even dimensions, we will be interested in the case when all the $a_{i}$ are taken equal, whereas in odd dimensions we will want explicit expressions for the case in which all but one of the $a_{i}$ are taken equal. In either situation, the full MP black hole solutions are cohomogeneity-2, and the corresponding near-horizon geometries are cohomogeneity-1. \par
From~(\ref{mpbhmet}) we find that the near-horizon geometry of the extremal MP black hole in $2n+2$ dimensions with $a_{i}=a$ is
\begin{eqnarray}\label{MPeqa}
 ds^2 &=& \frac{f_n(\theta)}{2n}\left(-\frac{(2n-1)^2}{2a^2n}\,r^2\,dv^2 + 2\,dvdr \right) + \frac{a^2\,f_n(\theta)}{2n-1}\, d\theta^2  + \frac{2\,n\,a^2\sin^2\theta}{2n-1}\, d\Sigma_{n-1}^2 \nonumber \\
 && \qquad  +\frac{4\,n^2\,a^2\,\sin^2\theta}{(2n-1)\, f_n(\theta)}\,\left(d\phi+A +\frac{(2n-1)^{3/2}}{2\,n^2\,a^2}\, r\, dv\right)^2
\end{eqnarray}
 where
\begin{equation}\label{ftheta}
f_n(\theta) = 1 + (2n-1)\, \cos^2\theta,
\end{equation}
and $d\Sigma_{n-1}^2$ is the Fubini-Study metric on $CP^{n-1}$ with K\"ahler form $J=\frac{1}{2}dA$, and $\phi$ is $2\pi$ periodic with $0 \leq \theta \leq \pi$.\footnote{We have written the squashed $\Sp^{2n}$ in terms of a polar angle $\theta$ and a round $\Sp^{2n-1}$. A round odd dimensional sphere with coordinates $(\nu_i,\phi_i)$ can be written as $\sum_{i=1}^{n} \, d\nu_i^2+\nu_i^2\,d\phi_i^2=(d\phi+A)^2+d\Sigma^2_{n-1}$ and $\sum_{i=1}^n \,\nu_i^2 \,d\phi_i = d\phi+A$, where $\nu_i$ are directions cosine.} The near-horizon geometry has $SO(2,1)\times U(1)\times SU(n)$ symmetry and the horizon is described by a cohomogeneity-1 metric on $\Sp^{2n}$. The $U(1) \times SU(n)$ has $\Sp^{2n-1}$ orbits. Note that the $n=1$ case corresponds to the near-horizon geometry of extremal Kerr discussed in~\cite{Bardeen:1999px, Kunduri:2007vf}.

Finally, we consider an extremal MP black hole in $D=2n+3$ dimensions, as this will prove useful for comparison to the black strings studied in the next section. Such a black hole has $U(1)^{n+1}$ rotational symmetries in general, and its near-horizon geometry for arbitrary angular momenta parameters $a_{i}$ can be easily read off from~(\ref{mpbhmet}) (one must shift $n\to n+1$). In the special case with $a_{i}=a_2$ for $i=2, \cdots ,n+1$, the black hole solution becomes cohomogeneity-2 with a cohomogeneity-1 near-horizon geometry given by
\begin{eqnarray}
ds^2 &=& F_+(\theta)\, \, \left( -\frac{\Pi''(r_+)}{2 \, \Pi(r_+)}
\, r^2\,dv^2 + 2\,dvdr \right)
\\ \nonumber  &+& \rho_+(\theta)^2d\theta^2+(r_+^2+a_1^2)\, \cos^2\theta \,d\psi^2+(r_+^2+a_2^2)\,\sin^2\theta \,d\Sigma_{n-1}^2\\ &+& \nonumber \frac{1}{F_+(\theta)}\left[a_2\sin^2\theta\left(d\phi + A+\frac{2\,r_+\,a_2}{(r_+^2+a_2^2)^2}\,r\,dv\right) + a_1\cos^2\theta \left(d\psi+\frac{2\,r_+\,a_1}{(r_+^2+a_1^2)^2}\,r\,dv\right) \right]^2
\label{mpbhmetc1}
\end{eqnarray}
with
\begin{equation}
\rho_+(\theta)^2 = r_+^2+a_2^2\,\cos^2\theta + a_1^2\,\sin^2\theta, \qquad
F_+(\theta)= \frac{r_+^2\, \rho_+(\theta)^2}{(r_+^2+a_1^2)(r_+^2+a_2^2)}
\label{}
\end{equation}
where $\phi$ is $2\pi$-periodic, $0 \leq \theta \leq \pi/2$ and we have defined $\psi \equiv \phi_1$.\footnote{Here we have written the squashed $\Sp^{2n+1}$ in terms of an angle $\psi$ and a polar angle $\theta$ leaving a round $\Sp^{2n-1}$ which we have then written in terms of $CP^{n-1}$ quantities as in the previous footnote.}
The near-horizon geometry has $SO(2,1)\times U(1)^2\times SU(n)$ symmetry and spatial cross sections of the horizon are equipped with a cohomogeneity-1 metric on $\Sp^{2n+1}$. One can of course, further specialize to $a_1 =a_2$ to obtain a homogeneous metric, which we will refrain from writing down explicitly. The $n=1$ case corresponds to the cohomogeneity-1 near-horizon geometry of the $D=5$ MP black hole discussed in~\cite{Bardeen:1999px, Kunduri:2007vf}.

\subsubsection{Black ring topology horizons}
\label{mpbsnh}

The only known exact vacuum solution with $\Sp^1 \times \Sp^{D-3}$ horizon
topology is the boosted MP string which is not asymptotically flat (there are of course  the perturbative singly spinning black ring solutions of \cite{Emparan:2007wm}).  We will consider only
odd dimensions in this section, as we will see only in this case can
the string near-horizon geometry correspond to that of an
asymptotically flat black ring. To construct the string we use the $d=2n+2$ MP
solution, so the string lives in $D=d+1$ dimensions. The string then
will have horizon topology $\Sp^1 \times \Sp^{2n}$ with symmetry
$U(1)^{n+1}$ (which is the correct number of rotational symmetries
in this dimension) and thus provide us with a candidate black ring
near-horizon geometry.

To proceed, we simply take the extremal MP black hole in $d=2n+2$ dimensions and add a flat direction $dz^2$ and boost $(t,z) \to ( \cosh \beta t -\sinh\beta z, -\sinh\beta t +\cosh\beta z)$. The explicit string metric is given in the appendix \req{mpbsstringsol}. We then take the near-horizon limit in the manner described in \App{Adetnh}. After some calculation one finds (denoting $\cosh\beta = c_\beta$  and $\sinh\beta = s_\beta$)
\bea \label{MPstringNH}
ds^2 &=& \frac{F_+}{c_\beta} \, \left(
-\frac{\Pi''(r_+)}{2\,c_\beta \,\Pi(r_+)} \,r^2 \, dv^2 + 2\, dv\, dr \right) +
\gamma_{\mu_i \mu_j} \, d\mu^i\, d\mu^j  \\
& & \qquad +\gamma_{ij} \, \left(
d\phi^i
 +\omega^{i}_{\beta}\, dz+ \frac{2\,r_+\, a_i}{(r_+^2+a_i^2)^2}\,r\,dv \right)\left(d\phi^j  +\omega^{j}_{\beta}\,dz +\frac{2\,r_+\, a_j}{(r_+^2+a_j^2)^2} \, r \, dv \right) + c_{\beta}^2\, dz^2 \nonumber
\eea
where
\begin{equation}
 \omega^{i}_{\beta} = \frac{s_\beta \, a_{i}}{r_{+}^2+a_{i}^2} \ ,
\end{equation}
and $F_+$ is defined in \req{fpiplus}. In this case, the metric on the horizon is
\bea\label{MPstringhor}
\gamma_{ab}dx^adx^b &=& \gamma_{\mu_i \mu_j}\,  d\mu^i d\mu^j + \gamma_{ij}\, \left(d\phi^i +\omega^{i}_{\beta}\, dz \right) \left(d\phi^j+\omega^{j}_{\beta}\, dz\right)   + c_{\beta}^2\, dz^2 \\
&=& r_+^2\, d\alpha^2 +\sum_{i=1}^n\, (r_+^2+a_i^2)\, \left[d\mu_i^2 +\mu_i^2 \, \left(d\phi^i+\omega^{i}_{\beta}\, dz\right)^2 \right] \nonumber \\
&& \qquad + \frac{1}{F_+}\, \sum_{i,j=1}^{n} \, a_i\,\mu_i^2 \,
a_j\, \mu_j^2 \, \left(d\phi^i+\omega^{i}_{\beta}\, dz\right)
\left(d\phi^j+\omega^{j}_{\beta}\, dz\right) + c_{\beta}^2\, dz^2  \
.\nonumber \eea
The coordinate ranges for the $\mu_{i}$ and $\phi^{i}$ are the same as that for the MP black hole \ie, $\phi_i$ are $2\pi$-periodic, $0\le \mu_i \le 1$, and
$-1\le \alpha \le 1$.   We denote the period of $z$ by $2\pi \,\ell$. The full near-horizon geometry \req{MPstringNH} is cohomogeniety-$n$ and has $SO(2,1)\times U(1)^{n+1}$ isometry. Spatial sections of the horizon have topology $\Sp^1 \times \Sp^{2n}$ and the metric on the horizon \req{MPstringhor}  describes a cohomogenity-$n$ metric on $\Sp^1 \times \Sp^{2n}$.

As in the black hole case discussed previously, the near-horizon of the extremal MP black string simplifies considerably in the case where all angular momenta parameters $a_{i}$ are taken to be equal. The full black string solution then becomes cohomogeneity-2 and the corresponding near-horizon geometry is cohomogeneity-1. The near-horizon geometry of the extremal MP string in $D=2n+3$ dimensions with $a_i=a\, \forall\;  i$ is
\begin{eqnarray}
 ds^2 &=& \frac{f_n(\theta)}{2\,n\,c_\beta}\left(-\frac{(2n-1)^2}{2\,a^2\,n\, c_\beta}\,r^2\,dv^2 + 2\,dvdr \right) + \frac{a^2\,f_n(\theta)}{2n-1}\,d\theta^2  + \frac{2n\,a^2\,\sin^2\theta}{2n-1}\,d\Sigma_{n-1}^2
  \nonumber \\ &+& \frac{4n^2\,a^2\,\sin^2\theta}{(2n-1)\,f_n(\theta)}\left( d\phi + A +\frac{s_\beta (2n-1)\,dz}{2n\,a} +\frac{(2n-1)^{3/2}}{2n^2\,a^2\,c_\beta}\, r\,dv \right)^2 + c_\beta^2 \,dz^2 \ ,
\end{eqnarray}
 where we follow the notation used in~(\ref{MPeqa}). The near-horizon geometry in this special case has $SO(2,1)\times U(1)^2 \times SU(n)$ symmetry and spatial cross sections of the horizon have topology $\Sp^{1}\times \Sp^{2n}$ and are equipped with a cohomogeneity-1 metric on $\Sp^{1}\times \Sp^{2n}$. The $n=1$ case corresponds to the Kerr string discussed in~\cite{Kunduri:2007vf}.

\subsection{What physical data can be extracted from  a near-horizon geometry?}
\label{nhphys}

In this section we will clarify which quantities can be
expected to be calculable from the near-horizon geometry alone,
without any knowledge of the full black hole solution. Apart from being of intrinsic interest our analysis should also clarify aspects related to the entropy function formalism \cite{Sen:2005wa} and the attractor mechanism -- see \cite{Sen:2007qy} for a review and \cite{Suryanarayana:2007rk,Hanaki:2007mb} for earlier discussions on obtaining conserved charges from the near-horizon.

One quantity which can be obviously computed from the near-horizon
geometry alone is the area of spatial sections of the horizon $\CA_H
=\int_{\CH} \sqrt{\gamma}$. However, this is only meaningful if it
is expressed as a function of the physical charges of the black
hole.

We will restrict our discussion to vacuum asymptotically flat black holes in
$D$ spacetime dimensions, possessing $U(1)^{n+1}$ spatial isometry
with generators $\frac{\partial}{ \partial \phi_i} $ (so
$i=1,\cdots,n+1$) where the angles $\phi_i$ are chosen to have
period $2\pi$. These generators are orthogonal at asymptotic
infinity.

While it is easy to calculate the near-horizon geometry of a known
solution, as was done for the MP black holes and black
strings in \sec{mpbhnh} and \sec{mpbsnh} respectively, our goal is
to try to use near-horizon geometries to learn about new
solutions which are yet to be constructed. In particular, we would
like to know whether \req{MPstringNH} is isometric to the
near-horizon geometry of an asymptotically flat black ring in $D=2n+3$ dimensions. We
will find that the Komar integrals evaluated on the horizon capture
some of the basic physical parameters, but not all.  Specifically,
determining the mass needs some additional data not present in the
near-horizon geometry. The basic issue is that the near-horizon limit loses
the information regarding the asymptotic stationary Killing field
and this prevents one from directly calculating the mass.

\paragraph{Angular momenta: }
\label{angmomenta}

Let us first investigate how the angular momenta can be calculated
from the near-horizon geometry. The angular
momentum of the black hole can be calculated from the Komar integrals
\begin{equation}
J_i = \frac{1}{16\, \pi \, G^{(D)}_N } \,  \int_{\CH_r} \, \star dm_i \,  .
\label{komaram}
\end{equation}
where $m_i$ denotes the metric dual one form to $\frac{\partial}{ \partial
\phi_i}$. The integral is taken over $\CH_r$ which is the compact
co-dimension two manifold defined in a neighbourhood of the horizon by the Gaussian null coordinates $v=$constant and $r=$ constant with $r \geq 0$ (so $\CH_0=\CH$). Usually the Komar integral is taken over the sphere at
infinity $\Sp_{\infty}^{D-2}$; the
difference is an integral over a manifold  whose boundary is the union of the two compact
manifolds which vanishes if $R_{\mu\nu}=0$ (\ie, for vacuum
solutions).

Now let us work in Gaussian null coordinates, \req{bhnhc}, {\it prior}
to taking  the near-horizon limit (thus all quantities depend on
$r$). It is useful to work in a non-coordinate basis, with the vielbeins
$(e^+,e^-,e^A)$ such that  $g_{\mu\nu}\, dx^{\mu}dx^{\nu} =
2\,e^+e^- +e^Ae^A$, defined by
\begin{equation}
e^+=dv, \qquad e^-=\frac{1}{2}\, r\, f(r,x)\, dv+dr+r\, h_a(r,x)\,
dx^a, \qquad e^A=\hat{e}^A \ . \label{gnullv}
\end{equation}
Here $\hat{e}^A$ are a set of vielbeins for the metric on $\CH_r$ so
$\gamma_{ab}(r,x)\,dx^a\,dx^b= \delta_{AB}\,\hat{e}^A\,\hat{e}^B$ . In this
basis
\begin{equation}
m_i = r \, h_i(r,x) \, e^+ + \, (m_i)_B \, e^B
\label{onefkill}
\end{equation}
where $h_i= h_a (\partial_{\phi_i})^a$.
Therefore,
\begin{equation}
\star dm_i \, |_{v,r = \const}=  \left[h_i(x)  +
\CO\left(r\right)\right] \, e^1 \wedge e^2 \cdots \wedge e^{D-2}
\label{donef}
\end{equation}
where as before $h_i(x) = h_i(0,x)$. Thus for small $r>0$
\begin{equation}
J_i=\frac{1}{16\pi\, G^{(D)}_N}\int_{\CH_r} \,\sqrt{\gamma}\; \left(
h_i(x) + \CO\left(r\right) \right)=\frac{1}{16\pi\,G^{(D)}_N}\, \int_{\CH}
\, \sqrt{\gamma}\big|_{r=0} \; h_i(x)+ \CO\left(r\right) \ ,
\end{equation}
and hence we must have
\begin{equation}
J_i= \frac{1}{16\pi\, G^{(D)}_N}\, \int_{\CH} \, \sqrt{\gamma}\;  h_i(x)
,
\end{equation}
where $\gamma_{mn}$ and $h_i$ are the metric $\gamma_{mn}(r,x)$ and the vector $h_i(r,x)$ evaluated at $r=0$, which coincide with the corresponding quantities appearing in the near-horizon limit metric \req{nhgeom}.

Thus one can indeed calculate the angular momenta $J_i$ from the near-horizon metric \req{nhgeom} alone. Note that this is not {\it a priori} guaranteed -- one could have had contributions from terms like $(\partial_r h_i)\big|_{r=0}$, which are inaccessible from the near-horizon data alone. We find it convenient to recast this formula into the coordinates introduced in \req{nhsym}:
\begin{equation}
J_i= \frac{1}{16\pi \, G^{(D)}_N }\, \int_{\CH} \, \sqrt{\gamma}\;
\Gamma^{-1} \, k_i \ . \label{intangkomar}
\end{equation}

There is however an important caveat in the
determination of the angular momenta. In
general, from the near-horizon geometry alone, it is not possible to know which Killing
fields on the horizon correspond to the generators
$\frac{\partial}{\partial \phi_i}$ chosen to be orthogonal
at asymptotic infinity. One can evade this problem for spherical
topology black holes since the Killing fields will have the same
number of fixed points on the horizon as at infinity and thus a
natural identification exists. However, for ring-like topology there
is not a unique way of identifying the generator of the $\Sp^1$.

\paragraph{Mass:}
Let us now consider the Komar integral associated to the Killing
vector $\frac{\partial}{\partial v}$. It turns out to be more illuminating to do this for a general black
hole \req{bhnhc} (not necessarily extremal). Let $V$ be the one form whose
metric dual is $\frac{\partial}{\partial v}$ so
\begin{equation}
V= r\, f(r,x)\,e^+ +dr+r \, h_A(r,x)\, e^A \label{onefV}
\end{equation}
and thus
\begin{equation}
\left(\star dV\right)\big|_{v,r= \textrm{const}} = \left( f(0,x) +
\CO\left(r\right)\right)\, e^1 \wedge e^2 \cdots \wedge e^{D-2} \ .
\label{onefVs}
\end{equation}
Hence employing the argument outlined above for computing the
angular momenta, we obtain:
\begin{equation}
Q_v \equiv -\left( \frac{D-2}{D-3} \right)\frac{1}{16\pi\, G^{(D)}_N} \,
\int_{\CH_r}\, \star dV = -\left( \frac{D-2}{D-3}
\right)\frac{1}{16\pi\, G^{(D)}_N}\, \int_{\CH} \,\sqrt{\gamma} \; f(0,x)
. \label{qint}
\end{equation}
For an extremal black hole we immediately learn that $Q_v=0$, as the
extremality condition ($\kappa=0$, see \req{surfgrav}) forces
$f(0,x) = 0$. The Killing vector $\frac{\partial}{\partial v}$ is
co-rotating and can be written as \be \frac{\partial}{\partial
v}=\frac{\partial}{\partial t}+\Omega_i \, \frac{\partial}{\partial
\phi_i} \label{corotatingKV} \ee where $\frac{\partial}{\partial t}$
is the stationary Killing field (asymptotically timelike). Therefore
\begin{equation}
M = \left( \frac{D-2}{D-3} \right) \, \Omega_i \, J_i
\label{extsmarr}
\end{equation}
where we define the mass $M$ through the usual Komar integral
associated with the stationary Killing field
$\frac{\partial}{\partial t}$. The near-horizon geometry has no
knowledge of the Killing field $\frac{\partial}{\partial t}$ and
hence there is no way of inferring the mass from the near-horizon
data alone. Put differently, one can evaluate  $J_i$ as argued above
from near-horizon data. However, to complete the determination of
the mass we need to know the angular velocities $\Omega_i$, which
are measured relative to a stationary observer at infinity. This is
the data one is missing from the near-horizon geometry.

As an aside, for a non-extremal black hole note the above provides an efficient proof of the Smarr relation \cite{Bardeen:1973gs} (see also \cite{Elvang:2007hg}), illustrating the advantage of using Gaussian null coordinates. The integral for $Q_v$ can be evaluated simply in terms of the surface gravity \req{surfgrav}. Since this is constant over the horizon we have
\begin{equation}
Q_v =   \left(\frac{D-2}{D-3}\right)\frac{\kappa \,\AH}{8\,\pi \,G^{(D)}_N}
\end{equation}
where $\AH=\int_{\CH} \,\sqrt{\gamma}$ is the area of the horizon.
Using this we can write
\begin{equation}
 M=\left( \frac{D-2}{D-3} \right) \left( \frac{\kappa \,\AH}{8\pi G^{(D)}_N}+ \Omega_i J_i \right)  ,
\label{smarr}
\end{equation}
for any vacuum asymptotically flat spacetime with a regular event
horizon, as is well known.

Finally, we remark that the existence of a near-horizon geometry
does not necessarily imply that there exists an extremal black hole
solution with such a near-horizon geometry and prescribed
asymptotics (\eg, asymptotically flat or AdS). For example, one
would not expect the near-horizon geometry of an extremal boosted
Kerr string to correspond to the near-horizon geometry of an
asymptotically flat black ring for generic boost values. This is
because we know from~\cite{Kunduri:2007vf} that the known vacuum
extremal black ring in 5d  \cite{Pomeransky:2006bd} has a
near-horizon geometry isometric to the boosted Kerr-string with a
particular value of boost parameter. Further, one might expect that
this known vacuum extremal black ring solution is the most general
solution with two rotational isometries in five dimensions \cf., \cite{Hollands:2007qf}.  It would be interesting to understand more generally what obstructions can occur when one tries to integrate out from the horizon and match to a solution with prescribed asymptotics.\footnote{This is important to understand the constraints on the entropy function formalism \cite{Sen:2005wa, Sen:2007qy}, where one works exclusively with the near-horizon geometry.}

\section{Extremal MP black holes}
\label{extrmpbh}
In this section we will consider the physical quantities of extremal
MP black holes in various dimensions.  We discuss the extremal locus for these solutions in terms of dimensionless parameters and the phase diagram for these solutions. A curious fact we observe is that in dimensions $D \ge 6$ it is possible to stay on the extremal locus while sending the angular momenta in certain planes to infinity. We show that it is therefore possible to obtain highly distorted extremal black hole horizons with spherical topology. Some of the results we obtain for the MP black holes will be useful for comparison with the conjectured black ring solution we propose in \sec{hdrings}. Some of these results have been previously discussed in
\cite{Kleihaus:2007kc, Emparan:2008eg}.

\subsection{Parameterization of MP black holes}
\label{mphyspar}

The physical quantities of interest \viz, the mass, the angular momenta and velocities and the area can be expressed in terms of the parameters $a_i$ appearing in the MP solution \req{mp2n1} and \req{mp2n2}. However, it is convenient to write the formulae in terms of $r_+$ (which is determined in terms of $a_i$ for extremal solutions) to keep the expressions compact. We have
\begin{equation}
\begin{aligned}
&{\underline{\rm Odd \;dimensional\; MP}}\;(D= 2n+3) \qquad &\qquad &{\underline{\rm Even \;dimensional \;MP}} \; (d = 2 n +2 ) \\
&\qquad M=\frac{A_{2n+1}}{16\pi\,G^{(D)}_N}\, (2n+1)\,\mu_{D}
&& \qquad M =\frac{A_{2n}}{16\pi\, G^{(d)}_N}\,(2n)\, \mu_{d} \\
& \qquad\Omega_i =\frac{a_i}{r_+^2+a_i^2}
&& \qquad
\Omega_i=\frac{a_i}{r_+^2+a_i^2} \\
& \qquad  J_i =\frac{A_{2n+1}}{8\pi\, G^{(D)}_N}\, a_i \, \mu_{D}
&& \qquad
J_i =\frac{A_{2n}}{8\pi\, G^{(d)}_N}\,a_i\,\mu_{d} \\
& \qquad \AH= A_{2n+1}\, \mu_{D}\, r_+
&& \qquad
\AH=A_{2n}\,\mu_{d}\,r_+\,, \\
& \qquad \mu_{D}= \frac{\Pi(r_+)}{r_+^2}
&& \qquad
\mu_{d}=\frac{\Pi(r_+)}{r_+}
\end{aligned}
\label{mpDpars}
\end{equation}
Note that the function $\Pi(r)$ is given in \req{Pidef} (with the
replacement $n \to n+1$ in the first column). Here $A_p$ denotes the
area of an unit $\Sp^{p}$, \ie\
$A_p=2\pi^{\frac{p+1}{2}}/\Gamma(\frac{p}{2})$. In writing
\req{mpDpars} we have separated the quantities that need information
about the asymptotic geometry ($M$ and $\Omega_i$) from those that
can be determined from the near-horizon alone ($J_i$ and $\AH$).

To discuss the physical behaviour of the solutions as a function of
the parameters, it is useful to define dimensionless variables. We
find it convenient to define reduced area and angular momenta by
fixing the mass of the solution as in \cite{Emparan:2007wm}. These
are defined by the following expressions in $D$ spacetime
dimensions:
\begin{equation}
\begin{split}
\aH &=\left[A_{D-2}\left(\frac{D-2}{16\pi \,G^{(D)}_N}\right)^{D-2}\right]^{\frac{1}{D-3}}
    \, \frac{\AH}{\left(M\right)^{\frac{D-2}{D-3}}} \\
j_i&=\frac{1}{2}\,\left[A_{D-2}\frac{(D-2)^{D-2}}{16\pi
\,G^{(D)}_N}\right]^{\frac{1}{D-3}}\,\frac{J_i}{\left(M\right)^{\frac{D-2}{D-3}}} \ .
\end{split}
\label{mpDred}
\end{equation}
The normalizations have been chosen such that the $D$-dimensional Schwarzschild black hole has unit reduced area $\aH$. For the reduced angular momenta we have chosen conventions that are natural generalizations of the definitions used in five dimensions.\footnote{While it is physical to measure the area with respect to a Schwarzschild black hole of the same mass, the normalization for the angular momenta are not uniquely characterized.  Our choice differs from the normalization chosen in \cite{Emparan:2007wm}, where the authors found it convenient to keep the formulae for black rings simple; we have simple expressions for MP black holes \req{redpars}.}
Using the explicit expressions for the physical parameters it is easy to check that
\begin{equation}
\begin{aligned}
&{\underline{\rm Odd \;dimensional\; MP}}\;(D= 2n+3) \qquad &\qquad
&{\underline{\rm Even \;dimensional \;MP}} \; (d = 2 n +2 ) \\
&\qquad j_i = \frac{a_i}{\left(\mu_D\right)^{\frac{1}{2n}}}
&& \qquad
j_i = \frac{a_i}{\left(\mu_d\right)^{\frac{1}{2n-1}}} \\
&\qquad \aH = \frac{r_+}{\left(\mu_D\right)^{\frac{1}{2n}}}
&& \qquad
\aH = \frac{r_+}{\left(\mu_d\right)^{\frac{1}{2n-1}}}
\end{aligned}
\label{redpars}
\end{equation}
%

\subsection{The extremal locus}
\label{mpextloc}

We now turn to the implications of extremality for MP black holes in
various dimensions. To begin with, recall that in four dimensions
the extremal Kerr black hole is given by a single point in the
reduced parameter space $j =1$ (recall that we fix the total mass).
The situation in five dimensions is already more interesting as we
have a ``moduli space'' of solutions; the extremality condition can be
written as
\begin{equation}
 a_1 +a_2 = \sqrt{\mu_{5}} \;\; \thus \;\; j_1 + j_2 =1 \ .
\label{fivedext}
\end{equation}
We therefore see that there is a one parameter family of solutions (labeled by say $j_1$) which takes values in a finite domain, $j_1 \in (0,1)$.

In $D>5$ there is also a non-trivial moduli space of solutions, albeit with one interesting feature -- these moduli spaces are non-compact. This arises because it is possible to attain extremality whilst sending some of the angular momenta to infinity. To understand this behaviour it is useful to characterize the extremal locus explicitly.

To understand the general picture in odd dimensions, realize that the equations \req{mphor} and \req{extcond}  are symmetric in the $a_i$ with both $r_+$ and $a_i$ having length dimension one. $\mu_D$ is therefore of scaling dimension $2n$. The simplest way to get the extremal locus is to eliminate $r_+$ between the two equations \req{mphor} and \req{extcond}; this will give
\begin{equation}
\mu_D = \CE_D(a_1,\cdots , a_{n+1}) \ , \qquad {\rm with} \qquad \CE_D(\lambda\, a_1, \cdots , \lambda \, a_{n+1}) = \lambda^{2n}\, \CE_D(a_1,\cdots , a_{n+1})
\label{}
\end{equation}
Likewise one can carry out the exercise for even spacetime dimensions where we get
\begin{equation}
\mu_d = \CE_d(a_1,\cdots , a_n) \ , \qquad {\rm with} \qquad \CE_d(\lambda\, a_1, \cdots , \lambda \, a_n) = \lambda^{2n-1}\, \CE_d(a_1,\cdots , a_n)
\label{mpevenextl}
\end{equation}
Using the above homogeneity property together with the definitions of the reduced parameters it is easy to see that the extremal locus can be expressed as a function of the reduced angular momenta alone
\begin{equation}
\CE_{p}(j_1, \cdots j_n) = 1 \ ,\qquad  p\in\{d, D\} \, .
\label{MPextgen}
\end{equation}
The precise expression is not that easy to determine in general, but it easy to get explicit formulae for low lying dimensions. We already know from \req{fivedext} that
\begin{equation}
\CE_5(j_1, j_2) = (j_1 + j_2)^2 \, .
\label{fdmpextl}
\end{equation}
Similarly, one can check that in six dimensions
\begin{equation}
\CE_6(j_1,j_2) = \frac{1}{3}\, \sqrt{\frac{2}{3}}\, \, \sqrt{-j_1^2 -j_2^2 + \sqrt{j_1^4 + j_2^4 + 14\, j_1^2 \,j_2^2}} \, \left(2\,j_1^2 +2\,j_2^2 + \sqrt{j_1^4 + j_2^4 + 14\, j_1^2 \,j_2^2}\right)
\label{}
\end{equation}
while the seven dimensional MP black holes give
\begin{equation}
\CE_7(j_1,j_2,j_3)
=\frac{\prod_{i=1}^3\left(\tilde{r}_+^2+j_i^2\right)}{\tilde{r}_+^2}
\ , \qquad \tilde{r}_+^2\equiv \frac{1}{3}\,\left(-\sum_{i=1}^{3}\,
j_i^2 + \sqrt{\sum_{i=1}^{3}\,j_i^4 - \sum_{i<j=2}^{3}\, j_i^2 \,
j_j^2 + 3} \right) \ .\label{}
\end{equation}

\begin{figure}[t!]
\begin{center}
\includegraphics[scale=0.72]{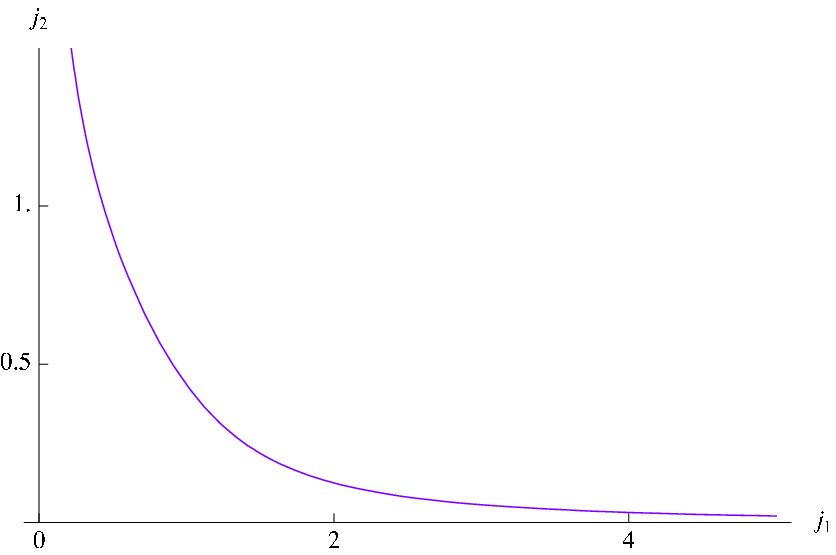}
\hspace{0.75cm}
 \includegraphics[scale=0.72]{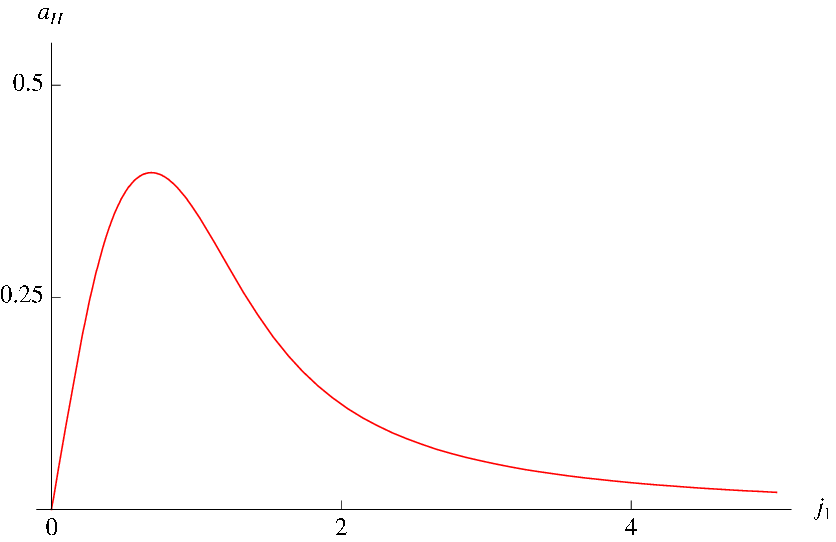}
\end{center}
\caption{\small{The phase space of extremal six dimensional MP black holes.
{\bf Left:} The extremal locus in the $(j_1, j_2)$ plane.
{\bf Right:} The area as a function of $j_1$; note that the maximum area configuration is the symmetric one, $j_1 = j_2$.  }}
\label{fig:sixdextl}
\end{figure}

\begin{figure}[h!]
\begin{center}
\includegraphics[scale=0.25]{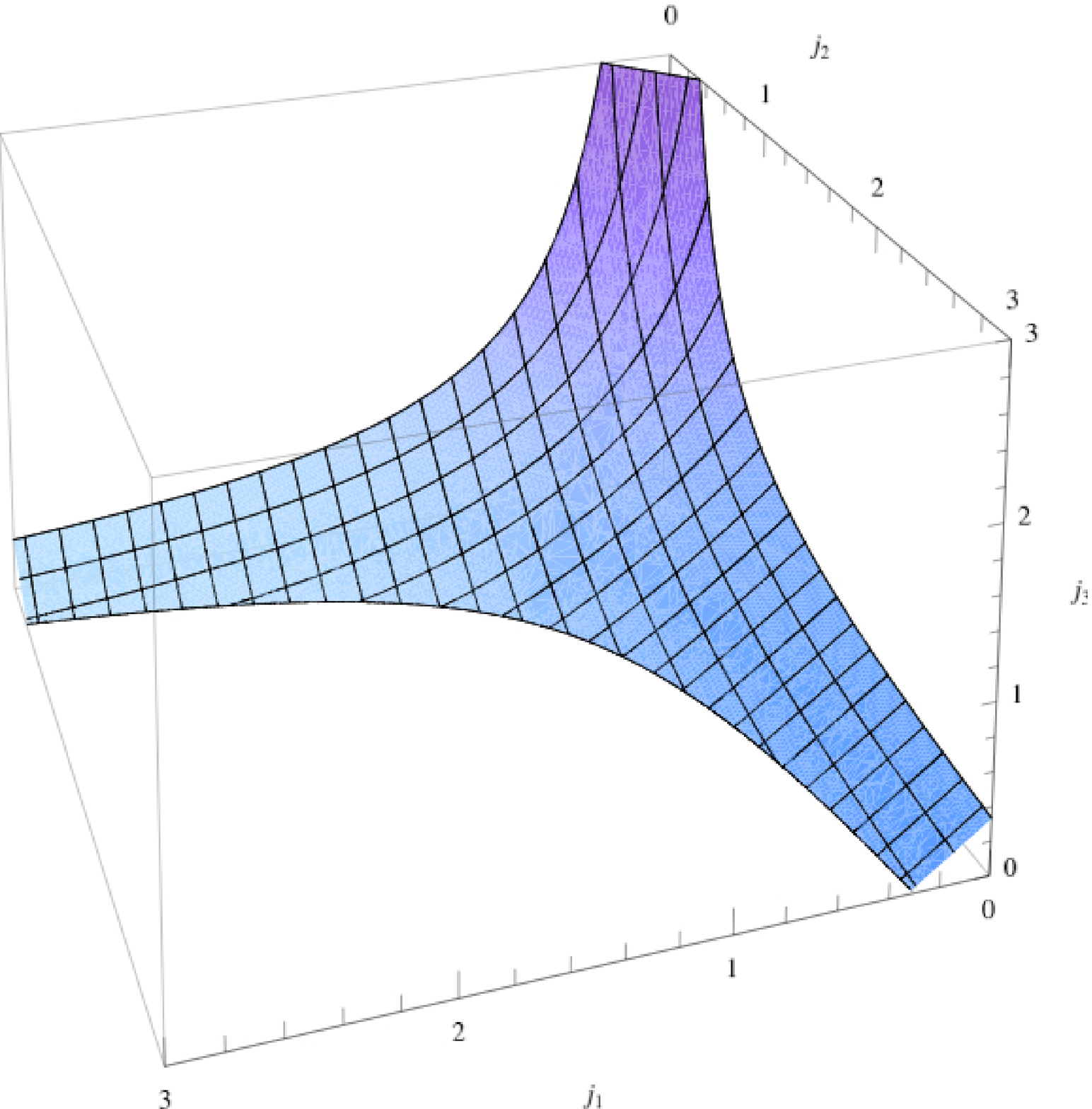}
\hspace{0.5cm}
 \includegraphics[scale=0.28]{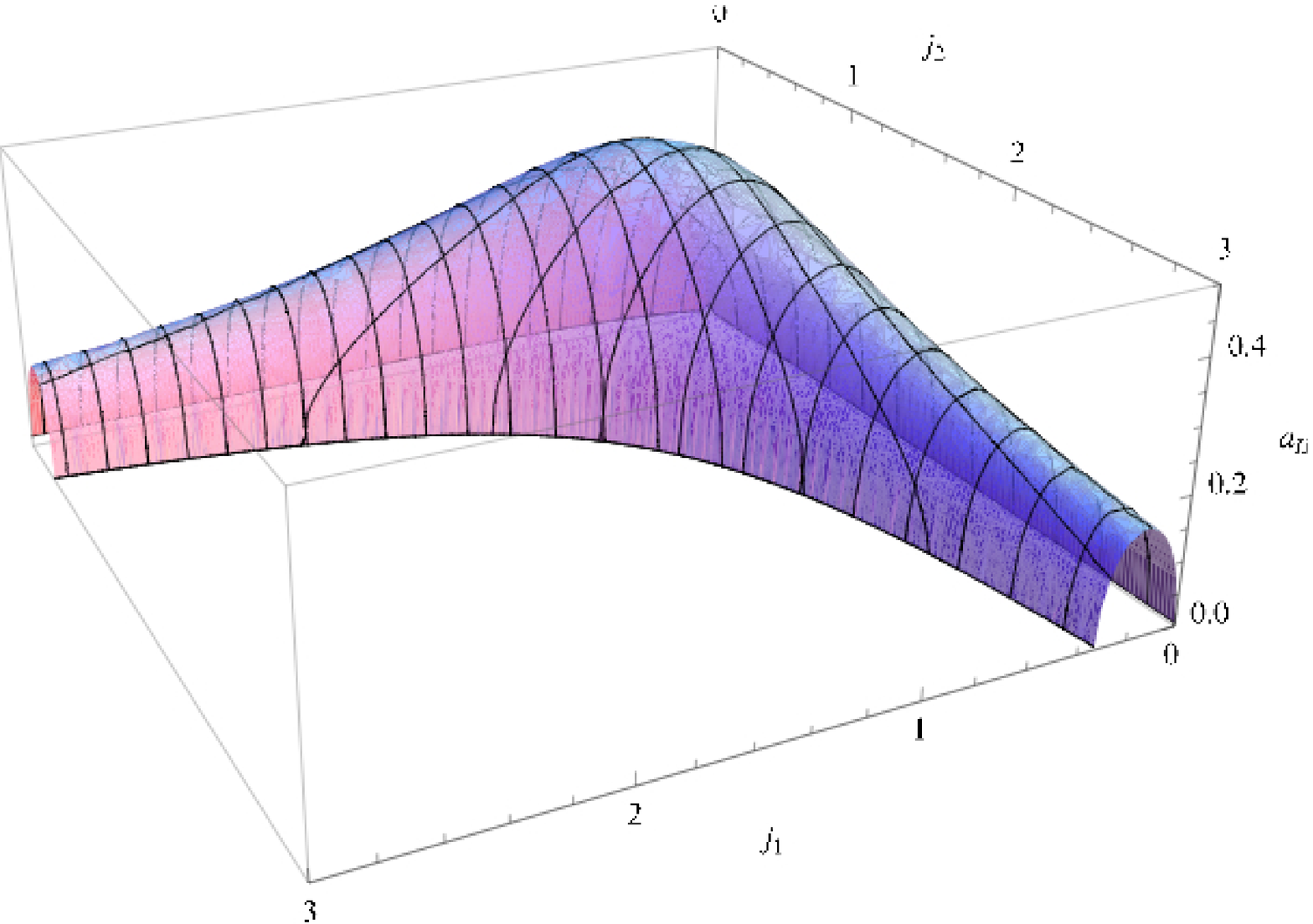}
\end{center}
\caption{\small{The phase space of extremal seven dimensional MP black holes.
{\bf Left:} The extremal locus as a function of the reduced angular momenta variables $(j_1, j_2, j_3)$. Here the allowed region is non-compact, with the arms of the fan extending off to infinity. Also note that as one of the $j_i$, say $j_1$ gets large, one the surface projected onto the $(j_2,j_3)$ plane starts to resemble a rescaled version of the five dimensional MP extremal locus, see \fig{fig:fivedphasediag}.
{\bf Right:} The phase plot $a_H(j_1,j_2)$. Note that the maximum occurs at the symmetric point $j_1 = j_2 = j_3$ and the limiting behaviour of the surface at large values of $j_i$ coincides with that of the five dimensional MP curve, see \fig{fig:fivedphasediag}}}.
\label{fig:sevendextl}
\end{figure}

\begin{figure}[t!]
\begin{center}
\includegraphics[scale=0.25]{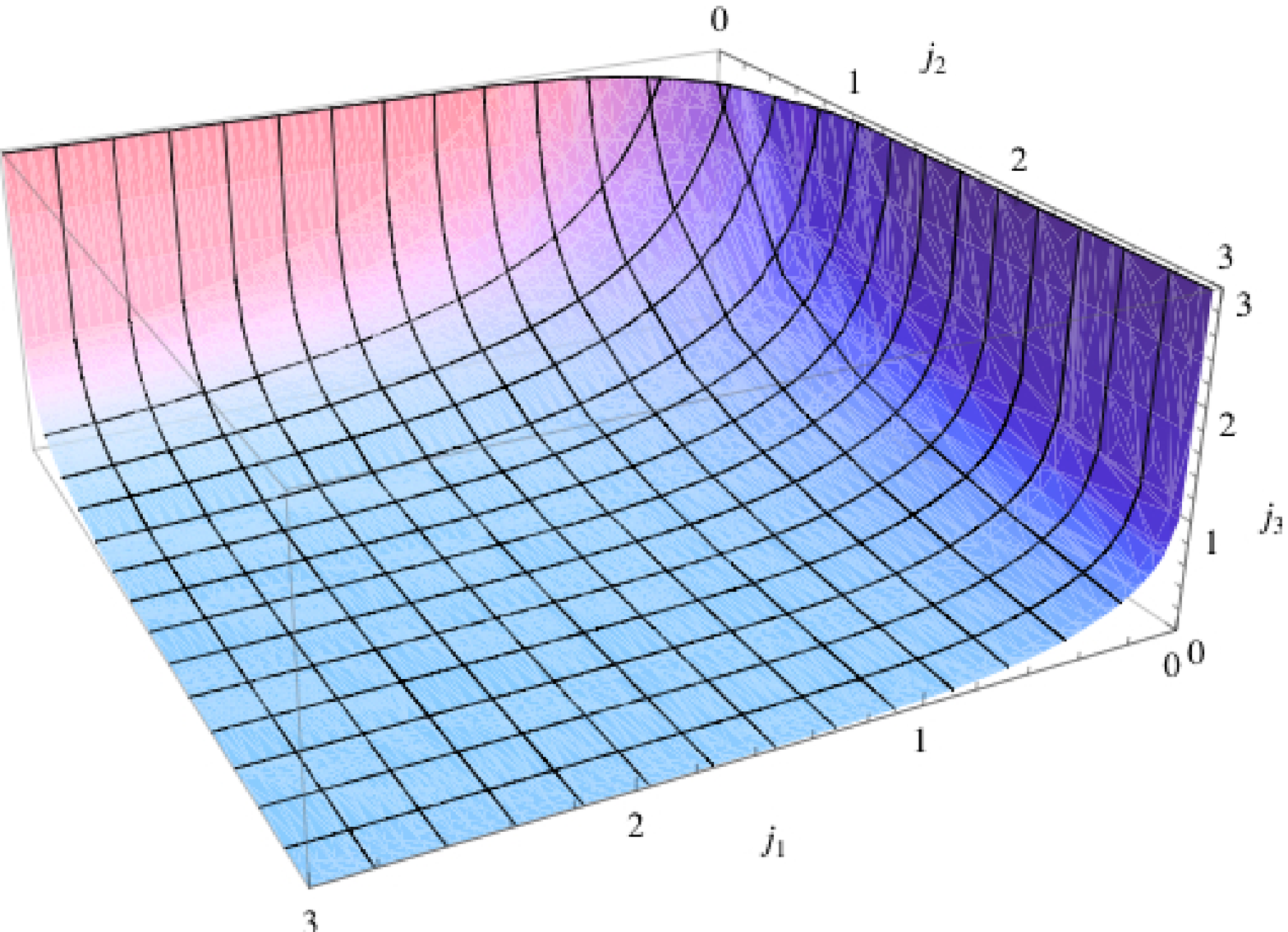}
\hspace{0.75cm}
 \includegraphics[scale=0.25]{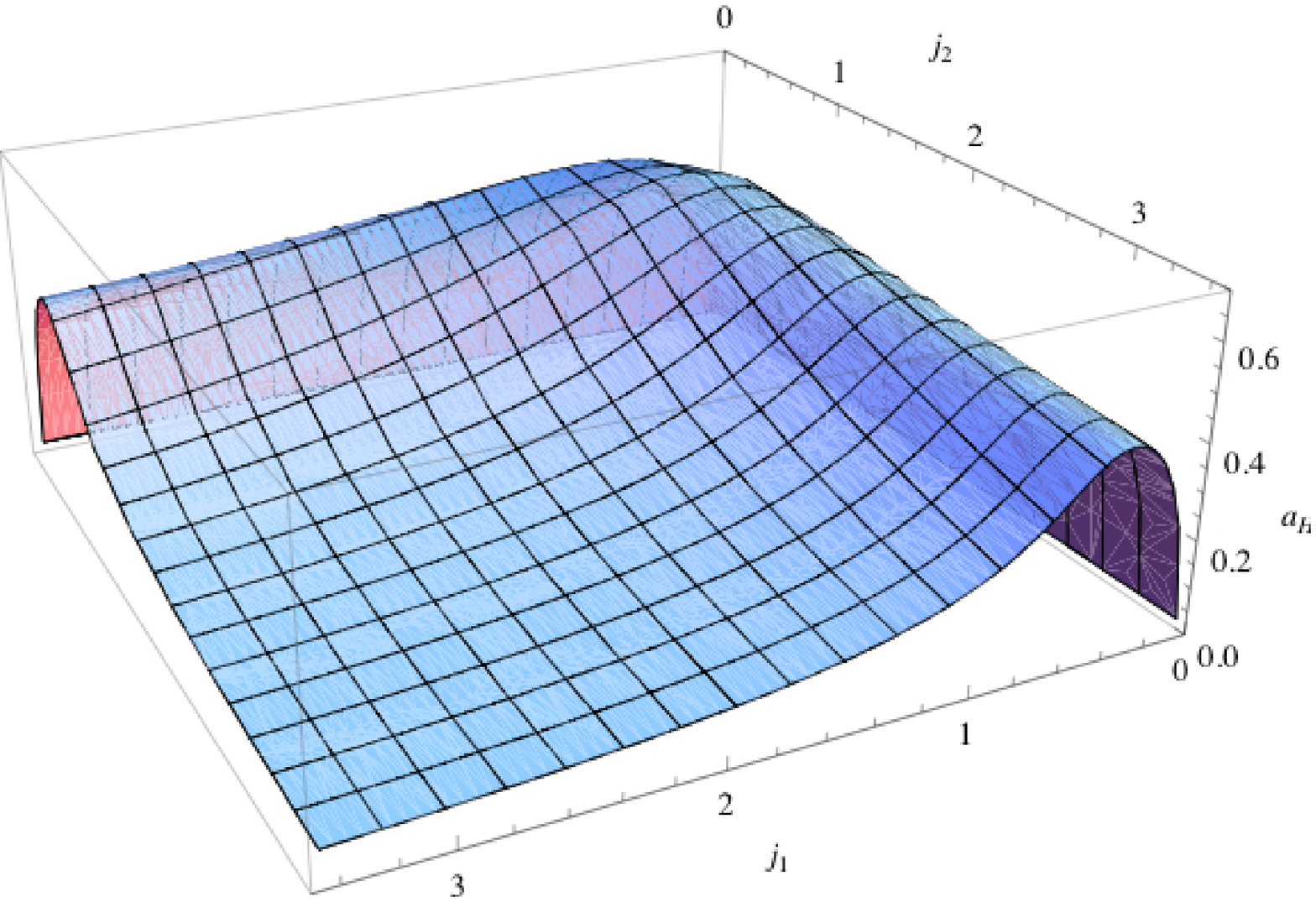}
\end{center}
\caption{\small{The phase space of extremal eight dimensional MP black holes.
{\bf Left:} The extremal locus as a function of the reduced angular momenta variables $(j_1, j_2, j_3)$.
{\bf Right:} The phase plot $a_H(j_1,j_2)$. The maximum occurs at the symmetric point $j_1 = j_2 = j_3$ and the limiting behaviour of the surface at large values of one of the $j_i$ coincides with that of the six dimensional MP curve. We can also take two angular momenta large in which case the solution behaves like a four dimensional Kerr black hole.}}
\label{fig:eightdextl}
\end{figure}

To check the assertion that the moduli spaces of extremal MP are
non-compact in higher dimensions it suffices to examine the
behaviour of $\CE_p$ as  the $j_i$ get large. In six dimensions it
is easy to check that $j_2 \gg 1$ implies $\CE_6(j_1,j_2) \sim 2 \,
j_1\,j_2^2$ which gives non-trivial solutions to \req{MPextgen}. The
extremal loci are illustrated in the \fig{fig:sixdextl},
\fig{fig:sevendextl}, and \fig{fig:eightdextl}, for six, seven and
eight dimensional MP black holes respectively.  The plots of the
extremal MP phase space as a function of the $j_i$s have been
described before in the literature: \fig{fig:sixdextl} (Left) has
previously appeared in \cite{Kleihaus:2007kc, Emparan:2008eg} while
\fig{fig:sevendextl} and  \fig{fig:eightdextl} (Left) were first
presented in \cite{Emparan:2008eg}. In all cases the non-compactness
of the moduli space is clearly visible.

A special case of interest for comparison to extremal black rings in \sec{hdrings} is the odd dimensional MP black holes with all but one angular momenta equal. In this case it is easy to show that (as before we take $a_i = a_2$ for $i =2, \cdots , n+1$)
\begin{equation}
r_+^2 = -\frac{n-1}{2\,n} \,a_1^2 + \frac{1}{2\,n}\,a_1\,\sqrt{(n-1)^2\, a_1^2 + 4\, n\, a_2^2}
\label{}
\end{equation}
which leads to
\begin{eqnarray}
\CE_D(j_1,j_2) &=& \left(\frac{2\,n\, j_2^2 - (n-1)\,j_1^2 \, + j_1 \,  \sqrt{(n-1)^2\, j_1^2 + 4\, n\, j_2^2}}{2\,n}\right)^{n-1} \nonumber \\
&& \qquad \times  \left( \frac{2\, n\, j_2^2 + (n^2 + 1)\, j_1^2 \, + (n+1)\, j_1 \, \sqrt{(n-1)^2\,j_1^2 + 4\, n\, j_2^2}}{2\, n}\right)
\label{extlocn1a2}
\end{eqnarray}
Even in this case the moduli space is non-compact; as $j_2 \to 0$ one has
\begin{equation}
 j_1 \sim\sqrt{\frac{(n-1)^{n-1}}{n^n} }\, \frac{1}{j_2^{n-1}} \, .
\label{alimsMP}
\end{equation}

Finally, there is another special case of interest; as is clear from the plots \fig{fig:sixdextl}, \fig{fig:sevendextl}, and \fig{fig:eightdextl}, the area for the MP black holes has a distinct maximum -- this occurs when all the angular momenta are equal. Physically this is  because the MP black hole is fattest at this point, by virtue of being uniformly distorted in all planes. Setting all the $a_i = a$ we have from \req{extcond} and \req{mpDpars}
\begin{equation}
\begin{aligned}
&{\underline{\rm Odd \;dimensional\; MP}}\;(D= 2n+3) \qquad &\qquad
&{\underline{\rm Even \;dimensional \;MP}} \; (d = 2 n +2 ) \\
&\qquad r_+ = \frac{a}{\sqrt{n}} & &\qquad r_+=\frac{a}{\sqrt{2n-1}}\\
& \qquad\mu_D = \frac{(n+1)^{n+1}}{n^n} \, a^{2n } && \qquad \mu_d = \frac{(2n)^n}{(2n-1)^{n-\frac{1}{2}}}\, a^{2n-1}\\
&\qquad j_\textrm{sym} = \frac{\!\!\sqrt{n}}{(n+1)^{\frac{n+1}{2n}}} & &\qquad  j_\textrm{sym} = \frac{\sqrt{2n -1}}{(2n)^{\frac{n}{2n-1}}}
\end{aligned}
\label{eqMPsym}
\end{equation}
where we have named the reduced angular momentum at this symmetric point $j_\textrm{sym}$.

\subsection{Membrane limits of MP black holes}
\label{membranelim}

The fact that the extremal MP black holes in $D>5$ have a non-compact parameter space, allows us to consider interesting limiting behaviour of the solutions.  In particular, following the discussion of \cite{Emparan:2003sy}, we can take some of the angular momentum variables to infinity and obtain an extremally spinning black brane. It is actually easy to see that this can be done at the level of the near--horizon geometry \req{mpbhmet}. Consider the following limit for the $2n+1$ and $2n+2$ dimensional extremal MP
\begin{equation}
a_j \to \infty \ , \qquad j=1, \cdots, p
\label{memlim}
\end{equation}
with $a_k$ (for $k=p+1, \cdots, n$), $r_+$, and define
\begin{equation}
\hat{\mu}= \frac{\mu}{ \prod_{j=1}^p a_j^2} \ , \qquad \sigma_j=a_j\,\mu_j \ , \qquad j=1, \cdots, p
\label{}
\end{equation}
which are all kept fixed. In this limit one can show:
\begin{equation}
\frac{\Pi'(r_+)}{\Pi(r_+)} \to \frac{\hat{\Pi}'(r_+)}{\hat{\Pi}(r_+)}, \qquad \frac{\Pi''(r_+)}{\Pi(r_+)} \to \frac{\hat{\Pi}''(r_+)}{\hat{\Pi}(r_+)} ,\qquad {\rm with} \;\;  \hat{\Pi}(r)= \prod_{k=p+1}^n(r^2+a_k^2) ,
\label{}
\end{equation}
and  $r_+$ now satisfies \req{mphor} and \req{extcond} with $\Pi(r) \to \hat{\Pi}(r)$.

\begin{figure}[t!]
\begin{center}
\includegraphics[scale=0.45]{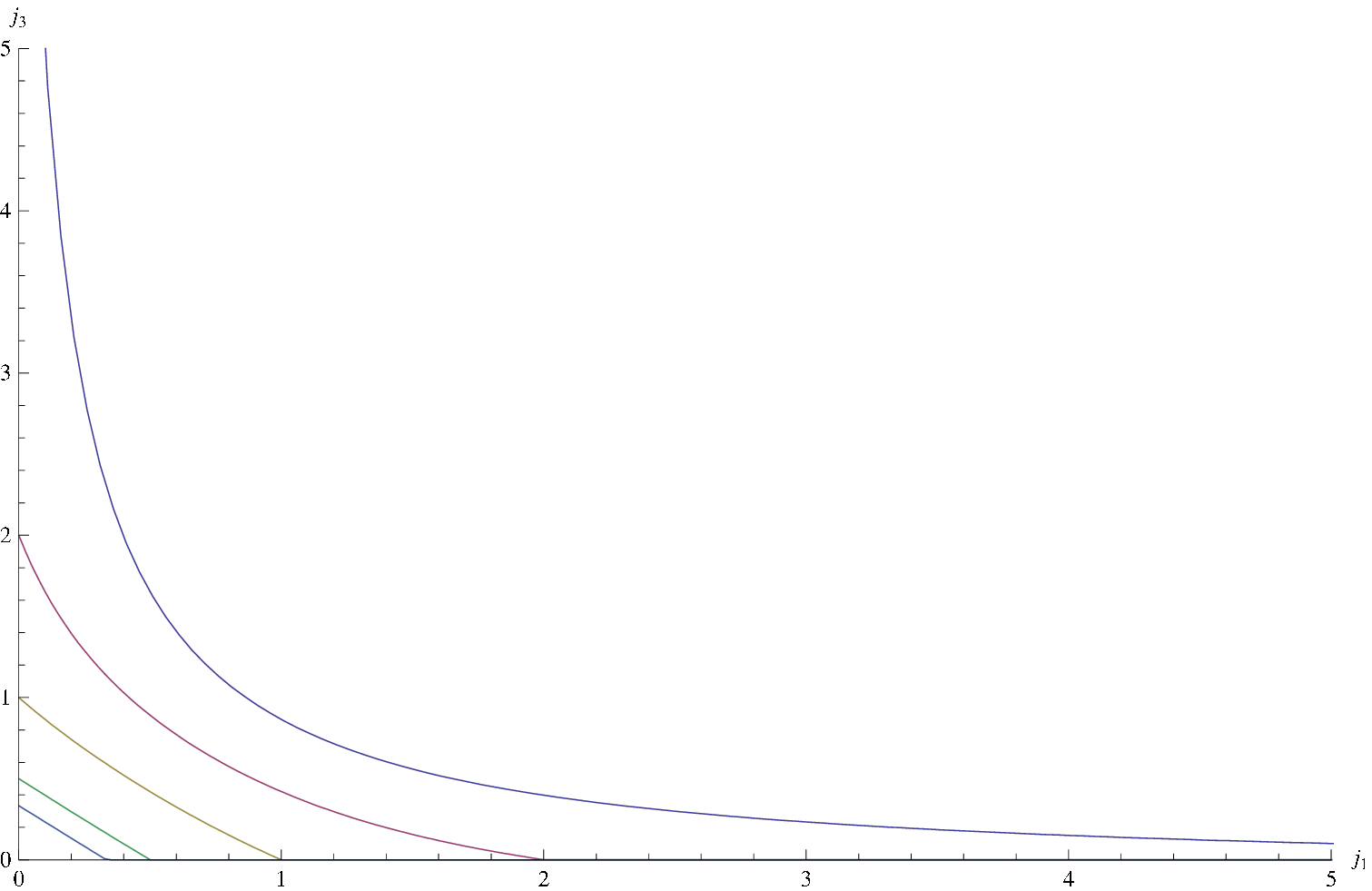}
\hspace{0.75cm}
 \includegraphics[scale=0.5]{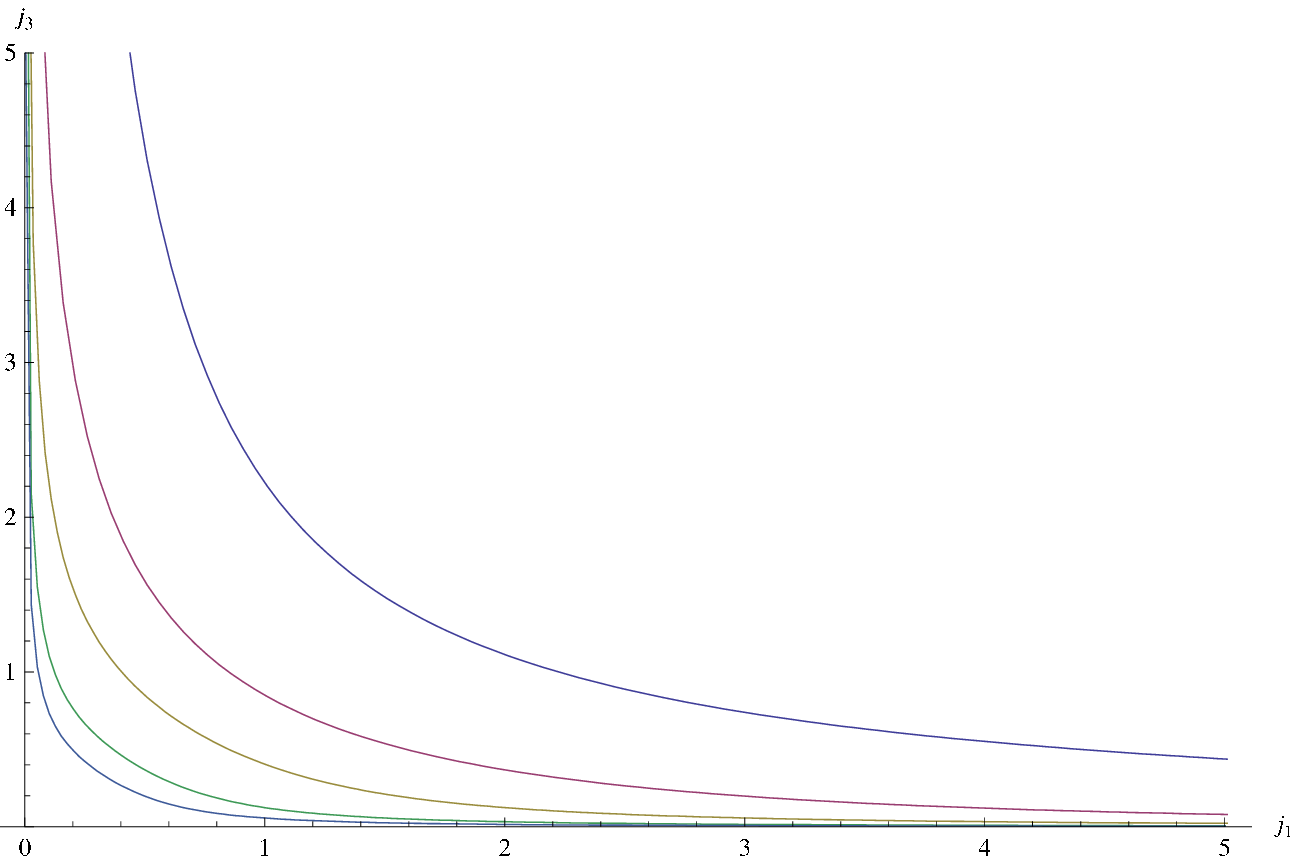}
\end{center}
\caption{\small{The extremal loci of MP black holes illustrating the membrane limit. {\bf Left:} The behaviour for 7d MP; the curves correspond to increasing values of $j_3$ moving from right to left. As $j_3$ gets large the extremal locus degenerates into a straight line as for the 5d MP, \cf, \req{fdmpextl} and \fig{fig:fivedphasediag}.  {\bf Right:} Eight dimensional MP where again we plot the curves are for different  $j_3$ values (increasing from right to left) -- compare the limiting behaviour with the 6d MP extremal locus, \fig{fig:sixdextl}.}}
\label{fig:membranelim}
\end{figure}

After taking this limit the near-horizon geometry of the MP metric becomes
\begin{eqnarray}
 ds^2 &=& \hat{F}_+ \left(
-\frac{\hat{\Pi}''(r_+)}{2 \hat{\Pi}(r_+)} \,r^2\,dv^2 + 2\,dvdr \right) + \sum_{k,l=p+1}^n \, \gamma_{\mu_k \mu_l} \,d\mu_k \,d\mu_l \\
&+& \sum_{k,l=p+1}^n\, \gamma_{k l} \, \left(d\phi^k
 +\frac{2\,r_+ \,a_k}{(r_+^2+a_k^2)^2}\,r\,dv \right)\left(d\phi^l+\frac{2\,r_+\, a_l}{(r_+^2+a_l^2)^2} \, r\,dv \right) + \sum_{j=1}^p \, d\sigma_j^2+\sigma_j^2 \,d\phi_j^2 \nonumber
 \end{eqnarray}
where
\begin{equation}
\gamma_{kl}= (r_+^2+a_k^2)\, \mu_k^2 \, \delta_{kl}+ \frac{1}{\hat{F}_+} \, a_k\mu_k^2 \, a_l \, \mu_l^2
\end{equation}
and
\begin{equation}
\hat{F}= 1- \sum_{k=p+1}^n \frac{a_k^2 \mu_k^2}{r_+^2+a_k^2}
\end{equation}
and
\begin{equation}
\begin{split}
\sum_{k,l=p+1}^n\,\gamma_{\mu_k \mu_l}\,d\mu_k\, d\mu_l &= \sum_{k=p+1}^n\,(r_+^2+a_k^2)\,d\mu_k^2, \\
 \sum_{k,l=p+1}^n\,\gamma_{\mu_k \mu_l}\, d\mu_k\, d\mu_l &= r_+^2\, d\alpha^2 + \sum_{k=p+1}^n\, (r_+^2+a_k^2)\, d\mu_k^2
 \end{split}
\end{equation}
in odd and even dimensions respectively. Note that $\sum_{k=p+1}^n \mu_k^2=1$ and $\alpha^2+ \sum_{k=p+1}^n \mu_k^2=1$ in odd and even dimensions respectively. Therefore, in the odd dimensional case we are left with exactly the near-horizon geometry of a $2n+1-2p$ dimensional extremal MP times $\R^{2p}$, and in the even dimensional case we are left with the near-horizon geometry of $2n+2-2p$ dimensional MP times $\R^{2p}$.

Therefore, we can reduce the even dimensional solution to a four dimensional extreme Kerr membrane ($p=n-1$) and the odd dimensional case to a five dimensional MP membrane ($p=n-2$).  This implies that in even dimensions one can take all but one of the angular momenta to infinity, while in odd dimensions we need to keep two of the angular momenta finite as we take the rest to infinity. This feature is clearly visible in \fig{fig:membranelim} where we show the projection of the extremal locus of MP black holes in 7d and 8d to the $(j_1,j_2)$ plane at different values of $j_3$; the extremal locus degenerates in the limit to that of 5d and 6d MP black holes, respectively.

\section{Extremal MP black strings}
\label{extrmpbstr}

One of the main motivations behind our exploration of higher dimensional vacuum extremal black holes is to learn about putative black ring solutions.
Before discussing the ring configurations we analyze some properties of boosted MP black strings in  $D=2n+3$ dimensions. This will be of use later, since one expects multiply spinning thin black rings to look like MP black strings.

The MP black string metrics in $D=2n+3$ dimensions are constructed
by taking a $d=2n+2$ dimensional MP black hole \req{mp2n2}, adding a
line with coordinate $z$ and boosting $(t,z) \to (\cosh\beta t-\sinh
\beta z, -\sinh\beta t +\cosh \beta z)$. As before we will denote
$s_\beta\equiv \sinh\beta$ and $c_\beta\equiv \cosh\beta$. In the MP
string metric we take $z$ to be a compact coordinate on an $\Sp^1$
with radius $\ell$ so $z \sim z + 2 \pi\, \ell$, and therefore the
spacetime is asymptotically $\R^{d-1,1} \times \Sp^1$.  The
resulting MP black string metric is parameterised by
$(\mu_d,a_i,\beta,\ell)$ where $i=1,\cdots n$ and $(\mu_d,a_i)$ are
the MP black hole parameters in $d=2n+2$ dimensions. Note that
extremality is achieved by taking
$\mu_d=\mathcal{E}_d(a_1,\cdots,a_n)$ as in (\ref{mpevenextl}). The
explicit metric is given in the Appendix in Boyer-Lindquist
coordinates, see \req{mpbsstringsol}. These MP black strings have a
regular horizon at $r=r_+$ (inherited from that of the $d$
dimensional MP black hole) whose spatial sections have topology
$\Sp^1 \times \Sp^{2n}$.

Recall that one can extend the ADM construction of the stress-energy
tensor in asymptotically flat spacetimes containing point-like
sources, to spacetimes with $p$-branes. Now, consider
$p$-dimensional extended sources in linearized gravity assuming that
the brane directions are translationally invariant. The
stress-tensor for the $p$-brane world-volume is then
\cite{Myers:1999ps}
\begin{equation}
 T_{ab}=\frac{1}{16\pi\, G^{(D)}_N}\int_{\Sp^{D-p-2}}\,d\Omega_{D-p-2}\; r^{D-p-2}\, \xi^i
\, \left[\eta_{ab}\,\big(\partial_ih^c_{\phantom c c}+\partial_ih^j_{\phantom j j}
    -\partial_jh^j_{\phantom j i}\big)
-\partial_ih_{ab}\right]\,,
\label{admst}
\end{equation}
where $a,b=0,\ldots, p$ denote the worldvolume directions, $i,j=1,\ldots,D-p-1$ denote the transverse directions to the brane, and $\xi^i$ is the unit normal to the transverse $(D-p-2)$-sphere. One should note that $h_{\mu\nu}=g_{\mu\nu}-\eta_{\mu\nu}$ is not gauge invariant, and in the expression above we have to use Cartesian coordinates.
For the boosted MP black strings in $D=2n+3$, the components of the ADM stress-tensor are given by
\begin{subequations}
\begin{align}
T_{tt}&=\frac{A_{2n}}{16\pi \,G^{(D)}_N}\,\mu_d\,\big[(2n-1)c_\beta^2+1\big]\,,\\
T_{tz}&=\frac{A_{2n}}{16\pi \,G^{(D)}_N}\,\mu_d\, (2n-1)c_\beta s_\beta\,,\\
T_{zz}&=\frac{A_{2n}}{16\pi \,G^{(D)}_N}\,\mu_d\, \big[(2n-1)s_\beta^2-1\big]\,. \label{pressure}
\end{align}
\end{subequations}
The mass and the momentum along the string can be computed by
integrating the energy and momentum densities\footnote{We will label
all string quantities by ``primed'' and reserve the un-primed
symbols for the corresponding black ring quantities.}
\begin{subequations}
\begin{align}
M'=&\int_0^{2\pi \ell} \,dz \,T_{tt}=2\pi \,\ell\, T_{tt}\,,\\
P'=&\int_0^{2\pi \ell} \,dz \,T_{tz}=2\pi \,\ell \, T_{tz}\,.
\end{align}
\end{subequations}
Defining an angular coordinate $\psi'$ as $\psi'=\frac{z}{\ell}$, so that $\psi'\sim \psi'+2\pi$, the ``angular momentum'' carried by the string along the $\psi'$ direction is simply given by
\begin{equation}
J_{\psi}'=\ell\,  P' \ .
\label{jprimedef}
\end{equation}
Of course, in addition to this ``angular momentum'' the string also carries angular momenta in each $\R^2 \subset \R^{d-1,1}$ which are parameterized by the $a_i$. These can be calculated using the standard Komar integrals near the horizon (\ie, using \req{komaram}) and can be shown to agree with the values for the $d$-dimensional MP black hole.

Given this data we can  write down the physical parameters, in terms of those of the
MP black hole in $d = 2n+2$ dimensions ($a_i$ and $\mu_d$)
and the period of the compact circle $\ell$:\footnote{Note that we present the momentum and boost along the string in terms of quantities adapted to rotational isometries rather than translational isometries for future use.}
 \begin{equation}
\begin{split}
M' &=\frac{A_{2n}}{8\,G^{(D)}_N}\, \left[(2n-1)\, c_{\beta}^2+1\right] \, \ell \, \mu_{d}\,,
\qquad \Omega_i'=~\frac{a_i}{c_{\beta}\, (r_+^2+a_i^2)}\, \qquad
\Omega'_{\psi}=\frac{s_\beta}{c_\beta\, \ell}\,,\\
J_i'&=\frac{A_{2n}}{4\,G^{(D)}_N}\,c_{\beta}~\ell a_i~\mu_{d}\,,\qquad
J_{\psi}'=\frac{A_{2n}}{8\,G^{(D)}_N}~(2n-1)\,s_{\beta}\,c_{\beta}~\ell^2~\mu_{d}\,,\\
\AH' &=2\pi\, c_{\beta}\,A_{2n}\, \ell\, r_+~\mu_{d}\,.
\end{split}
\label{mpstrpars}
\end{equation}
Note that these charges satisfy a Smarr-like relation (\cf, \cite{Townsend:2001rg, Harmark:2004ch, Kastor:2007wr} for discussions of first law and Smarr relation for black branes) :
\begin{equation}
\frac{D-3}{D-2}\, M' = \Omega_i'\, J_i'+\Omega_{\psi}'\, J'_{\psi}
-\frac{T_{zz}}{D-2} \; 2\pi \,\ell
\label{smarrstring}
\end{equation}
and first law
\begin{equation}
dM' = \Omega_i' \,dJ_i' +\Omega_{\psi}' \, dJ'_{\psi} -T_{zz}\,
d(2\pi \,\ell) \; .\label{flstring}
\end{equation}
Notice that both the first law and the Smarr relation incorporate explicit contributions coming from $T_{zz}$ which is the effective pressure or tension of the string. Also note that when $T_{zz}=0$ the Smarr relation and first law look like those for asymptotically flat black holes, provided one identifies the mass, \etc.

\section{Extremal black rings in five dimensions}
\label{fivedextr}

To set the stage for our analysis of extremal black rings we begin by recalling the situation in five dimensions where exact solutions are known.\footnote{See also the recent discussion of five dimensional extremal rings in \cite{Elvang:2007hs, Reall:2007jv}.} The original 2-parameter singly spinning black ring solution \cite{Emparan:2001wn} does not admit an extremal limit. However, its multiply spinning 3-parameter  generalization \cite{Pomeransky:2006bd} does (both solutions are balanced configurations) . Thus, in five dimensional vacuum gravity an exact solution representing an asymptotically flat extremal black ring is known. It is uniquely parameterized by its two angular momenta which must lie in the range $J_1>3\,J_2>0$ where $J_1$ is the angular momentum along which the $\Sp^1$ of the ring is aligned and $J_2$ is the angular momentum along the transverse $\Sp^2$.

Since this solution is extremal it admits a near-horizon limit. In~\cite{Kunduri:2007vf} it was shown that the resulting near-horizon geometry of the extremal black ring solution simplifies dramatically and it is given by a special case of the near-horizon geometry of the boosted Kerr-string. The near-horizon geometry of the extremal boosted Kerr-string is
\bea
\label{Kerrstring}
\nonumber ds^2 &=& \frac{1+\cos^2\theta}{2 \cosh\beta}\left(-\frac{1}{2a^2
\cosh\beta}\, r^2\,dv^2+2\,dvdr \right) + a^2\,(1+\cos^2\theta)\,d\theta^2 \\ &+& \frac{4a^2\,
\sin^2\theta}{1+\cos^2\theta} \,\left( d\phi' + \frac{\ell\sinh\beta}{2a}
d\psi' + \frac{r\,dv}{2a^2\,\cosh\beta} \right)^2 + \ell^2\, \cosh^2\beta \,d\psi'^2, \eea
where $\beta$ is the boost parameter, $a$ is the Kerr parameter, $\ell$ is the string radius and $\psi'$ and $\phi'$ are $2\pi$ periodic ($\phi'$ is the azimuthal
 angle of Kerr). This geometry is regular for any $a>0$ and $\ell>0$. From the results of~\cite{Kunduri:2007vf} it can be seen that the extremal black ring near-horizon geometry is given by (\ref{Kerrstring}) with the following constraints on the parameters:
\begin{equation}
\sinh^2\beta=1, \qquad \frac{\ell}{a}>4.
\label{fivedpar}
\end{equation}
The bounds on the parameter $ \frac{\ell}{a}$ can be better understood as follows. Noting that $\frac{(2+\lambda)^2}{2\lambda}$ is a monotonically decreasing function for $0 < \lambda < 2$ which ranges over the interval $(\infty, 4)$,  we can uniquely parameterize $\ell /a$ by
\begin{equation}
\frac{\ell}{a} = \frac{(2+\lambda)^2}{2\,\lambda} \ , \qquad {\rm for} \;\;  0< \lambda <2 .
\label{lampar}
\end{equation}
This choice of parameterization is such that $\lambda$ defined in \req{lampar}  is the same $\lambda$ with which the black ring solution is written in~\cite{Pomeransky:2006bd}.
Further, $(\phi',\psi')$ are related to $(\phi,\psi)$ of the extremal ring (defined as being orthogonal at infinity, with $\psi$ aligned along the $\Sp^1$ of the ring\footnote{We use conventional ring coordinates where $\psi$ refers to the angle in the plane of the ring in contrast to the choice made by \cite{Pomeransky:2006bd}.}) by:
\begin{eqnarray}
\label{psiphi}
\phi'&=&\phi+\psi, \qquad \psi'=\psi, \\
\partial_{\phi} &=&\partial_{\phi'}, \qquad \partial_{\psi}= \partial_{\psi'}+\partial_{\phi'} \, .
\end{eqnarray}

Given the near-horizon geometry of the boosted Kerr-string (\ref{Kerrstring}), the interpretation of the constraints on the parameters \req{fivedpar} is not apparent, as regularity of the geometry does not require them. However, they can be simply explained as follows. The value of the boost parameter actually corresponds to exactly the value which makes the Kerr-string tensionless. As a consequence, from the generalized Smarr relation \req{smarrstring} and first law for strings \req{flstring}, such solutions obey the standard Smarr relation and first law for asymptotically flat black holes (upon identifying the mass \etc.). Thus the tensionless condition for a string seems to be a necessary condition for its near-horizon geometry to correspond to that of an asymptotically flat black hole. The second constraint on the parameters $\ell /a >4$ can be written in terms of the radius of the $\Sp^1$ and $\Sp^2$ as follows \cite{Reall:2007jv}: $R_1=  \ell \sqrt{2}$ and $R_2= a \sqrt{2} $ and therefore $R_1/R_2>4$. This tells us that for a ring, in contrast to a string, one cannot take arbitrary values for the radii of the $\Sp^1$ and $\Sp^2$. There is yet another way of viewing this constraint on the parameters. The angular momenta for the extremal ring can be written in terms of the Kerr-string parameters:
\be
\label{5dringJ}
J_{\psi}= \frac{\pi\sqrt{2} a\ell}{G^{(5)}_N}(\ell+ 2a), \qquad J_{\phi}= \frac{2\sqrt{2} \pi}{G^{(5)}_N} a^2 \ell
\ee
leading to
\be
\frac{J_{\psi}}{J_{\phi}}= 1+\frac{\ell}{2a}>3.
\ee
Therefore the constraint on the parameters $(\ell,a)$ is exactly equivalent to the lower bound of $J_{\psi}$ of the ring. This provides a simple interpretation for the bounds on these parameters, as a black ring in asymptotically flat space needs a non-zero angular momentum along the ring to support it from collapsing. This is in contrast to a black string which may have arbitrarily small momentum along the string direction.

Thus, we have shown how the restriction on the parameters
\req{fivedpar} of the near-horizon geometry both originate from
properties of asymptotically flat black holes. However, we do not
have a good understanding of the origin of the coordinate change
(\ref{psiphi}) (although see \sec{jpsidetr}). Note that this
necessarily must contain information regarding how one can match the
near-horizon geometry to flat space at asymptopia.

Although certain physical properties can be computed from the near-horizon geometry alone, as argued earlier, quantities like the mass cannot in general. However, here we note that all the physical quantities of the extremal black ring are identical to those of the corresponding Kerr-string once (\ref{psiphi}) is taken into account (\ie, they do not receive $\ell^{-1}$ corrections to all orders).
Explicitly, the mass and angular velocities of the extremal black ring written in  terms of the string parameters are:
\begin{equation}
\label{mass5dr}
M=\frac{3\pi}{G^{(5)}_N}\,a\,\ell\,, \qquad
\Omega_\psi=\frac{1}{\sqrt{2}\, \ell}\,,\qquad \Omega_\phi=\frac{1}{2\,\sqrt{2}\, a} - \frac{1}{\sqrt 2\, \ell}\,.
\end{equation}
In fact the mass $M$, $J_{\phi}$, $\Omega_{\psi}$ do not actually depend on the coordinate change (\ref{psiphi}). The fact that the angular momenta match is not a surprise, since as we argued earlier these can be computed from the near-horizon data once one knows how to identify the angles. However, from this point of view, it is not clear why the mass and angular velocities should also coincide (of course it suffices to explain why the angular velocities match as then the mass follows from the Smarr relation).

\begin{figure}[t!]
 \begin{center}
 \includegraphics[scale=0.7]{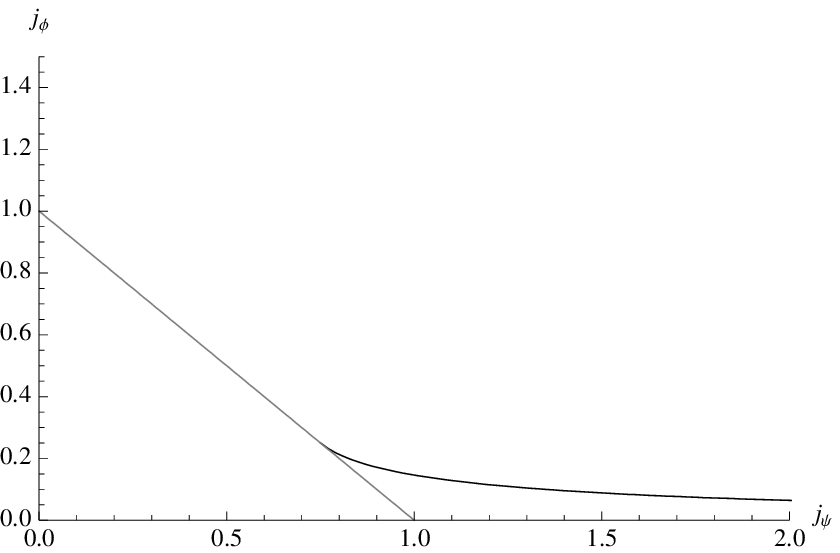}
\hspace{1cm}
 \includegraphics[scale=0.7]{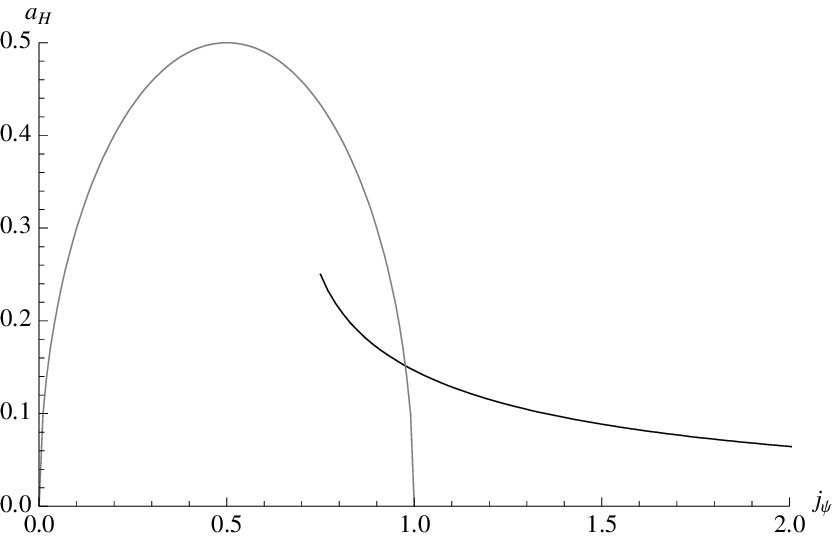}
\end{center}
\caption{Phase diagram for extremal MP black holes and doubly spinning black rings in $D=5$. The gray curve corresponds to the MP black hole and the black one to the black ring.  {\bf Left}: Plot of the $j_\phi$ vs. $j_\psi$ curve, where $j_\phi$ is the $\Sp^{2}$ angular momentum. Note that the point of intersection of the black ring and the black hole is excluded by the bounds on the angular momenta for the ring. {\bf Right}: Plot of the $a_\textrm{H}$ vs. $j_\psi$ curve. }
\label{fig:fivedphasediag}
\end{figure}

The physical parameters of this solution are best expressed in terms of reduced quantities
\begin{equation}
a_\textrm{H}=\sqrt{\frac{27}{256\, \pi}}\, \frac{\mathcal A_\textrm{H}}{(G^{(5)}_N\, M)^{3/2}} \ , \qquad
j_{\psi,\phi}=\sqrt{\frac{27\,\pi}{32\,G^{(5)}_N}}\, \frac{J_{\psi,\phi}}{M^{3/2}}
\label{ja5dring}
\end{equation}
which are used to plot the extremal locus of the solutions in
\fig{fig:fivedphasediag}. These plots have been described previously
in  \cite{Elvang:2007hs} (Right figure for the area) and
\cite{Emparan:2008eg} (Left figure for the moduli space of solutions
in five dimensions). We have a simple analytic expression for the
extremal locus using \req{5dringJ}, \req{mass5dr} and
\req{ja5dring}:
\begin{equation}
\label{5dphasering}
\CE^{BR}_5(j_\psi,j_\phi)= 8\, j_\phi \left( j_\psi - j_\phi\right) =1\ .
\end{equation}

Another interesting aspect of the five dimensional extremal
solutions is uniqueness \cite{Elvang:2007hs}. As is well known the
neutral singly spinning black ring solutions lead to a discrete
non-uniqueness; in a small window of the angular momentum $J_\psi$
(more precisely, $\sqrt{\frac{27}{32}} \le j_\psi < 1$) there are
three solutions (two black rings and a MP black hole) with the same
conserved charges. When we examine stationary extremal black holes
in five dimensions with two rotational Killing fields, we find that
uniqueness is restored. The situation is described in
\fig{fig:fivedphasediag}; on the Left plot the black ring curve
comes arbitrarily close to the MP curve but does not intersect. The
strict inequality $J_\psi > 3\, J_\phi$ for doubly spinning extremal
black rings excludes this intersection point.  A-priori there is no
reason why this should happen\footnote{We note that a similar
phenomenon occurs for supersymmetric black rings which must have
$J_{\psi}>J_{\phi}$ and the topologically spherical BMPV which has
$J_{\psi}=J_{\phi}$.}, however, this observation will prove useful
in our attempt to uncover properties of higher dimensional extremal
black ring geometries.

Therefore, to summarize, it appears that the extremal black ring
solution is related to the extremal boosted Kerr-string solution in
two logically distinct ways: the first coming from the equivalence
of the near-horizon geometries as explained above, and the second
coming from the fact that they have the same asymptotic charges (in
the sense that the ring quantities receive no corrections in
$\ell^{-1}$ relative to the string's to all orders).

\section{Extremal Black rings in higher dimensions}
\label{hdrings}

Having explored the space of known vacuum extremal solutions in diverse dimensions, we now turn to (yet to be found) black ring geometries in spacetime dimensions greater than five. Since exact vacuum solutions representing such objects are not known, our entire discussion will be based on the assumption that such solutions actually exist. Nonetheless, based upon the results presented previously, we will argue that one can still determine some important physical properties of such solutions.

\subsection{Extremal black rings as black strings}
\label{model}

To begin with it is useful to develop an intuitive picture for black rings in higher dimensions. As demonstrated in \cite{Emparan:2007wm} very thin black rings can be modeled as a bent black strings; this analysis was for black rings spinning only in the plane of the ring carrying no intrinsic angular momentum in the transverse directions. We expect that this class of solutions does not incorporate extremal black rings; this is because there are no known regular extremal vacuum solutions with vanishing angular momenta in any single plane. So we will concentrate exclusively on rings which spin in every available two plane.  This is different from the analysis of \cite{Emparan:2007wm}; nevertheless, very thin extremal rings can be thought of as multiply spinning black strings.

Moreover, for extremal solutions we have access to another well defined limit in which we expect black rings to look like black strings: the near-horizon limit.\footnote{This limit can be taken in a regime in which the gravitational self-interaction of the ring is strong \ie, not just for very thin rings.} After taking the near-horizon limit, we expect that the curvature of the ring disappears and is that of a straight string\footnote{This is because to move to co-rotating coordinates one needs to shift $\psi \to \psi -\Omega_{\psi}v$ and to take the near-horizon limit $v$ gets rescaled by a factor which is sent to $\infty$. Therefore this is like scaling the original $\psi$ by a factor which tends to $\infty$.}. This suggests that the near-horizon geometry of an extremal black ring is that of an extremal black string. Indeed this is exactly the case in five dimensions: the near-horizon geometry of the extremal black ring is isometric to that of a boosted Kerr black string~\cite{Kunduri:2007vf}. However, we will see this picture is only useful for odd dimensional black rings in $D=2n+3$ for $n \geq 1$.

As we move away from the near-horizon geometry, we do need to bend the $\Sp^{1}$ of the string to ensure that we have an asymptotically flat solution. This will not be possible for a generic boosted string. The reason is the tension carried by the string, determined by the $T_{zz}$ component of the effective stress tensor \req{pressure} . If we attempt to bend a tensile string we need to add energy-momentum into the system, taking us away from vacuum solutions. To ensure that we are able to wind the string into a ring, we demand that the string be tensionless. Using \req{pressure} this leads to the condition
\begin{equation}
T_{zz} = 0 \;\; \thus \;\;  \sinh\beta = \frac{1}{\sqrt{2n-1}}.
\label{pressless}
\end{equation}
Note that this the same as the condition derived in \cite{Emparan:2007wm} for the balance of singly spinning rings. The difference between our analysis and theirs is that they consider boosted Schwarzschild strings while we are interested in boosted extremal MP strings. While the two solutions are physically different, the leading contribution to the tension only arises from the mass of the solution and not the angular momentum, so the balance condition is unchanged.

We now describe some general expectations for extremal black ring solutions,  and  will then go on to describe the two limits mentioned above more precisely.

\paragraph{Symmetries:}
In $D>5$ we expect asymptotically flat black ring solutions to occur
with conserved charges $M,J_i$ with $i=1, \cdots n+1$ where $n=
[(D-1)/2]-1$, and symmetry $\R \times U(1)^{n+1}$. Such solutions
are presumably not always uniquely specified by their conserved
charges. Consider the subset of this family of solutions with zero
surface gravity. We expect such solutions to possess all $J_i$
non-zero (which is true for all known vacuum extremal black holes)
and further to be uniquely parameterized by these conserved charges,
as is the case in five dimensions \cite{Elvang:2007hs}. Let $J_i$,
for $i=2,\cdots,n+1$, be the angular momenta in the direction of the
transverse sphere $\Sp^{D-3}$ of the ring and $J_{\psi} \equiv
J_{1}$ be the angular momentum along the $\Sp^1$ of the ring
(orthogonal to $\Sp^{D-3}$ at infinity). The full solution will be
generically cohomogeneity $D-(n+2)$. However, if $J_i=J$ for
$i=2,\cdots,n+1$ we expect the solution to be cohomogeneity-2 in odd
dimensions with rotational symmetry $SU(n)\times U(1)^2$ (such a
symmetry enhancement occurs for all known solutions when one sets
all but one angular momenta equal).

As mentioned above there are two { \it distinct} limits in which we expect these geometries to simplify, namely the infinite radius limit and the near-horizon limit which we will now turn to. The first applies to both non-extremal and extremal black rings in any dimensions, while the second holds only for extremal black rings in odd dimensions.

\paragraph{Infinite radius limit:}
Let $R_{1}$ be the radius of the $\Sp^1$ of the ring\footnote{Since in general this can vary over the transverse sphere we will measure this at the poles of this sphere.} and $R_{2}$ be the effective radius of the $\Sp^{D-3}$ defined by its area. In the limit $\frac{R_1}{R_2} \to \infty $, we expect the geometry of the black ring to be given by that of a tensionless boosted MP black string, as is true for all known examples.\footnote{This was first observed for charged black rings \cite{Elvang:2003mj} and more recently \cite{Emparan:2007wm} has a detailed discussion of this issue.} Hence in the thin ring limit ($R_1 \gg R_2$),
the geometry of the black ring is well approximated by that of a straight MP black string with ${\cal{O}}(R_{2}/R_{1})$ corrections. This fact was recently exploited in~\cite{Emparan:2007wm}, where approximate solutions describing thin, singly spinning black rings for $D>5$ were obtained, by considering perturbations of Schwarzschild black strings. An important result obtained from this analysis was that approximate solutions could only be found provided that the strings were tensionless, \ie, $T_{zz}=0$, \req{pressless}, which agrees with the physical picture developed above. Note that the tensionless condition does not receive correction to order $R_2/R_1$ and therefore is valid not only in the strict infinite radius limit, but also for large but finite $R_1/R_2$.

\paragraph{Near-horizon limit:}
Now consider the near-horizon limit of such extremal black rings.
This will lead to near-horizon geometries specified by $n+1$
parameters and with spatial sections of the horizon of topology
$\Sp^1 \times \Sp^{D-3}$. When $J_i=J$ (for $i=2,\cdots n+1$), in odd
dimensions, we expect the near-horizon geometry to be cohomogeneity-1
with rotational symmetry enhanced to $SU(n)\times U(1)^2$. From the
theorem we proved in \sec{adssym} this implies that the near-horizon geometry should have $SO(2,1)$ symmetry. We expect such a symmetry enhancement in the
generic higher cohomogeneity case. Therefore the extremal black ring
near-horizon geometries should have $SO(2,1) \times U(1)^{n+1}$
symmetry.

Examples of near-horizon geometries satisfying such conditions are
easily constructed in odd $D$ as we now explain. Consider the
boosted MP-string in $D$ dimensions (\ie, one obtained by lifting
$D-1$ dimensional MP). Such a solution is specified by
$\mu,\,\ell,\,a_i,\,\beta$ where $\ell$ is the string radius,
$\beta$ is the boost parameter and $\mu,a_i$ are the mass and
angular momenta parameters (in the transverse directions to the
string) respectively, so $i=1,\cdots m=[(D-2)/2]$). The geometry
clearly has symmetry $\R \times U(1)^{m+1}$. Now take the extremal
limit which leads to a solution uniquely specified by $\ell,\, a_i,
\,\beta$ which are $m+2$ parameters. As we showed earlier, the
near-horizon geometry of this solution has $SO(2,1) \times
U(1)^{m+1}$ symmetry, is specified by $m+2$ parameters and spatial
sections of its horizon have $\Sp^1 \times \Sp^{D-3}$ topology. A
necessary condition required for these solutions to correspond to
the near-horizon limits of black rings is that the symmetry of the
solution matches, \ie, $m=n$. This occurs if and only if $D$ is odd
and thus $D=2n+3$. In this case the boosted extremal MP-string has
$n+2$ parameters. Thus, for each boost value one has an $n+1$
dimensional family of near-horizon geometries with the same
symmetry, topology and number of parameters as one would expect for
an extremal black ring. We derived the explicit form of these
near-horizon geometries in \sec{mpbsnh}, see equation
\req{MPstringNH}.

Therefore we expect the near-horizon geometry of odd dimensional
extremal rings to be given by that of the appropriate MP-string for
some particular value of the boost. In fact the boost must be such
that the MP string is tensionless, \ie, given by (\ref{pressless}).
The reasoning for this is as follows: we have argued that
(\ref{pressless}) must hold in the infinite radius limit. Further,
this condition (\ref{pressless}) does not receive corrections at
leading order in $R_2/R_1$ (at least in the  singly spinning limit),
and so we expect it to hold for all values of $R_1/R_2$. Therefore

\noindent {\bf Conjecture: } The near-horizon geometry of an
asymptotically flat extremal vacuum black ring in $D = 2n+3>5$
spacetime dimensions is globally isometric to the near-horizon
geometry of a boosted extremal MP black string carrying
non-vanishing angular momentum in all two planes $\R^2 \subset
\R^{2n}$ at a specific value of boost given by \req{pressless}.

Note that from the explicit near-horizon geometry we constructed earlier \req{MPstringNH}, this implies that:
\be
R_1 = \ell \sqrt{\frac{2n}{2n-1}}, \qquad R_2= \Pi(r_+)^{\frac{1}{2n}}.
\label{R1R2def}
\ee
In particular if $a_i=a$ (so $J_i=J$),  then $R_2$ simplifies and we have $R_1/R_2=\ell /a$. Also observe that this conjecture implies that the area of the horizon of the black ring $\CA_H$ is given by $\CA_H=\CA_H'$ where $\CA_H'$ is that of the MP black string \req{mpstrpars}.

\subsection{Conserved charges of extremal black rings}
\label{physparsbr}

We now turn to a discussion of the conserved charges of the ring, making use of the observations in \sec{nhphys} regarding which parameters  can be read off from the near-horizon.

First observe that the angular momenta of an asymptotically flat
vacuum black hole can be calculated from the near-horizon geometry
provided one knows how to identify the generators of the rotational
symmetries on the horizon with those at infinity (which are defined
as lying in orthogonal 2-planes). For a spherical topology black
hole with $U(1)^{n+1}$ symmetry the generators of rotational
symmetries are easily identified as the topology of the horizon is
the same as at infinity. However, for a black ring, such an easy
identification does not occur. This is because the generator of the
$\Sp^1$ is not uniquely defined as it has no fixed points. The
rotational symmetries of the $\Sp^{2n}$ do have fixed points and
these can be identified with the $n$ rotational Killing fields at infinity. Thus $J_i$
for $i=2,\cdots,n+1$ can be calculated from the near-horizon geometry,
but not $J_{\psi}\equiv J_{1}$. However, it must be the case that
\be
\partial_{\psi}=\partial_{\psi'}+ \sum_{i=2}^{n+1} \, c_i \partial_{\phi_i'} \;, \qquad \partial_{\phi_i}= \partial_{\phi_i'}\,\ee
where $\partial_{\psi'}$ is the generator of the $\Sp^1$ of the string in the near-horizon limit. As a coordinate change this reads: $\psi'=\psi$ and $\phi_i'=\phi_i+c_i\psi$, where $(\psi',\phi_i')$ are the $\Sp^1$ and $\Sp^{2n}$ coordinates of the string respectively. Further, $c_i$ must be integers to ensure the generators of
the $\Sp^1$ have closed orbits of period $2\pi$. It follows that
\begin{equation}
J_{\psi}= J_{\psi}'+ \sum_{i=2}^{n+1} \, c_i \, J_i' \; , \qquad J_{i}= J_i'
\label{cidefs}
\end{equation}
 where $J_{\psi}',J_i'$ can be
evaluated from the near-horizon geometry and are given by \eqref{mpstrpars}. Therefore one can
determine all angular momenta, up to the set of integers $c_i$. From
the explicit expressions for the angular momenta of the MP-string, \eqref{mpstrpars}
we find
\begin{equation}
J_{\psi} = J_{\psi}'\left( 1 + \sum_{i=2}^{n+1} c_{i}\frac{a_i}{\ell}
\frac{2}{\sqrt{2n-1}} \right) \ .
\label{jpsicorrl}
\end{equation}
These observations imply the following, using \req{R1R2def}: for the extremal black rings the $J_i$ (for $i=2,\cdots,n+1$) do not receive corrections in $R_2/R_1$ to any order, whereas $J_\psi$ can only receive a correction of order $R_2/R_1$ (iff any of the $c_i \neq 0$).

We will now turn to quantities which {\it cannot} be deduced from
knowledge of the near-horizon limit alone. The most important such
quantity is the ADM mass $M$ of the extremal black ring solution. We
will make the following assumption: the mass $M$ does not receive
any corrections in $R_2/R_1$ as compared to the mass of the string, \ie\ $M=M'$ where $M'$ is given in \eqref{mpstrpars}.
This fact is true in five dimensions as discussed earlier in \sec{fivedextr}.

Before closing this section we note the following: Consider the
co-rotating Killing vector of the extremal ring which is null on the
horizon $\partial_v = \partial_t + \Omega_{\psi}\, \partial_{\psi}+\Omega_i \,\partial_i$ where $\partial_t$ is the asymptotic stationary Killing vector. From this it follows that $\Omega_{\psi}=\Omega_{\psi}'$ and $\Omega'_i=\Omega_i+c_i\,
\Omega_{\psi}$. Therefore, from the Smarr relation for the tensionless MP string (\req{smarrstring} with $T_{zz} = 0$), we obtain
\be
\frac{D-2}{D-3}\,M'= \Omega_{\psi}'\,J_{\psi}'+\Omega'_i \,J'_i=\Omega_{\psi}\,J_{\psi}+\Omega_i \,J_i =\frac{D-2}{D-3}\, M\; .
\ee
The second equality follows from changing from $(\psi',\phi_i')$ (string) to $(\psi,\phi_i)$ (black ring) coordinates, whereas the third equality follows from the Smarr relation for asymptotically flat extremal black holes (which we know must hold in general).
This shows that the Smarr relation is actually insensitive to the knowledge of the integers $c_i$ and can therefore be used to determine $M$ given
$\Omega_i'$ and $\Omega_{\psi}'$. However, these angular velocities, like
the mass, are not encoded in the near-horizon limit. Thus, instead of
assuming the mass $M$ does not receive any $R_2/R_1$ corrections,
one could assume the $\Omega_i',\Omega_{\psi}'$ do not receive any
such corrections which then, via the Smarr relation, allows one to
deduce this fact for the mass as well. Note that this argument relies
crucially on the fact that the MP string being tensionless --
otherwise one would have an extra term in the Smarr relation coming
from that of the string \req{smarrstring}. Therefore, assuming the angular velocities
receive no corrections relative to the string is a {\it stronger}
condition than assuming this for the mass. We will not actually need
to make this stronger assumption to deduce the phase diagrams.

\paragraph{Summary:} We have argued that all conserved charges of the ring
are the same as those of the string, except for possibly $J_{\psi}$.
Further, the only way $J_{\psi}$ can differ from that of the string
is via a term of order $R_2/R_1$, \req{jpsicorrl},  if and only if any of the $c_i \neq
0$.

\subsection{Determining angular momentum in the plane of the ring}
\label{jpsidetr}

In this section we  will now present an argument which will allow us to deduce the set of integers $c_i$ discussed in the previous section and therefore $J_{\psi}$ for the conjectured extremal black rings. The upshot of our discussion will be that all the $c_i$ vanish for $D > 5$. Recall, from section \sec{fivedextr} we know that $c_i=1$ in $D=5$.

Following \cite{Emparan:2007wm}  consider thin extremal black rings, which corresponds to $R_1 \gg R_2$.  The authors of \cite{Emparan:2007wm} constructed perturbative solutions describing higher dimensional singly spinning black rings by computing the leading  $R_2/R_1 =\CO(\ell^{-1})$ correction to the boosted Schwarzschild black string and matching this onto a black ring solution valid in the weak field approximation. To repeat the analysis for extremal black rings, one would need to construct similar solutions by perturbing away from MP black strings. This requires knowledge of the appropriate sources for a multiply spinning black ring in the weak field approximation.\footnote{In order to construct the geometry of a black hole in a weak field approximation, we take $g_{\mu \nu} = \eta_{\mu \nu} + h_{\mu\nu}$ and consider appropriate effective sources for the stress tensor. $h_{\mu \nu}$ in transverse-traceless gauge $\nabla_\mu h^{\mu \nu} = 0$ and $h^\mu_\mu$ =0  satisfies  $\Box h_{\mu \nu} = - 16\, \pi \,G^{(D)}_N \, T_{\mu\nu}$.} The source encoding the angular momentum of a ring in the $\psi$ direction can be simply modeled as a current density along the ring. Intuitively, this is easy to understand once one thinks of the ring as a bent black string; in the string picture one has just momentum density which produces a current. For singly spinning rings, the rest of the source should then reduce to that of Schwarzschild at each point on the ring. Now, in the weak field approximation, a Schwarzschild black hole is flat space perturbed by a point mass for source; because these black holes possess only one length scale the weak field approximation is equivalent to the far-field behaviour.

If we wish to generalize this and write down appropriate sources for
the MP string, at each point on the ring one needs a source
corresponding to that of an MP black hole. However, such rotating
black holes possess other intrinsic length scales (associated to the
angular momenta) and therefore the weak field source is not given
simply by the far field solution. Indeed, the source for a Kerr
black hole is a complicated distribution of negative mass
density\footnote{A superluminally spinning disk of matter located on
the plane of rotation and bounded by the ring-singularity of Kerr
black hole \cite{Israel:1970ff}.} and one expects that MP black
holes to have similarly complicated sources. It is possible however
that one may need to only focus on slowly spinning black holes $ a
\ll r_+$. In this case one can utilize a point source of a spinning
particle, see \cite{Emparan:2008eg}, because we have established a
hierarchy of scales and are ignoring physics at sub-horizon
scale.\footnote{We thank Roberto Emparan for emphasizing this point
to us.}  In any case, we will present an argument that sidesteps the
precise knowledge of the sources of multiply spinning black rings.
To do so we will appeal to some results of the analysis of the
singly spinning case \cite{Emparan:2007wm}, which we now recall.

For thin singly spinning black rings, the leading $1/\ell$
correction to the geometry display interesting distinctions between
$D = 5$ and $D >5 $. In particular, the physical parameters of the
ring receive $1/\ell$ corrections in five dimensions, but remain
uncorrected at this order in higher dimensions. The reason for this
can be traced to regularity of the solutions. To see this in more detail consider the linearized solutions constructed in \cite{Emparan:2007wm} which are valid in the overlap region $r_0 \ll r \ll \ell$ where $r_0$ is the characteristic scale of
the transverse $\Sp^{D-3}$ (so $r_0 \sim R_2$) and $r$ is a radial
coordinate. The relevant component of the metric looks like
\begin{equation}
g_{t \psi'} = C \, \ell\,\left(\frac{r_0}{r}\right)^{D-4} \, \left( 1 + \frac{r}{\ell} \,  \cos\theta + \CO\left(\ell^{-2}\right) \right)
\label{schwfalloff}
\end{equation}
where $(\theta,r)$ are a set of ring adapted coordinates introduced
in~\cite{Emparan:2007wm} and recall $z = \ell\psi'$. This perturbation is regular in $D > 5$
but is not regular in $D = 5$. This can only be seen by comparing to
the regular linearized solution for a ring in asymptotically flat
space which is valid for $r_0\ll r$; in the overlap region this
solution possesses a constant term in $g_{t\psi}$ at
$\CO\left(\ell^{-1}\right)$ which ensures that $\partial_{\psi}$ has
a fixed point in the correct place at infinity ($\psi$ here refers to true $\Sp^1$ direction at infinity). To cure this
pathology in \req{schwfalloff} one needs to shift $t \to t -
\alpha \, r_0\, \psi'$. This results in a shift of the
physical parameters measured at infinity at order
$\CO\left(\ell^{-1}\right)$ . In higher dimensions the constant term
in the expansion of the weak field solution ($r_0 \ll r$) occurs at a
higher order $\CO\left(\ell^{4-D}\right)$ and thus the physical
parameters are not affected at $\CO\left(\ell^{-1}\right)$.

Now, consider constructing a multiply spinning ring solution in asymptotically flat space in the weak field approximation. The weak field source for such a solution must depend on $R_1$ (the ring radius), $R_2$ the radius of the transverse sphere, and the angular momenta $J_i$ in the transverse sphere. Such a solution is valid for $R_2\ll r$. Second, we take a linearized ring solution about the MP-string; this will depend on $\ell$, $r_+$ and the MP rotation parameters $a_i$; note that for extremal solutions $r_+=r_+(a_i) \sim a_i$. The regime of validity for such a solution is $r_+ \ll \ell $. Now, we would like to match these two solutions in an analogous manner to the analysis in \cite{Emparan:2007wm} which led to \req{schwfalloff} for singly spinning rings.   From our near-horizon analysis $R_1 \sim \ell$ and $R_2 \sim r_+$. Since there are two scales in the problem $\ell,r_+$ the trick is to work in a region where both of the approximate solutions above are valid. This occurs when $r_+ \ll r \ll \ell$. By continuity with the singly spinning case, we expect the linearized solution in this overlap region to look like
\bea
g_{t\psi'} &=& C \, \ell \left( \frac{r_+}{r} \right)^{D-4} \left[ 1+ \sum_{p=1}^{\infty}  F_p\left(\frac{a_i}{\ell},\mu_i \right) \left(\frac{r}{\ell} \right)^p \right],  \nonumber \\
{\rm with}&& F_p\left( \frac{a_i}{\ell}, \mu_i \right)= F_p(0,\mu_i)+ \mathcal{O}\left(\frac{a_i}{\ell} \right)
\label{mpstringexp}
\eea
where $F_p(0,\mu_i)$ is a function of the direction cosines $\mu_i$, and  is equal to $F_p(\theta)$ in the solution \req{schwfalloff} and $C$ is a constant independent of $\ell$.

The important thing to note is that the constant term in $g_{t\psi'}$ in \req{mpstringexp} is contained in $F_{D-4}= \CO(1)$ as $\ell \to \infty$. This constant term is affected  by gauge transformations: under $t \to t+ \delta\, \ell \, \psi'$ we find that  $F_{D-4} \to F_{D-4} + \ell^{D-4}\, \delta $ and thus $\delta =\CO(\ell^{4-D})$ for a finite limit. Such transformations do not change $\partial_t$ but do shift $\partial_{\psi'} \to \partial_{\psi'}- \delta \,\ell\, \partial_t$. We want to find the value of $\delta $ which shifts $\partial_{\psi'}$ to the vector which matches onto $\partial_{\psi}$. However, even without its knowledge we see that such a matching predicts that $J_{\psi}=J_{\psi}'+ \left(\frac{D-3}{D-2} \right)\delta \, \ell\, M$. Therefore, using $M=\mathcal{O}(\ell)$ and $J_{\psi}'=\mathcal{O}(\ell^2)$ (which follow from \req{mpstrpars}) we learn that $J_{\psi}=J_{\psi}'\, \left[1+ \CO(\ell^{4-D})\right]$.\footnote{Note that such a matching also predicts that $M$ receives no correction to this order, which is consistent with our assumption that it receives no corrections to all orders made in \sec{physparsbr}.}

However, in the previous section we argued that $J_{\psi}=J_{\psi}'[1+ O(\ell^{-1})]$ (see \req{jpsicorrl}), if and only if any of the $c_i \neq 0$, or $J_{\psi}=J_{\psi'}$ if all $c_i=0$. Therefore we see that for $D>5$ one must have all $c_i=0$ which implies
\begin{equation}
 J_\psi=J'_{\psi}
\end{equation}
and so in fact $\partial_{\psi}=\partial_{\psi'}$. Observe this argument is only valid for $D>5$. Indeed, in $D=5$ one does in fact get a correction to $J_{\psi}$ as can be seen explicitly from (\ref{5dringJ}).

The main point is that despite the lack of knowledge of the precise sources, the leading fall-off at large distances is given by the mass term -- for a string in $D$ dimensions, this is $r^{4-D}$. This then implies that any corrections to the physical parameters occur at $\CO\left(\ell^{4-D}\right)$, which is ruled out by the near-horizon analysis discussed in \sec{physparsbr}.  We may summarise the results of this section by the following

\noindent {\bf Claim: } For $D=2n+3\geq 7$, the Killing field that
generates translations along the string direction in the
near-horizon limit is proportional to the Killing field that
generates rotations along the $\Sp^1$ of the ring defined to be in a
plane orthogonal to the transverse $\Sp^{2n}$ at asymptotic
infinity.

\subsection{Bounds on black ring parameters and uniqueness}
\label{ringbdds}

We have determined the physical parameters of extremal black rings as explained above. To fully specify the solution, we must provide bounds on these parameters.  In general, for a black ring in asymptotically flat space one expects the angular momentum along the $\Sp^1$ of the ring to be bounded from below since it provides the centrifugal force to compensate the ring's tension and gravitational self-attraction. This is of course in contrast to a black string whose linear momentum can be arbitrarily small.

In \sec{fivedextr} we have seen that uniqueness is not violated for
five dimensional extremal black objects with a single connected
horizon. Rather, the bounds on the ring parameters admit extremal
ring solutions whose conserved charges are arbitrarily close to the
conserved charges of the MP black hole, but never equal. \textit{If}
we assume that this phenomenon extends to higher dimensions we can
determine a lower bound for $J_\psi$; this would be defined to be
the value where the extremal ring locus intersects the extremal
$D$-dimensional MP locus. It turns out that the MP locus always
intersects with our conjectured extremal black ring
locus\footnote{This is not the case when $c_i \neq 0$. In these
cases the conserved charges of the spherical black hole solutions
cannot become arbitrarily close to black ring ones. We have checked
this explicitly for the equal angular momenta  case, but believe it
to be generic.} as we will show in the next section. The upshot of
this proposal to constrain the parameters is that uniqueness would
not be violated.

Rather than working directly with the conserved charges and the area we will revert to reduced variables. Based upon our arguments, the reduced quantities \req{mpDred} for the conjectured extremal black rings are\footnote{Having argued that the angular direction in the plane of the ring $\psi$ is the same as direction of the string $ \psi' = \frac{z}{\ell}$ we will henceforth drop the distinction between the two and use $j_\psi$ to indicate the reduced angular momentum along the ring.} (recall $D = 2n +3$):
\begin{equation}
\begin{aligned}
j_i &=   \CN_n \,
\sqrt{\frac{2n}{2n-1}} \, \frac{a_i}{(\mu_d \, \ell)^{1/2n}}  = \CN_n \, \sqrt{\frac{2n}{2n-1}} \, \frac{q_i}{\CE_d(q_i)^{\frac{1}{2n}}}\\
j_\psi  &=
 \CN_n \, \frac{\sqrt{2n}}{2}\,\frac{\ell}{(\mu_d \, \ell)^{1/2n}} = \frac{1}{2}\, \CN_n \, \sqrt{2n}\, \frac{1}{\CE_d(q_i)^{\frac{1}{2n}}} \\
a_H &=  \CN_n \, \sqrt{\frac{2n}{2n-1}}\, \frac{r_+}{(\mu_d \, \ell)^{1/2n}}
\end{aligned}
\label{bringred}
\end{equation}
where
\begin{equation}
\CN_n = \left(\frac{1}{2\sqrt{\pi}}\, \frac{\Gamma\left(n+\frac{1}{2} \right)}{\Gamma(n+1)}\right)^{\frac{1}{2n}} \ .
\label{nndefs}
\end{equation}
The second equality follows from using the extremal locus of the $d$-dimensional MP black hole \req{mpevenextl} together with its homogeneity properties and $q_i = a_i/\ell$.
The dimension dependent normalization is the same as what we used earlier for the MP black hole in \req{mpDred} which was chosen to keep  reduced quantities for MP black holes simple. This convention differs from those used in~\cite{Emparan:2007wm}, which are chosen to simplify expressions for black rings. Explicitly, the reduced quantities $\j_a$ of~\cite{Emparan:2007wm} are related to ours by $
\j_a = \frac{1}{\sqrt{2n}\; \CN_n} \, \left(\frac{1}{2}\right)^{\frac{1}{2n}} \, j_a \ .
$  Note that due to the arbitrariness in the choice of the normalization of the reduced quantities it is only meaningful to compare ratios of reduced quantities in a given dimension.

The task now is to compare the reduced quantities for the rings to that of the MP black holes and plot the resulting phase diagrams.

\subsection{Phase diagrams of extremal black rings}
\label{ringphas}

We are now ready to draw the phase diagram for the higher (odd-)dimensional extremal black objects following our proposal. First of all we need to find the extremal locus for black rings generalizing \req{5dphasering}, analogous to the expression obtained in \req{MPextgen} for MP black holes. Essentially,  we are after the relation between $j_\psi$, the angular momentum in the $\psi$ direction and the $j_i$, the transverse angular momenta.  The data necessary to carry out this exercise is the reduced angular momenta for the ring, given explicitly in \req{bringred} and the extremal locus for even-dimensional MP black holes specified by the function $\CE_d(j_i)$. Using the homogeneity properties of the functions $\CE_d(j_i)$, it is easy to show that the desired extremal locus for the black rings is given by:
\begin{equation}
j_\psi = \frac{\alpha_n}{\CE_d(j_i)}\equiv \frac{\CN_n^{2n}}{2}\, \frac{\left(2n\right)^{n}}{\left(2n-1\right)^{n-\frac{1}{2}}} \, \frac{1}{\CE_d(j_i)}
\label{ringextl}
\end{equation}
which can be re-expressed in a terms of a homogeneous function $\CE^{\textrm{BR}}(j_\psi, j_i)$ as:
\begin{equation}
\CE^{\textrm{BR}}_D\left(j_\psi, j_i\right) \equiv  \frac{j_{\psi} \, \CE_d(j_i)}{\alpha_n} = 1  \qquad {\rm with} \qquad \CE^{\textrm{BR}}_D(\lambda\, j_\psi, \lambda \, j_i) = \lambda^{2n}\, \CE^{\textrm{BR}}_D(j_\psi,j_i)  \ .
\label{}
\end{equation}

Since the general expression for $\CE_d(j_i)$ for MP strings is
complicated, it is  useful to consider the simple case of equal
rotational parameters ($j_i = j$) on the $\Sp^{2n}$.  Exploiting the
results for extremal MP black holes in $d=2n+2$ dimensions with
equal angular momenta on $\Sp^{2n}$, \req{eqMPsym}, the black ring
extremal locus simplifies to
\begin{equation}
j_\psi = \frac{\CN_n^{2n}}{2} \, \frac{1}{j^{2n-1}} \, .
\label{ringjext}
\end{equation}

\begin{figure}[t!]
 \begin{center}
\includegraphics[scale=0.8]{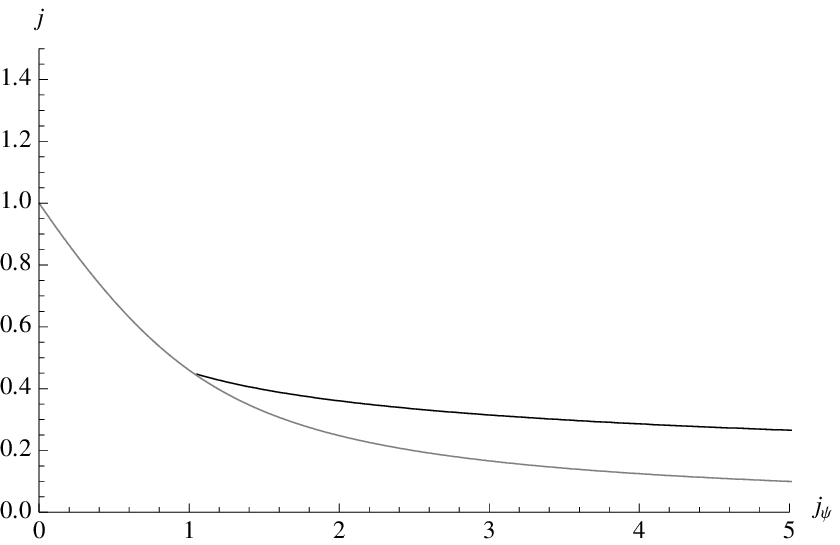}
\hspace{1cm}
 \includegraphics[scale=0.8]{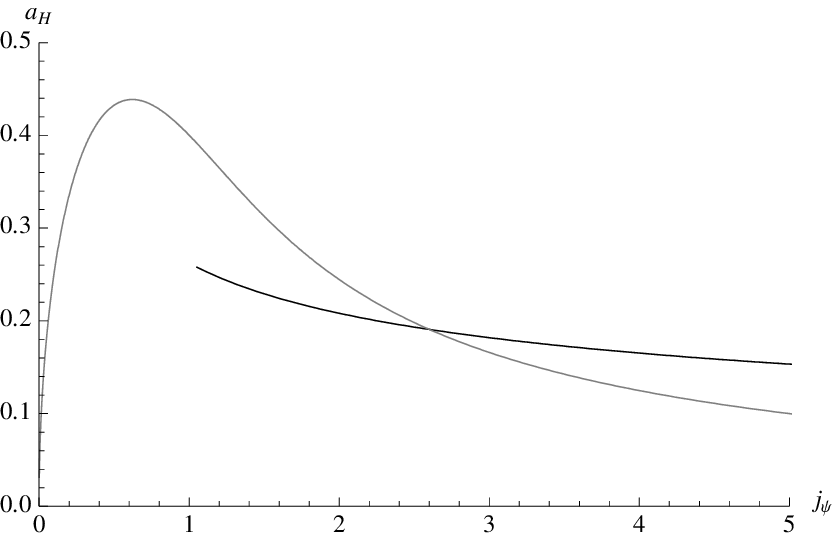} \\
\vspace{0.5cm}
\includegraphics[scale=0.8]{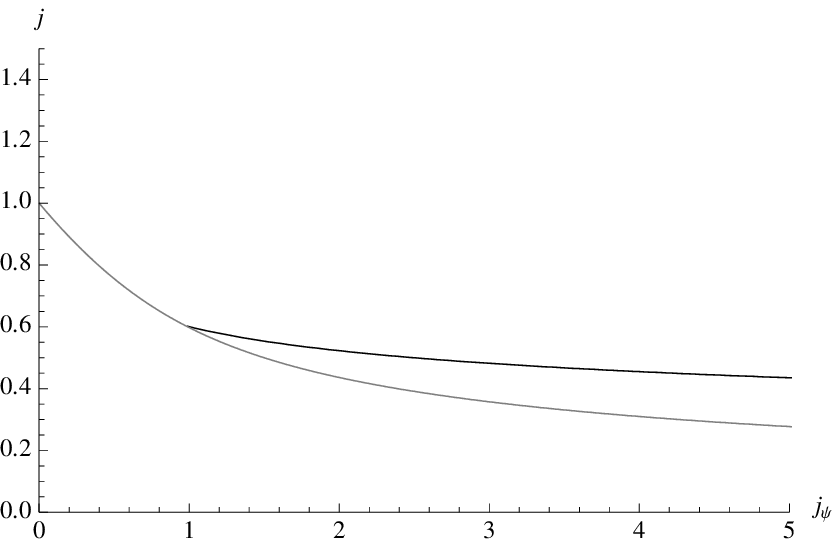}
\hspace{1cm}
 \includegraphics[scale=0.8]{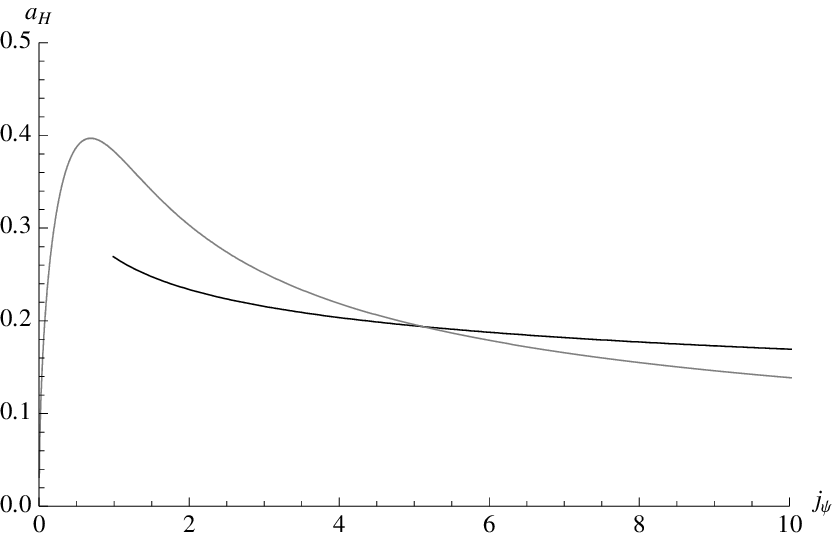} \\
\vspace{0.5cm}
\includegraphics[scale=0.8]{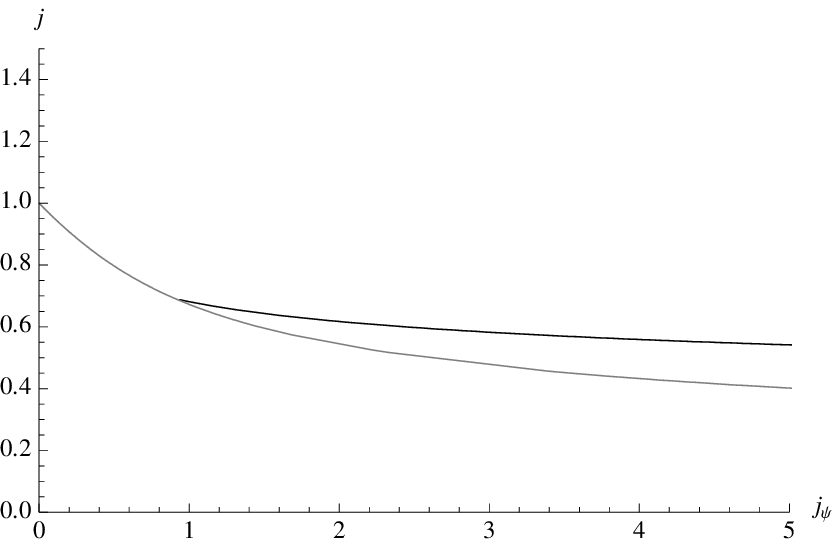}
\hspace{1cm}
 \includegraphics[scale=0.8]{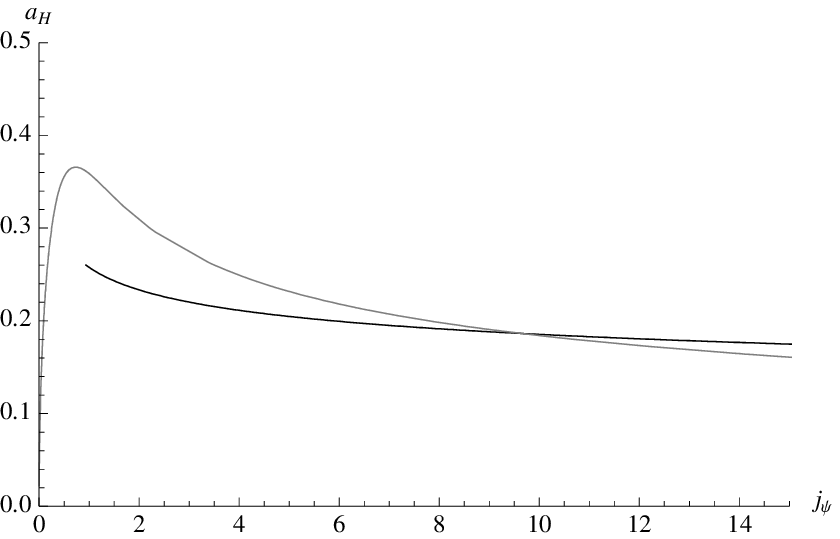}
\end{center}
\caption{Phase diagram for extremal black holes and black rings with  equal angular momenta on the transverse $\Sp^{2n}$  in $D=7,9,11$ (in this order.) In all cases, the gray curve corresponds to the MP black hole and the black one to the black ring. The corresponding phase diagram for $D=5$ is given in \fig{fig:fivedphasediag}.  The ring curve is terminated at the point of intersection with the MP curve; this preserves uniqueness and provides a lower bound on the allowed value of $j_\psi$. {\bf Left}: Plot of the $j$ vs. $j_\psi$ curve, where $j=j_i$ is any of the angular momenta on the $\Sp^{2n}$. {\bf Right}: Plot of the $a_\textrm{H}$ vs. $j_\psi$ curve.}
\label{fig:phasediag}
\end{figure}
This can now be compared with the behaviour of the extremal MP black
hole in $D = 2n+3$ dimensions with all but one equal angular
momenta. The extremal locus in this case is given for the black
holes by \req{extlocn1a2} with $j_1 = j_\psi$ and $j_2 = j$. The
resulting phase diagrams for $D= 7, 9,11$ are depicted in figure
\ref{fig:phasediag}, where we plot the extremal loci for the MP
black holes and the extremal black rings. An important feature,
consistent with uniqueness, is demanding that the potential
intersection point of the MP black hole and black ring curves is a
strict lower bound for $j_\psi$ of the black ring, called
$\left(j_\psi\right)_\textrm{min}$; for these solutions we predict
$j_\psi > \left(j_\psi\right)_\textrm{min}$ (\ie, strict inequality
as in the 5d case) thus avoiding an intersection.

\begin{table}[h]
\begin{center}
\begin{tabular}{cc}
\hline\hline $D$ & $(j_\psi)_\textrm{min}/j_\textrm{sym}$ \\
 \hline
7 & 1.657\\
9 & 1.421 \\
11 & 1.241\\
\hline\hline
\end{tabular}
\end{center}
\caption{\small{The minimal value of $j_\psi$ in various dimensions.}}
\label{lowbnds}
\end{table}

\begin{figure}[t!]
\begin{center}
\includegraphics[scale=0.37]{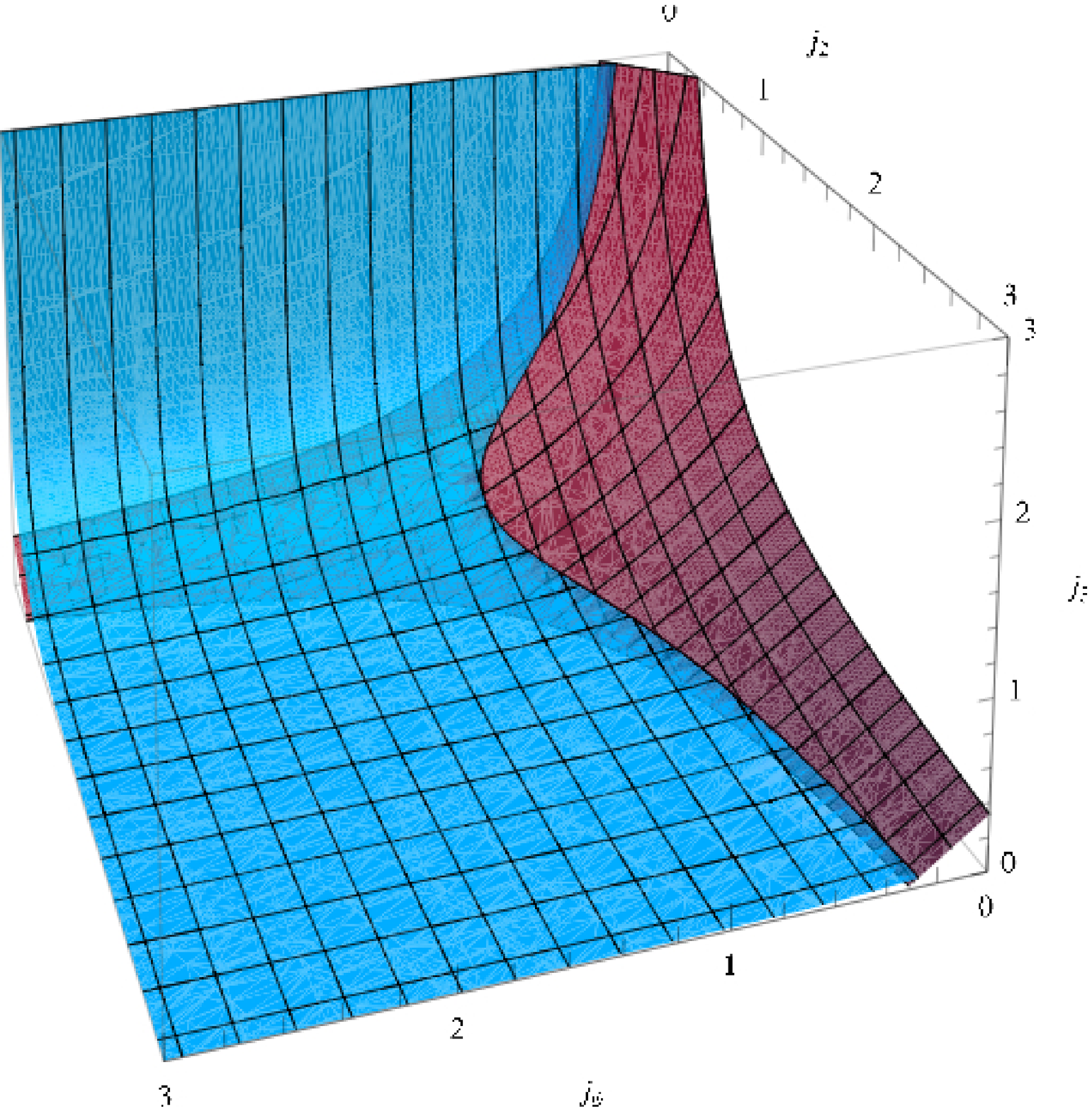}
\hspace{0.75cm}
 \includegraphics[scale=0.42]{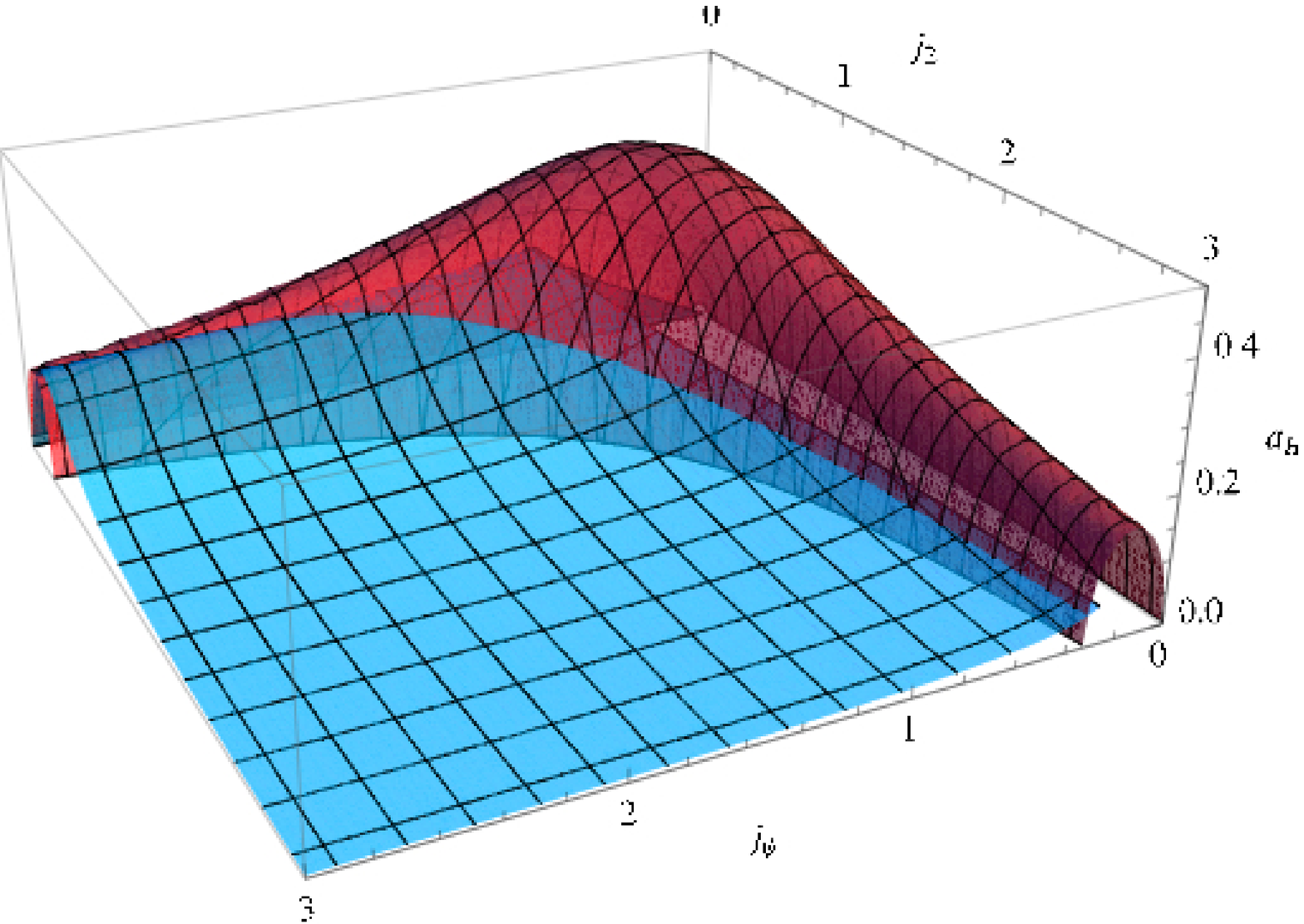}
\end{center}
\caption{\small{Our prediction for the phase diagram of seven
dimensional extremal black rings. We have superposed on this plot
the extremal MP solutions for comparison. {\bf Left:} The extremal
loci plotted as a surface in the reduced angular momenta
$(j_\psi,j_2,j_3)$ space. We have cut-off
the ring surface along the intersection curve with the MP extremal
locus. Nevertheless, there exist solutions with arbitrarily small
$j_\psi$. {\bf Right:}  Reduced area of the black rings and black
holes.}} \label{fig:phasediag7dp}
\end{figure}

In table \ref{lowbnds} we give the values of
$\left(j_\psi\right)_\textrm{min}$. To remove the effects of
dimension dependent normalization coefficients involved in defining
the reduced parameters \req{mpDred} and \req{bringred}, we present
these values compared to the value of the maximal area MP black
hole in $D=2n+3$ dimensions, \ie, $j_\textrm{sym}$ of \req{eqMPsym}.
Curiously, this minimal value decreases with dimension, lending
credence to the lore that gravity is weaker in higher dimensions
(and thus one ought to be able to balance rings more
easily).\footnote{For $D=5$ the minimal value of the true $j_\psi$
as in \req{ja5dring} measured thus is $1.5$, while the string result
of using $j_\psi$ from \req{bringred} (\ie, not accounting for the
mixing of angles) gives 1.707. This is a consequence of the $1/\ell$
effects in five dimensions.} In the general situation \ie, for MP
black holes with arbitrary rotation in the transverse $\Sp^{2n}$, we
continue to have an intersection between the ring and black hole
extremal loci. We expect this intersection always happens along a
co-dimension one surface; in seven dimensions the surfaces intersect
along a connected curve while in nine dimensions the intersection occurs along a connected surface.\footnote{It is interesting that the
intersection happens along a connected co-dimension one hypersurface
of the extremal loci. A-priori it was not guaranteed that the
surfaces intersect; at best one could have expected them to meet at
a disjoint union of lower dimensional hypersurfaces.} This is seen
clearly in \fig{fig:phasediag7dp}, where we see the intersection of
the surfaces of extremal MP and black rings in  $D=7$. The
intersection happens along a connected curve, which extends off to
large values of $j_2$ or $j_3$ and thereby allows extremal rings
with arbitrarily small values of $j_\psi$; hence there exist
extremal rings with $\left(j_\psi\right)_\textrm{min} \to 0$. This
feature is a novel prediction of our analysis for black rings in $D
> 5$.  

\begin{figure}[t]
\begin{center}
\includegraphics[scale=0.8]{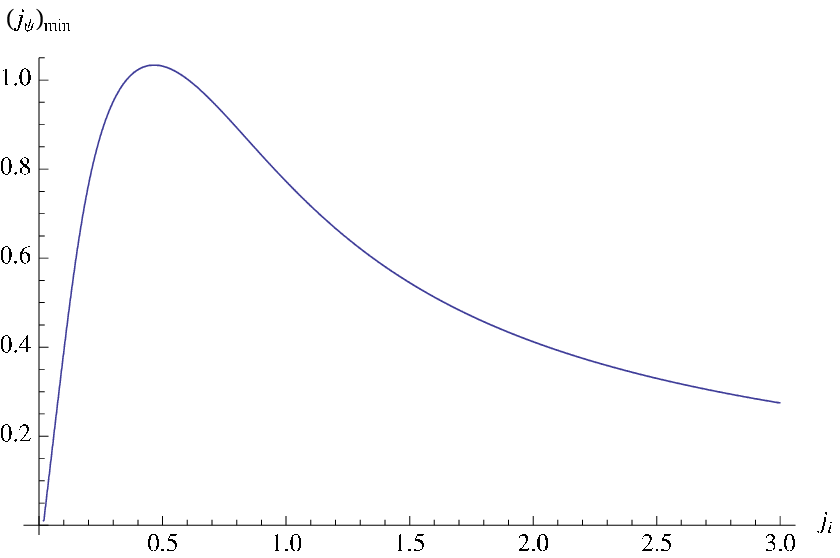}
\hspace{0.75cm}
\includegraphics[scale=0.37]{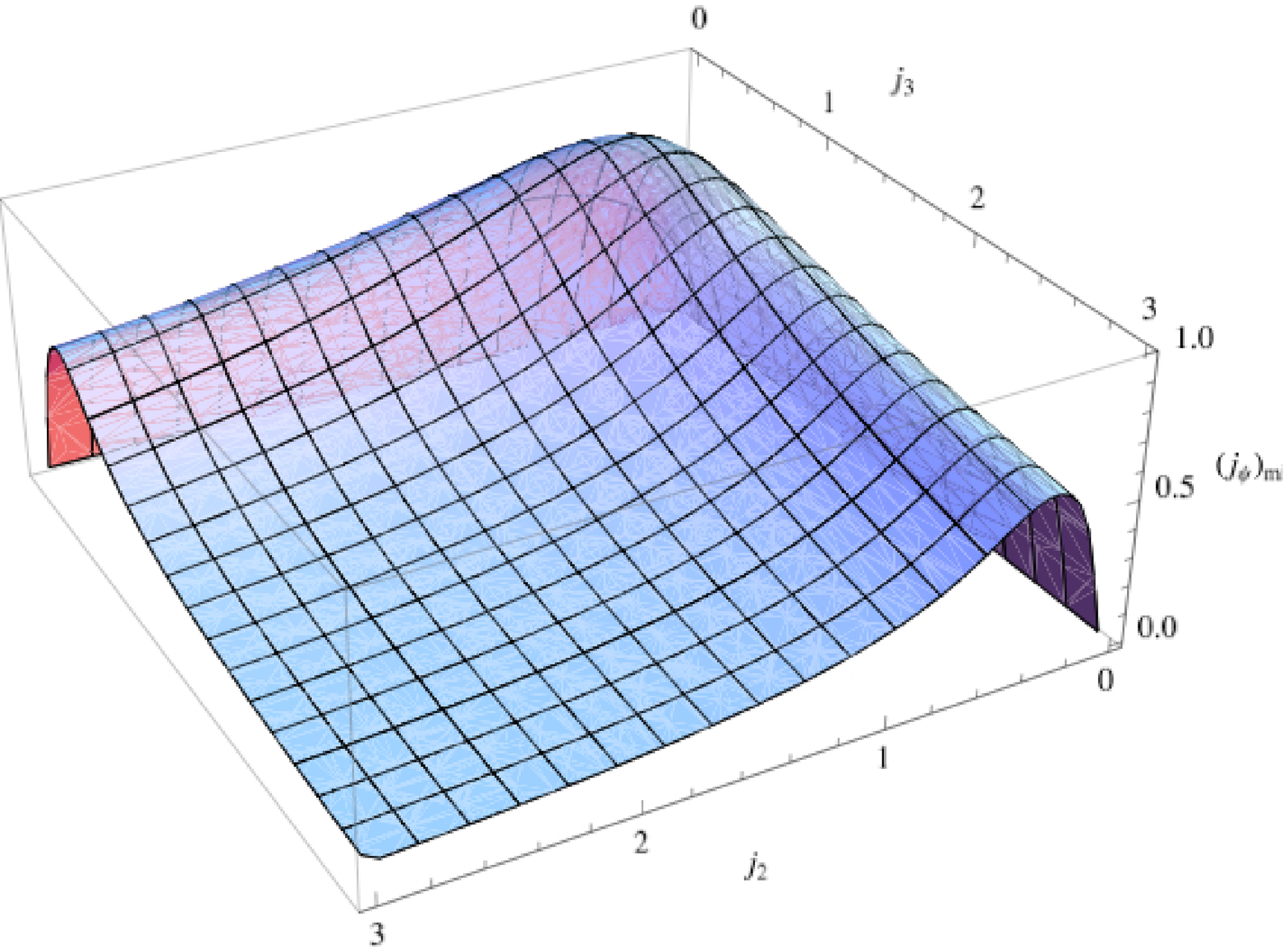}
\end{center}
\caption{\small{ {\bf Left:} The value of  $\left(j_\psi\right)_\textrm{min}$ as a function of $j_i$ for the seven dimensional extremal black rings. As emphasized in the text there are rings with arbitrarily small values of $\left(j_\psi\right)_\textrm{min}$ owing to the possibility of taking $j_2 \to \infty$ or  $j_3 \to \infty$. {\bf Right:} The value of $\left(j_\psi\right)_\textrm{min}$ as a function of $(j_2,j_3)$ for the nine dimensional black ring. Note that now both $j_2$ and $j_3$ can become large simultaneously.}}
\label{fig:jpsimin7d}
\end{figure}

\subsection{Predictions for extremal black rings}
\label{ringprops}

We now discuss a few features of  extremal black rings that our analysis reveals. We have already mentioned that higher dimensional extremal rings can stay balanced with arbitrarily small angular momentum in the plane of the ring (essentially by being spun up in the transverse directions), see \fig{fig:jpsimin7d}.  This is in fact in line with the observation made for doubly spinning (non-extremal) rings five dimensions in \cite{Elvang:2007hs} that increasing the transverse spin tends to reduce $\left(j_\psi\right)_\textrm{min}$. Intuitively, this might be attributed to a spin-spin interaction of anti-podal components of the ring along the $\Sp^1$.\footnote{We thank Veronika Hubeny for emphasizing this point to us.}

\paragraph{Entropy:} For large $j_\psi$, the black rings are entropically more favourable (\cf, \fig{fig:phasediag}) than the MP black holes. While such behaviour is expected for singly spinning rings (necessarily non-extremal), due to the ultra-spinning instability of the MP black holes \cite{Emparan:2003sy}, in the extremal context its origins are different. One heuristic reason for this is that thin rings are like hula-hoops, while ultra-spinning black holes (even extremal) are like disks; for given mass the disks tend to be thinner than rings and hence have smaller area \cite{Emparan:2007wm}.  Of course, we are far from Newtonian physics in this strong gravity regime of extremal black holes, but the intuitive argument serves to illustrate the essential distinction between the objects.

 From \fig{fig:phasediag} it is clear that the black ring area is monotonically decreasing with $j_\psi$ for the special case of all the transverse angular momenta being equal, unlike the MP black hole where the area is maximized at the symmetric point ($j_\psi = j$). This is because the ring picks out a preferred direction and so the area has no characteristic feature as a function of $j_\psi$. On the contrary, the   absence preferred choice for MP black holes results in an asymmetric behaviour of $a_H(j_\psi)$, with a characteristic maximum.  This monotonicity property for the rings is lost when we consider generic rotations in the transverse $\Sp^{2n}$.

\paragraph{Membrane limit:} In \sec{membranelim} we showed that the extremal MP black holes have an ultra-spinning limit following \cite{Emparan:2003sy}; this limit appears to be more subtle for extremal black rings. To see this recall that above we demonstrated that the membrane limit commutes with the near-horizon limit for MP black holes.  Now consider taking the limit in the near-horizon of an extremal black string with horizon topology $\Sp^{1} \times \Sp^{2n}$ in $D$ spacetime dimensions. We expect to obtain a black brane with horizon topology $\Sp^{1} \times \Sp^{2n-2p}\times \R^{2p}$ by the arguments of \sec{membranelim}. Ignoring the flat space factors thus generated (arising from the $\R^2$s where we sent $a_i \to \infty$), we have a black string in some $D-2p$ dimensions. This string is however tensile -- the pressurelessness condition \req{pressless}  being dimension dependent, leads to differing values of the boost for the string and the membrane (in the former case we have the boost as in \req{pressless}, while in the latter the boost would have to be $ \sinh^2\beta = \frac{1}{2\, n -2\, p -1}$).   So assuming that a potential membrane limit commutes with the  near-horizon limits for the extremal black rings, we are led to a contradiction.  On the other hand, inspection of the phase diagram \fig{fig:phasediag7dp} suggests that as one of the $j_i$ gets large, the ring `behaves' like a lower dimensional ring.  However, upon closer inspection, it appears that in addition to taking the $j_i \to \infty$ we need to  rescale $j_\psi$ by a finite amount to precisely recover the lower dimensional ring's extremal locus. This is consistent with the differing values of boost described above. Putting these things together suggests that there is no simple analog of the membrane limit of MP black holes for the black  rings.

\section{Discussion}
\label{discuss}

The main focus of the present paper has been an exploration of vacuum extremal black holes in dimensions greater than five. As motivated in \sec{intro} the restriction to  extremal solutions is non-trivial and as we have discussed there is a rich moduli space of such solutions even in vacuum gravity. Apart from examining aspects of known examples of extremal black objects based on the MP black holes, we have also obtained predictions of certain properties of hitherto unknown extremal black ring solutions.

On the mathematical side, we have shown that vacuum solutions with
degenerate horizons have an enhanced near-horizon \ads{2} isometry
assuming the rotational symmetry of the solution enhances such that
spatial sections of the horizon are cohomogeneity-1. From the known
solutions to date, MP black holes (and strings) with equal angular
momenta in even dimensions and all but one equal angular momenta in
odd dimensions fall into this class of geometries. It is natural to
expect that this symmetry enhancement extends to situations where
the spatial sections of the horizon have less symmetry given the
fact that we proved this was the case for the generic extremal MP
black holes with unequal angular momenta. However, proving the
general statement requires new techniques beyond those developed in
\cite{Kunduri:2007vf} which we exploited. The primary complication
is that the method used is well adapted to cohomogeneity-1 metrics
on the horizon; higher cohomogeneity metrics occurring in the
generic case are troublesome due to absence of well-adapted
coordinates for the analysis. Nevertheless, given that there are
high cohomogeneity metrics displaying the enhanced symmetry, it
would be interesting to see if the Theorem proved in \sec{adssym}
can be generalized.

We have also presented explicit examples of vacuum near-horizon geometries with horizon topology $\Sp^{D-2}$ and $\Sp^{1} \times \Sp^{D-3}$. Of course, in higher than five dimensions one might expect other topologies. Although we have concentrated on black rings, it would be very interesting to find solutions with even more exotic horizon topology.\footnote{Of course one can trivially construct vacuum near-horizon geometries by taking the direct product of $\R^{1,1}$ with a Ricci flat metric on $\CH$, $ds_{NH}^2=dv\,dr+ds^2(\CH)$. While this gives examples with non-trivial horizon topology they are in a sense trivial and we do not expect them to correspond to near-horizon limits of asymptotically flat black holes. It is the non-trivial examples which take the form of fibrations over AdS$_2$ which we are interested in.}

Given the near-horizon geometry alone, an important question is how much of the physical characteristics of the solution, such as the conserved charges \etc, can be determined. We have argued that while it possible using Komar integrals to capture certain angular momenta, the near-horizon metric does not capture the mass of the black object or the angular velocities (both of which depend on the asymptotic stationary frame). More generally, as discussed at end of \sec{nhphys}, a given near-horizon geometry might not always be extended to a full solution with prescribed asymptotics. Understanding the constraints on when this can be done is an interesting open question.

The new physical result in the paper is the prediction of the phase structure for extremal black rings in higher dimensions. As we have emphasized, higher dimensional black holes/rings are not easily amenable to analytic treatment owing to the lack of solution generating techniques. Given this, our strategy has been to exploit the mathematical result above, combined with certain plausible physical assumptions, to construct the phase diagram for black rings.

The logical argument for our construction of extremal black rings
can be summarized as follows: given the near-horizon symmetry
enhancement for degenerate horizons, extremal black rings should
have a near-horizon that has three main characteristics (i) \ads{2}
isometry, (ii) ring topology, and (iii) correct rotational symmetry,
\ie, must be a subgroup of $SO(D-1)$.  We already know of a solution
in odd spacetime dimensions with these characteristics -- the
near-horizon geometry of the extremal MP black string. The MP black
string is not asymptotically flat, but heuristically, can be made so
if the string is tensionless, by bending it. The tensionlessness
condition fixes the boost parameter of the string to a numerical
(dimension dependent) constant, which is independent of the
transverse rotation. Our main claim is that the black string
near-horizon geometry at this particular value of the boost
parameter is {\it isometric} to the near-horizon geometry of an
extremal black ring. The rest of the analysis is then aimed at
determining the physical parameters of the black ring. We have
argued that the conserved quantities of the extremal ring are the
same as those for the corresponding black string.

We infer the angular momenta in the transverse sphere of the ring using the Komar integrals -- these are therefore the same as the corresponding values of the MP black string. We assume the mass of the asymptotically flat black ring receives no corrections relative to the MP black string. The remaining physical parameter, the angular momentum in the plane of the ring we also argue agrees with the string. Deducing this involves realizing that the only admissible correction to the string's angular momentum occurs at leading order in  $1/\ell$ (recall that $\ell$ size of the circle wrapped by the string) and showing that such corrections do not arise in $D > 5 $. This then completes the determination of the physical parameters. Finally, we put bounds on the physical parameters by requiring that  black ring extremal locus is unbounded above and bounded below by the extremal locus of the MP black holes. This requirement implies uniqueness is not violated, but this is not demanded by any known theorem. Curiously, uniqueness is satisfied by the known extremal black objects in five dimensions. It would be very interesting to understand the origin of this uniqueness and explore its consequences further.

Of course, a generic feature of higher dimensional non-extremal black holes is violation of uniqueness.\footnote{In the context of generalized Weyl solutions, \cite{Hollands:2007qf} it is shown that the rod-structure of a solution together with its conserved charges fully specify the solution.} Therefore, from this point of view, it seems remarkable that uniqueness is restored for extremal black holes in five dimensions. Curiously, note that the extremal doubly spinning rings do come arbitrarily close to violating uniqueness \cite{Elvang:2007hs}. It is interesting to observe that based on our proposal, if any $c_i\neq 0$ (in \eqref{jpsicorrl}) it seems one cannot get extremal rings arbitrarily close to MP black holes. It is thus tempting to speculate that this suggests all $c_i=0$ (consistent with our final proposal), so that extremal higher dimensional rings can come arbitrarily close to violating uniqueness as in five dimensions.

Our analysis throws up some interesting features of five dimensional solutions which do not seem to be shared by their higher dimensional counterparts. Apart from the fact that extremal rings in five dimensions have slightly different characteristics (in terms of the mixing of the ring and the transverse directions), there is a curious fact about the entropy of five dimensional black objects. The area formula for extremal five dimensional MP or doubly spinning black rings takes an extremely simple form. For example, for the extremal black ring solution of \cite{Pomeransky:2006bd} one has
\begin{equation}
\CA_H = 8 \pi \, J_\phi  \ .
\label{}
\end{equation}
The simplicity of this and analogous expressions for four and five dimensional black holes have been used to motivate a microscopic counting of the entropy of these objects \cite{Emparan:2006it,Emparan:2007en,Horowitz:2007xq,Reall:2007jv,Emparan:2008qn}. These results rely on locally supersymmetric D-brane intersections (with supersymmetry being globally broken due to different supercharges being preserved at different intersections). It is well known that in dimensions greater than five there are no localized supersymmetric black holes. Further, the known extremal solutions in vacuum gravity have complicated area formulae. For example for the 6d extremal MP one has
\be
\CA_H= 8\pi \sqrt{ \sqrt{ \frac{(J_1^2+J_2^2)^2}{36} + \frac{J_1^2J_2^2}{3}} - \frac{J_1^2+J_2^2}{6}}.
\ee
This seems suggestive that the mechanism for micro-state counting of black holes in four and five dimensions is a happy accident of local supersymmetry, while higher dimensional extremal solutions are intrinsically more complex.

Although the focus of this paper has been on asymptotically flat
vacuum black holes let us comment on AdS black holes, as some of the
techniques used in this paper can easily accommodate a cosmological
constant. For example, the theorem we proved on symmetry enhancement
in the near-horizon limit applies equally to AdS vacuum gravity. Our
discussion of extremal black holes has relied heavily on the use of
Gaussian null coordinates and the `double-scaling' limit which
defines the near-horizon geometry \req{nhgeom}. Indeed such
coordinates have been used previously in the context of
supersymmetric AdS$_5$ black holes. In particular, under some
assumptions, supersymmetric AdS$_5$ black rings were ruled out. This
was deduced from a classification of all possible near-horizon
geometries of supersymmetric AdS$_5$ black holes with $R \times
U(1)^2$ symmetry~\cite{Kunduri:2006uh,Kunduri:2007qy}.

More recently, again in the AdS/CFT context, a map relating
gravitational solutions in asymptotically AdS spacetimes to
solutions of conformal fluid dynamics has been constructed in a
long-wavelength approximation \cite{Bhattacharyya:2007vs,
Bhattacharyya:2007jc,VanRaamsdonk:2008fp}. The discussion so far has
been for uncharged fluids just carrying energy-momentum
corresponding to AdS vacuum gravity. Note that extremal solutions in
pure AdS gravity cannot be supersymmetric. However, the techniques
used in this paper, which rely only on extremality, could be applied
to the study of ``extremal'' fluids. For instance, it would be
interesting to exploit the Gaussian null coordinates to analyze the
entropy current for such degenerate horizons, generalizing the
recent analysis of \cite{Bhattacharyya:2008ta}. It would also be
interesting to develop techniques to analyze the behaviour of
extremal fluids in confining gauge theories as discussed in
\cite{Aharony:2005bm,Lahiri:2007ae}. The latter analysis predicts a
phase diagram for black holes and rings in AdS spacetimes using the
dual field theory; this could provide an interesting test for the
phase diagram proposed in this paper.

\subsection*{Acknowledgements}
\label{acks}

It is a great pleasure to thank Henriette Elvang, Roberto Emparan, Harvey Reall,  Simon Ross and especially Veronika Hubeny for very fruitful discussions on aspects of black holes in higher dimensions.  PF, JL and MR are supported by an STFC Rolling grant, while HKK is supported by a STFC postdoctoral fellowship.

\appendix
\section{Calculating near-horizon limits}
\label{Adetnh}

In this appendix we describe how to construct the near-horizon geometries for the MP black holes and strings given in \sec{mpbhnh} and \sec{mpbsnh}, following the approach developed in~\cite{Kunduri:2007vf}. The starting point for any of these calculations are the MP solutions which, in Boyer-Lindquist type coordinates, are:
\begin{equation}
ds^2=-dt^2+ \sum_{i=1}^n(r^2+a_i^2)\,(d\mu_i^2+\mu_i^2\,d\phi_i^2)+ \frac{\mu \,r^2}{\Pi\, F}\,\left(dt- \sum_{i=1}^na_i\,\mu_i^2\, d\phi_i \right)^2 + \frac{\Pi \,F}{\Pi-\mu \,r^2}\,dr^2
\label{mp2n1}
\end{equation}
in $2n+1$ dimensions and
\begin{equation}
ds^2=-dt^2+ r^2\,d\alpha^2+ \sum_{i=1}^n(r^2+a_i^2)(d\mu_i^2+\mu_i^2\,d\phi_i^2)+ \frac{\mu \,r}{\Pi\, F}\left(dt-\sum_{i=1}^na_i\,\mu_i^2\, d\phi_i \right)^2 + \frac{\Pi\, F}{\Pi-\mu \,r}\,dr^2
\label{mp2n2}
\end{equation}
in $2n+2$ dimensions, where
\begin{equation}
F=1-\sum_{i=1}^{n}\, \frac{a_i^2\, \mu_i^2}{r^2+a_i^2}
\label{Fmpdef}
\end{equation}
and the rest of the functions and coordinates are defined as in the
main text\footnote{We have taken $a_i \to -a_i$ relative to the
original MP metric in~\cite{Myers:1986un}. Without loss of
generality we will assume our $a_i>0$.}. We are also interested in
the MP black strings in $D=2n+3$. These are constructed by taking
the $2n+2$ dimensional MP metric \req{mp2n2} adding $dz^2$ and
boosting $(t,z) \to ( c_{\beta}t-s_{\beta}z \, , - s_{\beta} t+
c_{\beta} z)$ where $c_{\beta} \equiv \cosh \beta$ and $s_{\beta}
\equiv \sinh \beta$.\footnote{Note this is equivalent to changing
coordinates to $(t',z')=( c_{\beta}t+s_{\beta}z \, ,  s_{\beta} t+
c_{\beta} z)$ and then subsequently dropping the ``primed''.} The
explicit metric for this thus reads
\begin{eqnarray}
ds^2 &=&-dt^2+ dz^2+ r^2\,d\alpha^2+ \sum_{i=1}^n(r^2+a_i^2)(d\mu_i^2+\mu_i^2\,d\phi_i^2) \nonumber \\ &&+ \frac{\mu \,r}{\Pi\, F}\left( c_{\beta} dt-s_{\beta} dz- \sum_{i=1}^na_i\,\mu_i^2\, d\phi_i \right)^2 + \frac{\Pi\, F}{\Pi-\mu \,r}\,dr^2 \, .
\label{mpbsstringsol}
\end{eqnarray}

In fact it is convenient to illustrate the method of computing a near-horizon limit for a more general class of ``black-hole like'' metrics. This generalizes the discussion in
\cite{Kunduri:2007vf}. Consider solutions of the form
\begin{eqnarray}\label{genmet}
ds^2 &=& g_{tt}(R,\rho)\, dt^2 + 2\, g_{ti}(R,\rho)\, dt\,d\Phi^i + g_{RR}(R,\rho)\, dR^2 \\
&&  \qquad + g_{pq}(R,\rho)\, d\rho^{p}\,d\rho^{q} + g_{ij}(R,\rho)\, d\Phi^{i}\,d\Phi^{j}, \nonumber
\end{eqnarray}
where $p,q=1, \cdots m$ and $i,j=1, \cdots, n$.\footnote{Of course the integer $m$ is related to $n$ in the solutions we are dealing with but we will leave it free.}.The $\rho^p$ coordinates are a set of ``polar'' angles and the $\Phi^i$ are a set of azimuthal angles. Here $R$ is a radial coordinate and the event horizon is located at $R=0$. We emphasize that all known black hole solutions (in any dimension), including black rings, can be written in the above form.  By shifting the $\Phi^{i}$ by constant multiples of $t$, we may always choose a co-rotating frame in which the Killing vector $\partial_t$ is null on the horizon, which we assume henceforth. For all known extremal black hole solutions it is the case that
\begin{equation}
g_{tt} = f_{t}(\rho)\, R^2 + \CO\left(R^3\right) ,  \qquad
g_{ti} = f_{i}(\rho)\, R + \CO\left(R^2\right) ,\qquad
g_{RR} = f_{R}(\rho)\, R^{-2} + \CO\left(R^{-1}\right)
\end{equation}
 for functions $f_{\mu}$ that are determined from the given solution. To construct the near-horizon limit, we proceed by introducing coordinates valid on the horizon, $(v,r,\phi^{i})$ via
\begin{equation}\label{NHtrans}
R = r, \qquad
dt = dv + \left(\frac{a_{0}}{r^2} + \frac{a_{1}}{r}\right) dr\ , \qquad
d\Phi^{i} = d\phi^{i} + \frac{b_{0}^i}{r} \, dr \ .
\end{equation}
 The constants $a_{0},a_{1},b_{0}$ are fixed by requiring the metric and its inverse be analytic at the horizon $r=0$. Now take the near-horizon limit defined by $v \to v / \epsilon, r\to \epsilon r$ with $\epsilon \to 0$. Referring to the metric that results after taking the limit as $\hat{g}_{\mu\nu}$, we easily obtain the following components, as they are not affected by the transformation~(\ref{NHtrans}):
\begin{equation} \hat{g}_{vv} = f_{t}(\rho)\,r^2, \qquad \hat{g}_{pq} = g_{pq}(0,\rho), \qquad \hat{g}_{vi} = f_{i}(\rho)\,r, \qquad \hat{g}_{ij} = g_{ij}(0,\rho) .
\end{equation}
 Now define $k_{i}=f_{i}(\rho)$. Elimination of the divergent $1/r$  terms in $g_{ri}$ requires
\begin{equation}
b_{0}^i = -a_{0}\,\gamma^{ij}(\rho)\,k_{i} = -a_{0}\,k^{i}.
\end{equation} Consistency obviously requires that the $k^{i}$ be constants. This is certainly  true for the examples we are dealing with and ultimately follows from the fact that we are dealing with solutions to the field equations. Next, it is straightforward to check that, after the taking the near-horizon limit,
\begin{equation}
\Gamma \equiv \hat{g}_{vr} = a_{0}f_{t}(\rho) + b_{0}^{i}k_{i}= a_0(f_t(\rho)-k^ik_i),
\end{equation} and hence now we need only  $a_{0}$ to determine the full near-horizon metric. We now turn to $g_{rr}$. Eliminating the $1/r^2$ term yields the condition
\begin{equation}
a_{0}^2 = \frac{f_{R}(\rho)}{k^i k_i - f_t(\rho)}.
\end{equation} This equation is only consistent if the RHS is a constant and further, since $f_R(\rho)>0$ a second consistency condition is that $k^ik_i-f_t>0$; again, these consistency requirements are met for the solutions we are dealing with. We could now determine $a_{1}$ by removing the $1/r$ divergence in $g_{rr}$. However, note that $dr^2/r$ vanishes in the near-horizon limit, and hence to construct the near-horizon geometry we do not need $a_{1}$. Finally, observe that we may write $\hat{g}_{vv} = f_{t}(\rho)r^2 = (a_{0}^{-1}\Gamma + k^{i}k_{i})r^2$. Putting this all together, we have the following cohomogeneity-$m$ near-horizon geometry:
\begin{equation}\label{genNH}
ds^2 = \Gamma(\rho)[a_{0}^{-1}r^2 dv^2 + 2dvdr ] + \hat{g}_{pq}(\rho)d\rho^{p}d\rho^{q} + \gamma_{ij}(\rho)(d\phi^{i} + rk^{i}dv)(d\phi^{j} + rk^{j}dv).
\end{equation} We choose signs such that $\Gamma >0$, so we are dealing with a future horizon. This implies $a_{0} < 0$.  In this form, it is clear that the near-horizon geometry has an $SO(2,1)\times U(1)^n$ isometry.


\begin{thebibliography}{10}

\bibitem{Emparan:2001wk}
R.~Emparan and H.~S. Reall, ``Generalized Weyl solutions,'' {\em Phys. Rev.}
  {\bf D65} (2002) 084025,
\href{http://www.arXiv.org/abs/hep-th/0110258}{{\tt hep-th/0110258}}.

\bibitem{Emparan:2001wn}
R.~Emparan and H.~S. Reall, ``A rotating black ring in five dimensions,'' {\em
  Phys. Rev. Lett.} {\bf 88} (2002) 101101,
\href{http://www.arXiv.org/abs/hep-th/0110260}{{\tt hep-th/0110260}}.

\bibitem{Emparan:2006mm}
R.~Emparan and H.~S. Reall, ``Black rings,'' {\em Class. Quant. Grav.} {\bf 23}
  (2006) R169,
\href{http://www.arXiv.org/abs/hep-th/0608012}{{\tt hep-th/0608012}}.

\bibitem{Belinsky:1971nt}
V.~A. Belinsky and V.~E. Zakharov, ``Integration of the Einstein Equations by
  the Inverse Scattering Problem Technique and the Calculation of the Exact
  Soliton Solutions,'' {\em Sov. Phys. JETP} {\bf 48} (1978)
985--994.

\bibitem{Belinsky:1979mh}
V.~A. Belinsky and V.~E. Sakharov, ``Stationary Gravitational Solitons with
  Axial Symmetry,'' {\em Sov. Phys. JETP} {\bf 50} (1979)
1--9.

\bibitem{Mishima:2005id}
T.~Mishima and H.~Iguchi, ``{New axisymmetric stationary solutions of
  five-dimensional vacuum Einstein equations with asymptotic flatness},'' {\em
  Phys. Rev.} {\bf D73} (2006) 044030,
\href{http://www.arXiv.org/abs/hep-th/0504018}{{\tt hep-th/0504018}}.

\bibitem{Figueras:2005zp}
P.~Figueras, ``A black ring with a rotating 2-sphere,'' {\em JHEP} {\bf 07}
  (2005) 039,
\href{http://www.arXiv.org/abs/hep-th/0505244}{{\tt hep-th/0505244}}.

\bibitem{Tomizawa:2005wv}
S.~Tomizawa, Y.~Morisawa, and Y.~Yasui, ``{Vacuum solutions of five dimensional
  Einstein equations generated by inverse scattering method},'' {\em Phys.
  Rev.} {\bf D73} (2006) 064009,
\href{http://www.arXiv.org/abs/hep-th/0512252}{{\tt hep-th/0512252}}.

\bibitem{Pomeransky:2006bd}
A.~A. Pomeransky and R.~A. Sen'kov, ``Black ring with two angular momenta,''
\href{http://www.arXiv.org/abs/hep-th/0612005}{{\tt hep-th/0612005}}.

\bibitem{Elvang:2007rd}
H.~Elvang and P.~Figueras, ``Black Saturn,'' {\em JHEP} {\bf 05} (2007) 050,
\href{http://www.arXiv.org/abs/hep-th/0701035}{{\tt hep-th/0701035}}.

\bibitem{Iguchi:2007is}
H.~Iguchi and T.~Mishima, ``Black di-ring and infinite nonuniqueness,'' {\em
  Phys. Rev.} {\bf D75} (2007) 064018,
\href{http://www.arXiv.org/abs/hep-th/0701043}{{\tt hep-th/0701043}}.

\bibitem{Evslin:2007fv}
J.~Evslin and C.~Krishnan, ``{The Black Di-Ring: An Inverse Scattering
  Construction},''
\href{http://www.arXiv.org/abs/arXiv:0706.1231 [hep-th]}{{\tt arXiv:0706.1231
  [hep-th]}}.

\bibitem{Izumi:2007qx}
K.~Izumi, ``Orthogonal black di-ring solution,''
\href{http://www.arXiv.org/abs/arXiv:0712.0902 [hep-th]}{{\tt arXiv:0712.0902
  [hep-th]}}.

\bibitem{Elvang:2007hs}
H.~Elvang and M.~J. Rodriguez, ``Bicycling Black Rings,''
\href{http://www.arXiv.org/abs/0712.2425}{{\tt 0712.2425}}.

\bibitem{Emparan:2008eg}
R.~Emparan and H.~S. Reall, ``{Black Holes in Higher Dimensions},''
\href{http://www.arXiv.org/abs/0801.3471}{{\tt 0801.3471}}.

\bibitem{Obers:2008pj}
N.~A. Obers, ``{Black Holes in Higher-Dimensional Gravity},''
\href{http://www.arXiv.org/abs/0802.0519}{{\tt 0802.0519}}.

\bibitem{Harmark:2004rm}
T.~Harmark, ``{Stationary and axisymmetric solutions of higher- dimensional
  general relativity},'' {\em Phys. Rev.} {\bf D70} (2004) 124002,
\href{http://www.arXiv.org/abs/hep-th/0408141}{{\tt hep-th/0408141}}.

\bibitem{Myers:1986un}
R.~C. Myers and M.~J. Perry, ``Black Holes in Higher Dimensional Space-Times,''
  {\em Ann. Phys.} {\bf 172} (1986)
304.

\bibitem{Emparan:2007wm}
R.~Emparan, T.~Harmark, V.~Niarchos, N.~A. Obers, and M.~J. Rodriguez, ``The
  Phase Structure of Higher-Dimensional Black Rings and Black Holes,'' {\em
  JHEP} {\bf 10} (2007) 110,
\href{http://www.arXiv.org/abs/0708.2181}{{\tt 0708.2181}}.

\bibitem{Hovdebo:2006jy}
J.~L. Hovdebo and R.~C. Myers, ``Black rings, boosted strings and
  Gregory-Laflamme,'' {\em Phys. Rev.} {\bf D73} (2006) 084013,
\href{http://www.arXiv.org/abs/hep-th/0601079}{{\tt hep-th/0601079}}.

\bibitem{Elvang:2006dd}
H.~Elvang, R.~Emparan, and A.~Virmani, ``Dynamics and stability of black
  rings,'' {\em JHEP} {\bf 12} (2006) 074,
\href{http://www.arXiv.org/abs/hep-th/0608076}{{\tt hep-th/0608076}}.

\bibitem{Gauntlett:2002nw}
J.~P. Gauntlett, J.~B. Gutowski, C.~M. Hull, S.~Pakis, and H.~S. Reall, ``All
  supersymmetric solutions of minimal supergravity in five dimensions,'' {\em
  Class. Quant. Grav.} {\bf 20} (2003) 4587--4634,
\href{http://www.arXiv.org/abs/hep-th/0209114}{{\tt hep-th/0209114}}.

\bibitem{Reall:2002bh}
H.~S. Reall, ``Higher dimensional black holes and supersymmetry,'' {\em Phys.
  Rev.} {\bf D68} (2003) 024024,
\href{http://www.arXiv.org/abs/hep-th/0211290}{{\tt hep-th/0211290}}.

\bibitem{Gauntlett:2003fk}
J.~P. Gauntlett and J.~B. Gutowski, ``{All supersymmetric solutions of minimal
  gauged supergravity in five dimensions},'' {\em Phys. Rev.} {\bf D68} (2003)
  105009,
\href{http://www.arXiv.org/abs/hep-th/0304064}{{\tt hep-th/0304064}}.

\bibitem{Gutowski:2004ez}
J.~B. Gutowski and H.~S. Reall, ``{Supersymmetric AdS(5) black holes},'' {\em
  JHEP} {\bf 02} (2004) 006,
\href{http://www.arXiv.org/abs/hep-th/0401042}{{\tt hep-th/0401042}}.

\bibitem{Gutowski:2004yv}
J.~B. Gutowski and H.~S. Reall, ``{General supersymmetric AdS(5) black
  holes},'' {\em JHEP} {\bf 04} (2004) 048,
\href{http://www.arXiv.org/abs/hep-th/0401129}{{\tt hep-th/0401129}}.

\bibitem{Gutowski:2004bj}
J.~B. Gutowski, ``{Uniqueness of five-dimensional supersymmetric black
  holes},'' {\em JHEP} {\bf 08} (2004) 049,
\href{http://www.arXiv.org/abs/hep-th/0404079}{{\tt hep-th/0404079}}.

\bibitem{Gaiotto:2005gf}
D.~Gaiotto, A.~Strominger, and X.~Yin, ``{New connections between 4D and 5D
  black holes},'' {\em JHEP} {\bf 02} (2006) 024,
\href{http://www.arXiv.org/abs/hep-th/0503217}{{\tt hep-th/0503217}}.

\bibitem{Elvang:2004rt}
H.~Elvang, R.~Emparan, D.~Mateos, and H.~S. Reall, ``{A supersymmetric black
  ring},'' {\em Phys. Rev. Lett.} {\bf 93} (2004) 211302,
\href{http://www.arXiv.org/abs/hep-th/0407065}{{\tt hep-th/0407065}}.

\bibitem{Gaiotto:2005xt}
D.~Gaiotto, A.~Strominger, and X.~Yin, ``{5D black rings and 4D black holes},''
  {\em JHEP} {\bf 02} (2006) 023,
\href{http://www.arXiv.org/abs/hep-th/0504126}{{\tt hep-th/0504126}}.

\bibitem{Bena:2004de}
I.~Bena and N.~P. Warner, ``{One ring to rule them all ... and in the darkness
  bind them?},'' {\em Adv. Theor. Math. Phys.} {\bf 9} (2005) 667--701,
\href{http://www.arXiv.org/abs/hep-th/0408106}{{\tt hep-th/0408106}}.

\bibitem{Gauntlett:2004wh}
J.~P. Gauntlett and J.~B. Gutowski, ``{Concentric black rings},'' {\em Phys.
  Rev.} {\bf D71} (2005) 025013,
\href{http://www.arXiv.org/abs/hep-th/0408010}{{\tt hep-th/0408010}}.

\bibitem{Elvang:2004ds}
H.~Elvang, R.~Emparan, D.~Mateos, and H.~S. Reall, ``{Supersymmetric black
  rings and three-charge supertubes},'' {\em Phys. Rev.} {\bf D71} (2005)
  024033,
\href{http://www.arXiv.org/abs/hep-th/0408120}{{\tt hep-th/0408120}}.

\bibitem{Gauntlett:2004qy}
J.~P. Gauntlett and J.~B. Gutowski, ``{General concentric black rings},'' {\em
  Phys. Rev.} {\bf D71} (2005) 045002,
\href{http://www.arXiv.org/abs/hep-th/0408122}{{\tt hep-th/0408122}}.

\bibitem{Elvang:2005sa}
H.~Elvang, R.~Emparan, D.~Mateos, and H.~S. Reall, ``{Supersymmetric 4D
  rotating black holes from 5D black rings},'' {\em JHEP} {\bf 08} (2005) 042,
\href{http://www.arXiv.org/abs/hep-th/0504125}{{\tt hep-th/0504125}}.

\bibitem{Bena:2005ni}
I.~Bena, P.~Kraus, and N.~P. Warner, ``{Black rings in Taub-NUT},'' {\em Phys.
  Rev.} {\bf D72} (2005) 084019,
\href{http://www.arXiv.org/abs/hep-th/0504142}{{\tt hep-th/0504142}}.

\bibitem{Kunduri:2006ek}
H.~K. Kunduri, J.~Lucietti, and H.~S. Reall, ``{Supersymmetric multi-charge
  AdS(5) black holes},'' {\em JHEP} {\bf 04} (2006) 036,
\href{http://www.arXiv.org/abs/hep-th/0601156}{{\tt hep-th/0601156}}.

\bibitem{Kunduri:2006uh}
H.~K. Kunduri, J.~Lucietti, and H.~S. Reall, ``{Do supersymmetric anti-de
  Sitter black rings exist?},'' {\em JHEP} {\bf 02} (2007) 026,
\href{http://www.arXiv.org/abs/hep-th/0611351}{{\tt hep-th/0611351}}.

\bibitem{Kunduri:2007qy}
H.~K. Kunduri and J.~Lucietti, ``{Near-horizon geometries of supersymmetric
  AdS(5) black holes},'' {\em JHEP} {\bf 12} (2007) 015,
\href{http://www.arXiv.org/abs/arXiv:0708.3695 [hep-th]}{{\tt arXiv:0708.3695
  [hep-th]}}.

\bibitem{Sen:2007qy}
A.~Sen, ``Black Hole Entropy Function, Attractors and Precision Counting of
  Microstates,''
\href{http://www.arXiv.org/abs/arXiv:0708.1270 [hep-th]}{{\tt arXiv:0708.1270
  [hep-th]}}.

\bibitem{Bardeen:1999px}
J.~M. Bardeen and G.~T. Horowitz, ``The extreme Kerr throat geometry: A vacuum
  analog of AdS(2) x S(2),'' {\em Phys. Rev.} {\bf D60} (1999) 104030,
\href{http://www.arXiv.org/abs/hep-th/9905099}{{\tt hep-th/9905099}}.

\bibitem{Kunduri:2007vf}
H.~K. Kunduri, J.~Lucietti, and H.~S. Reall, ``{Near-horizon symmetries of
  extremal black holes},'' {\em Class. Quant. Grav.} {\bf 24} (2007)
  4169--4190,
\href{http://www.arXiv.org/abs/arXiv:0705.4214 [hep-th]}{{\tt arXiv:0705.4214
  [hep-th]}}.

\bibitem{Emparan:2004wy}
R.~Emparan, ``{Rotating circular strings, and infinite non-uniqueness of black
  rings},'' {\em JHEP} {\bf 03} (2004) 064,
\href{http://www.arXiv.org/abs/hep-th/0402149}{{\tt hep-th/0402149}}.

\bibitem{Emparan:2006it}
R.~Emparan and G.~T. Horowitz, ``Microstates of a neutral black hole in M
  theory,'' {\em Phys. Rev. Lett.} {\bf 97} (2006) 141601,
\href{http://www.arXiv.org/abs/hep-th/0607023}{{\tt hep-th/0607023}}.

\bibitem{Emparan:2007en}
R.~Emparan and A.~Maccarrone, ``Statistical description of rotating
  Kaluza-Klein black holes,'' {\em Phys. Rev.} {\bf D75} (2007) 084006,
\href{http://www.arXiv.org/abs/hep-th/0701150}{{\tt hep-th/0701150}}.

\bibitem{Horowitz:2007xq}
G.~T. Horowitz and M.~M. Roberts, ``{Counting the Microstates of a Kerr Black
  Hole},'' {\em Phys. Rev. Lett.} {\bf 99} (2007) 221601,
\href{http://www.arXiv.org/abs/arXiv:0708.1346 [hep-th]}{{\tt arXiv:0708.1346
  [hep-th]}}.

\bibitem{Reall:2007jv}
H.~S. Reall, ``Counting the microstates of a vacuum black ring,''
\href{http://www.arXiv.org/abs/arXiv:0712.3226 [hep-th]}{{\tt arXiv:0712.3226
  [hep-th]}}.

\bibitem{Emparan:2008qn}
R.~Emparan, ``{Exact Microscopic Entropy of Non-Supersymmetric Extremal Black
  Rings},''
\href{http://www.arXiv.org/abs/arXiv:0803.1801 [hep-th]}{{\tt arXiv:0803.1801
  [hep-th]}}.

\bibitem{Dabholkar:2006tb}
A.~Dabholkar, A.~Sen, and S.~P. Trivedi, ``Black hole microstates and attractor
  without supersymmetry,'' {\em JHEP} {\bf 01} (2007) 096,
\href{http://www.arXiv.org/abs/hep-th/0611143}{{\tt hep-th/0611143}}.

\bibitem{Astefanesei:2006sy}
  D.~Astefanesei, K.~Goldstein and S.~Mahapatra,
  ``Moduli and (un)attractor black hole thermodynamics,''
  \href{http://www.arXiv.org/abs/hep-th/0611140}{{\tt hep-th/0611140}}.

\bibitem{Astefanesei:2006dd}
D.~Astefanesei, K.~Goldstein, R.~P. Jena, A.~Sen, and S.~P. Trivedi,
  ``{Rotating attractors},'' {\em JHEP} {\bf 10} (2006) 058,
\href{http://www.arXiv.org/abs/hep-th/0606244}{{\tt hep-th/0606244}}.

\bibitem{Astefanesei:2007bf}
  D.~Astefanesei and H.~Yavartanoo,
  ``{Stationary black holes and attractor mechanism},''
  {\em Nucl.\ Phys.\  B} {\bf 794} (2008) 13,
  \href{http://www.arXiv.org/abs/arXiv:0706.1847  [hep-th]}{{\tt arXiv:0706.1847 [hep-th]}}.

\bibitem{Suryanarayana:2007rk}
N.~V. Suryanarayana and M.~C. Wapler, ``{Charges from Attractors},'' {\em
  Class. Quant. Grav.} {\bf 24} (2007) 5047--5072,
\href{http://www.arXiv.org/abs/arXiv:0704.0955 [hep-th]}{{\tt arXiv:0704.0955
  [hep-th]}}.

\bibitem{Hanaki:2007mb}
K.~Hanaki, K.~Ohashi, and Y.~Tachikawa, ``{Comments on charges and near-horizon
  data of black rings},'' {\em JHEP} {\bf 12} (2007) 057,
\href{http://www.arXiv.org/abs/arXiv:0704.1819 [hep-th]}{{\tt arXiv:0704.1819
  [hep-th]}}.

\bibitem{Isenberg:1983cc}
J.~Isenberg and V.~Moncrief, ``Symmetries of Cosmological Cauchy Horizons,''
  {\em Commun. Math. Phys.} {\bf 89} (1983) 387--413.

\bibitem{Hollands:2006rj}
S.~Hollands, A.~Ishibashi, and R.~M. Wald, ``{A higher dimensional stationary
  rotating black hole must be axisymmetric},'' {\em Commun. Math. Phys.} {\bf
  271} (2007) 699--722,
\href{http://www.arXiv.org/abs/gr-qc/0605106}{{\tt gr-qc/0605106}}.

\bibitem{Chrusciel:2005pa}
P.~T. Chrusciel, H.~S. Reall, and P.~Tod, ``{On non-existence of static vacuum
  black holes with degenerate components of the event horizon},'' {\em Class.
  Quant. Grav.} {\bf 23} (2006) 549--554,
\href{http://www.arXiv.org/abs/gr-qc/0512041}{{\tt gr-qc/0512041}}.

\bibitem{Sen:2005wa}
A.~Sen, ``{Black hole entropy function and the attractor mechanism in higher
  derivative gravity},'' {\em JHEP} {\bf 09} (2005) 038,
\href{http://www.arXiv.org/abs/hep-th/0506177}{{\tt hep-th/0506177}}.

\bibitem{Bardeen:1973gs}
J.~M. Bardeen, B.~Carter, and S.~W. Hawking, ``{The Four laws of black hole
  mechanics},'' {\em Commun. Math. Phys.} {\bf 31} (1973)
161--170.

\bibitem{Elvang:2007hg}
H.~Elvang, R.~Emparan, and P.~Figueras, ``{Phases of Five-Dimensional Black
  Holes},'' {\em JHEP} {\bf 05} (2007) 056,
\href{http://www.arXiv.org/abs/hep-th/0702111}{{\tt hep-th/0702111}}.

\bibitem{Hollands:2007qf}
S.~Hollands and S.~Yazadjiev, ``{A Uniqueness theorem for 5-dimensional
  Einstein-Maxwell black holes},''
\href{http://www.arXiv.org/abs/arXiv:0711.1722 [gr-qc]}{{\tt arXiv:0711.1722
  [gr-qc]}}.

\bibitem{Kleihaus:2007kc}
B.~Kleihaus, J.~Kunz, and F.~Navarro-Lerida, ``{Rotating Black Holes in Higher
  Dimensions},''
\href{http://www.arXiv.org/abs/arXiv:0710.2291 [hep-th]}{{\tt arXiv:0710.2291
  [hep-th]}}.

\bibitem{Emparan:2003sy}
R.~Emparan and R.~C. Myers, ``Instability of ultra-spinning black holes,'' {\em
  JHEP} {\bf 09} (2003) 025,
\href{http://www.arXiv.org/abs/hep-th/0308056}{{\tt hep-th/0308056}}.

\bibitem{Myers:1999ps}
R.~C. Myers, ``Stress tensors and Casimir energies in the AdS/CFT
  correspondence,'' {\em Phys. Rev.} {\bf D60} (1999) 046002,
\href{http://www.arXiv.org/abs/hep-th/9903203}{{\tt hep-th/9903203}}.

\bibitem{Townsend:2001rg}
P.~K. Townsend and M.~Zamaklar, ``The first law of black brane mechanics,''
  {\em Class. Quant. Grav.} {\bf 18} (2001) 5269--5286,
\href{http://www.arXiv.org/abs/hep-th/0107228}{{\tt hep-th/0107228}}.

\bibitem{Harmark:2004ch}
T.~Harmark and N.~A. Obers, ``{General definition of gravitational tension},''
  {\em JHEP} {\bf 05} (2004) 043,
\href{http://www.arXiv.org/abs/hep-th/0403103}{{\tt hep-th/0403103}}.

\bibitem{Kastor:2007wr}
D.~Kastor, S.~Ray, and J.~Traschen, ``The First Law for Boosted Kaluza-Klein
  Black Holes,'' {\em JHEP} {\bf 06} (2007) 026,
\href{http://www.arXiv.org/abs/arXiv:0704.0729 [hep-th]}{{\tt arXiv:0704.0729
  [hep-th]}}.

\bibitem{Elvang:2003mj}
H.~Elvang and R.~Emparan, ``Black rings, supertubes, and a stringy resolution
  of black hole non-uniqueness,'' {\em JHEP} {\bf 11} (2003) 035,
\href{http://www.arXiv.org/abs/hep-th/0310008}{{\tt hep-th/0310008}}.

\bibitem{Israel:1970ff}
W.~Israel, ``Source of the Kerr Metric,'' {\em Phys. Rev.} {\bf D2} (1970),
  no.~4, 641--646.

\bibitem{Bhattacharyya:2007vs}
S.~Bhattacharyya, S.~Lahiri, R.~Loganayagam, and S.~Minwalla, ``{Large rotating
  AdS black holes from fluid mechanics},''
\href{http://www.arXiv.org/abs/arXiv:0708.1770 [hep-th]}{{\tt arXiv:0708.1770
  [hep-th]}}.

\bibitem{Bhattacharyya:2007jc}
S.~Bhattacharyya, V.~E. Hubeny, S.~Minwalla, and M.~Rangamani, ``{Nonlinear
  Fluid Dynamics from Gravity},''
\href{http://www.arXiv.org/abs/arXiv:0712.2456 [hep-th]}{{\tt arXiv:0712.2456
  [hep-th]}}.

\bibitem{VanRaamsdonk:2008fp}
M.~Van~Raamsdonk, ``{Black Hole Dynamics From Atmospheric Science},''
\href{http://www.arXiv.org/abs/arXiv:0802.3224 [hep-th]}{{\tt arXiv:0802.3224
  [hep-th]}}.

\bibitem{Bhattacharyya:2008ta}
S.~Bhattacharyya, V.~E. Hubeny, R.~Loganayagam, G.~Mandal, S.~Minwalla,
  T.~Morita, M.~Rangamani, and H.~S. Reall, ``Local fluid dynamical entropy
  from gravity,'' \href{http://www.arXiv.org/abs/arXiv:0803.2526 [hep-th]}{{\tt
  arXiv:0803.2526 [hep-th]}}.

\bibitem{Aharony:2005bm}
O.~Aharony, S.~Minwalla, and T.~Wiseman, ``{Plasma-balls in large N gauge
  theories and localized black holes},'' {\em Class. Quant. Grav.} {\bf 23}
  (2006) 2171--2210,
\href{http://www.arXiv.org/abs/hep-th/0507219}{{\tt hep-th/0507219}}.

\bibitem{Lahiri:2007ae}
S.~Lahiri and S.~Minwalla, ``{Plasmarings as dual black rings},''
\href{http://www.arXiv.org/abs/arXiv:0705.3404 [hep-th]}{{\tt arXiv:0705.3404
  [hep-th]}}.

\end{thebibliography}
\bibliographystyle{utphys}

\providecommand{\href}[2]{#2}\begingroup\raggedright\endgroup

\end{document}